\documentclass{article}

\usepackage{PRIMEarxiv}

\usepackage[utf8]{inputenc} % allow utf-8 input
\usepackage[T1]{fontenc}    % use 8-bit T1 fonts
\usepackage{hyperref}       % hyperlinks
\usepackage{url}            % simple URL typesetting
\usepackage{booktabs}       % professional-quality tables
\usepackage{amsfonts}       % blackboard math symbols
\usepackage{amsmath}
\usepackage{amssymb}
\usepackage{nicefrac}       % compact symbols for 1/2, etc.
\usepackage{microtype}      % microtypography
\usepackage{lipsum}
\usepackage{graphicx}
\usepackage{subfigure}
\usepackage{mathpazo}
\usepackage{sectsty}
\usepackage{tikz}

\usepackage{booktabs}
\usepackage{subcaption}
\usepackage{multirow}
\usepackage{array}
\usepackage{color}
\newcommand{\change}[1]{\textcolor{black}{#1}}

\usepackage{pifont}% http://ctan.org/pkg/pifont

\graphicspath{{media/}}     % organize your images and other figures under media/ folder

\title{A Machine Learning Enabled MDO for Bio-Inspired Autonomous Underwater Gliders
}

\author{
  A. Serani$^{1,\star}$, G. Palma$^1$, J. Wackers$^2$, and M. Diez$^1$\\
  $^1$National Research Council-Institute of Marine Engineering, Rome, Italy\\
  $^2$Ecole Centrale de Nantes, Nantes, France\\
  $^\star$\texttt{andrea.serani@cnr.it} \\
}

\begin{document}

\begin{tikzpicture}[remember picture,overlay]
   % Nodo per il riempimento con trasparenza
   \node [rectangle, fill=cyan, fill opacity=0.5, anchor=north, minimum width=\paperwidth, minimum height=3cm] at (current page.north) {};

   % Nodo separato per il testo, senza trasparenza
   \node [anchor=north, minimum width=\paperwidth, minimum height=3cm, text width=\textwidth, align=center, text height=5ex, text depth=15ex, align=left] at (current page.north) {
     \sffamily\small
     \textbf{This is a preprint submitted to:} \textit{Structural and Multidisciplinary Optimization}
     % \textbf{This is a preprint of the following article:}\\
     % A. Serani and M. Diez, A Scoping Review on Simulation-based Design Optimization in Marine Engineering: Trends, Best Practices, and Gaps. \textit{Archives of Computational Methods in Engineering}, 2024.\\
     % \textbf{The published article is available by following the DOI: \texttt{10.1007/s11831-024-10127-1}, which may differ from this preprint.}
   };
\end{tikzpicture}

\maketitle

\begin{abstract}
The preliminary design of autonomous underwater gliders (AUGs) is intrinsically challenging due to the strong coupling between the external hydrodynamic shape, the hydrostatic balance, the structural integrity, and internal packaging constraints. This complexity is further amplified for bio-inspired configurations, whose rich geometric parametrizations lead to high-dimensional design spaces that are difficult to explore using conventional optimization approaches.
This work presents a machine learning enabled bi-level multidisciplinary design optimization (MDO) framework for the performance-driven design of a manta-ray-inspired AUG. At the upper level, hydrodynamically efficient external geometries are explored in a reduced design space obtained through physics-driven parametric model embedding, which identifies a low-dimensional latent representation directly correlated with the lift, drag, and pressure distributions. At the lower level, a constrained internal sizing problem determines the minimum feasible empty weight by accounting for structural, hydrostatic, geometric, and payload constraints.
To render the resulting bi-level problem computationally tractable, a multi-fidelity surrogate-based optimization strategy is adopted, combining low- and high-fidelity hydrodynamic models with stochastic radial basis function surrogates and adaptive Bayesian sampling. The framework enables efficient exploration of the coupled design space while rigorously managing model uncertainty and computational cost.
The optimized configurations exhibit a 14.7\% improvement in maximum hydrodynamic efficiency and a 12.8\% reduction in empty weight relative to the baseline design, while satisfying all disciplinary constraints. These results demonstrate that the integration of physics-driven dimensionality reduction and multi-fidelity machine learning enables scalable and physically consistent MDO of complex bio-inspired underwater vehicles.
\end{abstract}

% keywords can be removed

%========================================================================
\section{Introduction}
Preliminary design plays a central role in the development of complex engineered systems, as it establishes the architectural layout, geometric configuration, and disciplinary interactions that drive downstream performance and feasibility \cite{crawley2015system}.

In the marine domain, autonomous underwater gliders (AUGs) represent a class of long-endurance vehicles that rely on buoyancy-driven propulsion and extremely low energy consumption \cite{Stephen09,petritoli2024autonomous}. Their design involves reconciling stringent mission, performance, and physical constraints, including hydrodynamic efficiency, structural robustness under high external pressure, neutral buoyancy at depth, internal packaging of sensors and batteries, and stability in unsteady or harsh underwater environments. Traditional conceptual-design strategies for AUGs largely rely on simplified parametric studies, single-fidelity simulations, or expert-driven heuristics \cite{eriksen2001seaglider,sherman2002autonomous,gafurov2015autonomous,sun2015parametric}. Such approaches often lead to suboptimal or overly conservative designs and offer limited capability for systematically exploring high-dimensional, highly nonlinear design spaces.

In contrast, modern aeronautical \cite{martins2013multidisciplinary,martins2021engineering} and naval \cite{serani2022hull,serani2024scoping} design practices have increasingly embraced advanced multidisciplinary design optimization (MDO) workflows, characterized by tight coupling among disciplines, automated exploration of the design space, and extensive use of surrogate modeling, multi-fidelity strategies, and uncertainty quantification \cite{peherstorfer2018survey}. Despite their proven effectiveness, these methods have only recently begun to penetrate the underwater-glider community \cite{li2021shape,he2024hydrodynamic,wang2025wing}. The gap between current AUG design workflows \cite{sun2017shape,wang2024model} and the state-of-the-art practices in aeronautics motivates the development of more advanced, data-driven, and fully integrated frameworks for early-stage marine underwater vehicle design.

Recent progress in machine learning (ML) has introduced several enabling techniques for simulation-based design \cite{li2022machine}. Surrogate modeling approaches---including radial basis functions (RBF), Gaussian processes, and neural networks---allow high-fidelity performance metrics to be approximated rapidly and accurately, enabling global exploration without the need for repeated calls to expensive simulations. Multi-fidelity modeling combines the strengths of both low- and high-fidelity analyses: low-fidelity tools provide inexpensive trend estimates, while high-fidelity models supply accuracy where needed \cite{beachy2021emulator,beachy2024epistemic,beachy2025adaptive}. %Within this context, stochastic RBF (SRBF) surrogates have emerged as an effective option thanks to their flexibility, robustness to noise, and natural ability to provide uncertainty estimates. 
Active learning and Bayesian optimization exploit estimated uncertainty fields of the surrogates to guide sampling toward high-value regions of the design space \cite{serani2019adaptive,difiore2024active}.
Dimensionality reduction represents a third enabling pillar of modern simulation-driven design \cite{serani2025survey}. Approaches such as principal component analysis \cite{d2020design,ccelik2021reduced,li2022low,serani2023parametric}, autoencoders \cite{d2017nonlinear,liu2025cnn,seo2024study}, and physics-informed model embeddings \cite{serani2025extending,gaggero2026physics} can drastically reduce the dimension of the design space for complex shape parameterizations, improving interpretability, reducing surrogate complexity, and mitigating the \textit{curse of dimensionality} \cite{bellman1966dynamic}. This is particularly relevant for biomimetic geometries \cite{fattepur2024bio} such as manta-inspired wings or fuselages, which pose unique challenges \cite{honaryar2018design,hernandez2023design,liu2026large,sun2021design}: their rich geometric variability, especially when paired with internal arrangement constraints and section-wise parametrizations, easily leads to large design spaces that are difficult to sample effectively. %The physics-driven Parametric Model Embedding (PD--PME) method used in this work addresses this challenge by constructing a reduced latent space directly from combined geometric and physical information, ensuring that the retained deformation modes remain strongly correlated with hydrodynamic performance.

The AUG design problem addressed in this study is inherently multidisciplinary. Hydrodynamics governs lift, drag, and flow separation phenomena at low-Reynolds numbers; hydrostatics determines buoyancy and trim; structural mechanics imposes strict limitations on the pressure hull, which must survive external loads up to 1000 m depth; and geometric packaging ensures that the scientific payload, batteries, actuators, and buoyancy engines fit within the available volume while respecting center-of-gravity and stability requirements. 
The strong coupling among these disciplines motivates an MDO architecture based on the bi-level integrated system synthesis (BLISS) \cite{sobieszczanski2002bilevel}: the upper level optimizes the external geometry to maximize hydrodynamic efficiency and minimize empty weight, while a lower-level internal-sizing problem determines the optimal pressure-hull configuration for a given external geometry, subject to structural, geometric, hydrostatic, and packaging constraints. 
In this setting, the lower level acts as a disciplinary feasibility and sizing problem, returning the minimum admissible empty weight associated with each external shape. 
This organization follows the BLISS paradigm for large-scale system synthesis, enhanced through multi-fidelity surrogate coordination and integration (BLISS-2000) \cite{gray2013standard}.

\begin{figure*}[!b]
    \centering
    \includegraphics[width=0.9\linewidth]{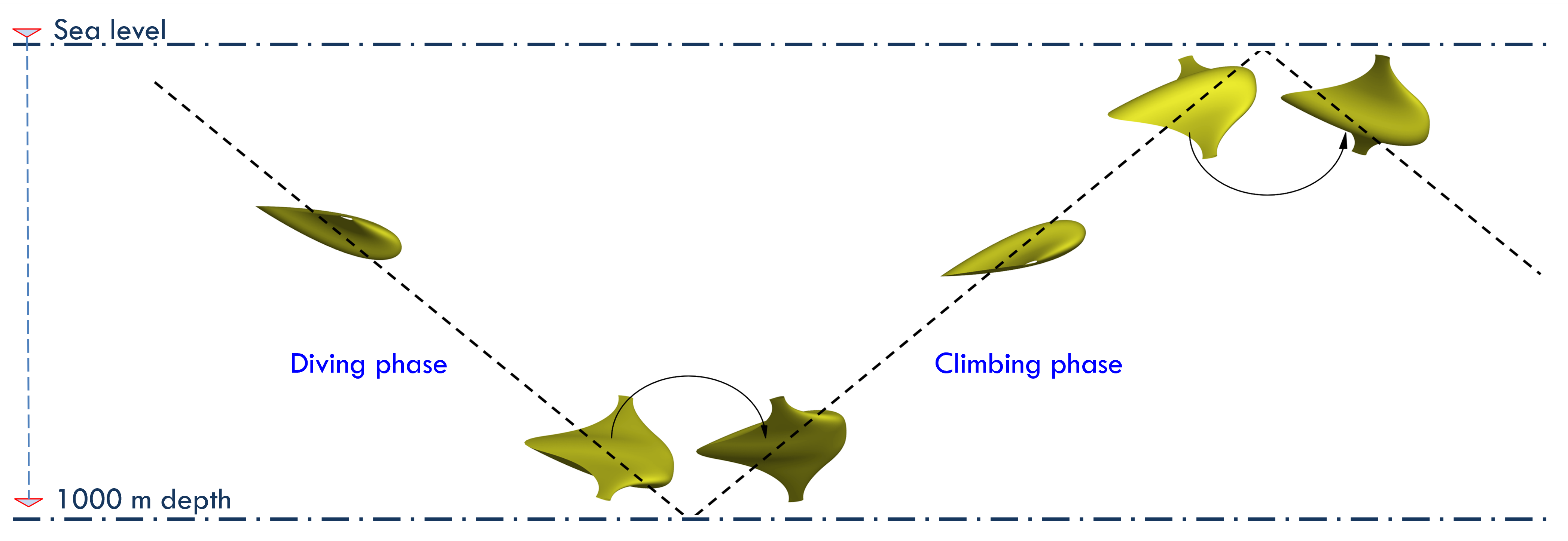}
    \caption{Sketch of AUG saw‑tooth trajectories}
    \label{fig:diveclimb}
\end{figure*}
The present work builds upon preliminary studies on the surrogate-based design of bio-inspired gliders \cite{serani2025preliminary} and extends the analysis and assessment of the results. The main  contributions of this paper are:
\begin{enumerate}
\item A complete mathematical formulation of a bi-level, multi-objective, multi-constraint optimization problem for bio-inspired AUGs, integrating hydrodynamics, hydrostatics, structural mechanics, and internal packaging \change{(Section 3}).
\item A dimensionality reduction strategy based on physics-driven parametric model embedding (PD--PME) \cite{serani2025extending} that constructs a low-dimensional latent space only from physical quantities, improving surrogate accuracy and sampling efficiency \change{(Section 4.1)}.
\item A unified multi-fidelity surrogate modeling framework based on stochastic RBFs (SRBF) \cite{volpi2015development,serani2019adaptive,pellegrini2023multi}, equipped with uncertainty quantification and trained on both low- and high-fidelity flow solvers (potential flow and RANS) \change{(Section 4.2)}.
\item A batch Bayesian optimization strategy tailored for multi-objective problems, which uses expected hypervolume improvement \cite{abdolshah2018expected} in subspaces created through clustering in an augmented feature space, for diverse and informative sampling along the Pareto front \change{(Section 4.3)}.
\item A comprehensive numerical demonstration on a manta-inspired AUG, including Pareto-front evolution, surrogate uncertainty analysis, solver cross-validation, and hydrodynamic interpretation of the optimal design \change{(Section 5)}.
\end{enumerate}

The proposed methodology yields a non-dominated set of bio-inspired designs with significant improvements in both hydrodynamic efficiency and structural weight. \change{In terms of computational effort, the multi-fidelity strategy reduces the estimated high-fidelity cost by more than 95\% with respect to a reference scenario in which the same number of optimization evaluations is performed exclusively using the high-fidelity RANS solver.} This work thus presents a fully integrated, ML-enhanced MDO framework tailored to underwater-glider design and, more broadly, contributes to early-stage shape optimization problems involving complex physics and strong multidisciplinary coupling.

\section{Design problem and physical models}
The AUG considered in this work is designed for long-endurance oceanographic operations in the Mediterranean Sea, whose average depth is about 1500 m (with the deepest point about 5100 m below sea level). 
Specifically, the primary mission profile consists of repeated saw-tooth trajectories (see Fig.~\ref{fig:diveclimb}) between the surface and depths up to 1000 m, at a nominal forward speed $U_\infty = 0.25$ m/s. 
During these vertical cycles, the vehicle must maintain stable trim, robust roll characteristics, and sufficient hydrodynamic efficiency to minimize energy expenditure per unit distance.
As shown in Fig.~\ref{fig:diveclimb}, phase switching (diving to climbing) is modeled by assuming that the vehicle can roll by approximately 180$^\circ$ between descending and ascending legs, thereby flying both legs with the same hydrodynamic incidence and trim attitude. 
Under this operational assumption, the flow solution for the ascent is obtained by exchanging suction and pressure sides, and the MDO problem can be formulated and optimized for a single reference glide condition.

\begin{figure*}[!t]
    \centering
    \includegraphics[width=1\linewidth]{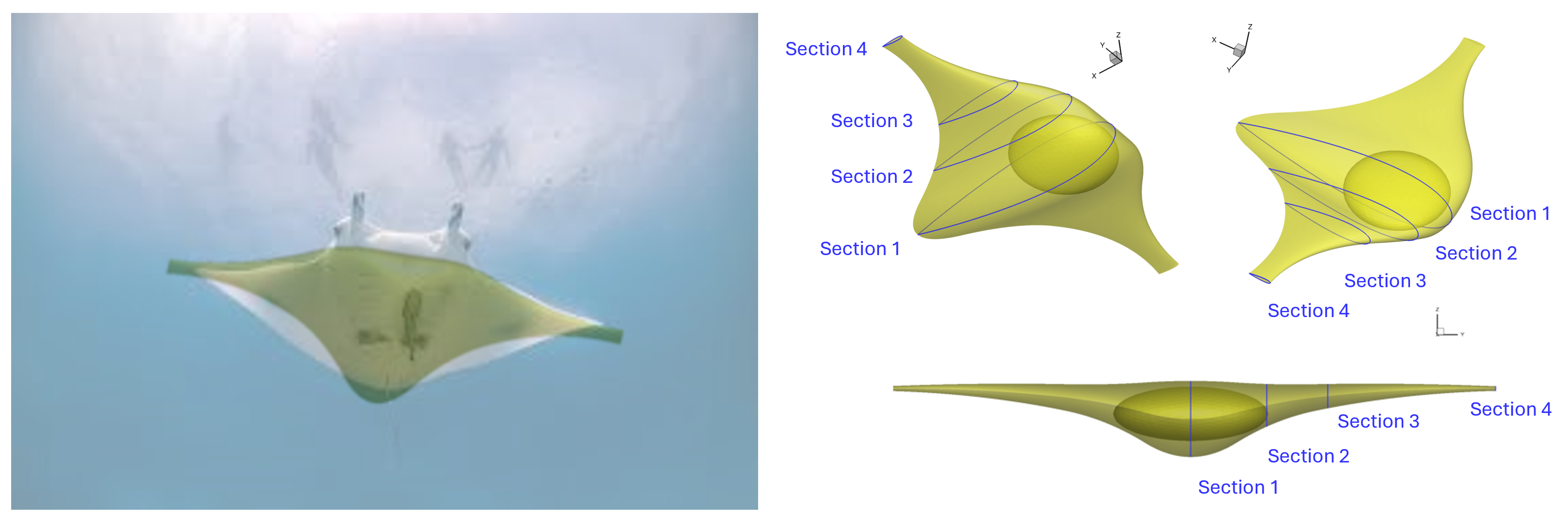}
    \caption{Comparison of AUG shape with manta-ray (left), along with details (right) of the airfoil sections describing the CAD parametrization of the AUG shell and the pressure hull.}
    \label{fig:manta_shape}
\end{figure*}

The targeted scientific payload includes CTD (conductivity, temperature, and density) sensors, dissolved-oxygen modules, optical instruments, navigation units, and battery packs, selected to be representative of typical essential ocean variables (e.g., physical, biogeochemical, ecosystem) \cite{ioc-goos}. 
Electronic units, batteries, and navigation systems are housed within the pressurized hull, while sensing probes (e.g., CTD, dissolved oxygen, and optical sensors) are assumed to be pressure-tolerant and externally mounted in flooded compartments.

From a design perspective, the mission imposes a series of coupled and sometimes conflicting requirements:
\begin{itemize}
\item \textbf{Hydrodynamic performance}, maximize the lift‑to‑drag ratio (L/D) at the operating conditions to reduce energy consumption during vertical motion.
\item \textbf{Structural integrity}, ensure that the pressurized hull withstands hydrostatic loads up to about 10 MPa (corresponding to the target operational depth of 1000 m) without buckling or excessive deformation, as a limit-state requirement appropriate for preliminary design.
\item \textbf{Hydrostatic balance}, guarantee neutral buoyancy at operational depth and positive buoyancy at the surface with empty bladders.
\item \textbf{Packaging feasibility}, accommodate payload and actuation systems within the internal pressure vessel while respecting geometric constraints imposed by the external fuselage.
\item \textbf{Stability}, preserve favorable static and dynamic stability characteristics, particularly roll stability during descending and ascending legs.
\end{itemize}
These requirements frame the multidisciplinary design task addressed in the remainder of the paper.

\subsection{Geometry and design variables}
The external shape is inspired by the manta-ray morphology, characterized by a broad central body and smoothly tapered wings (see Fig.~\ref{fig:manta_shape}, left). The geometry is parameterized through four transverse sections defined via NACA 4-digit based airfoil shapes and spatial transformations (see Fig.~\ref{fig:manta_shape}, right).
Each section includes:
\begin{itemize}
\item airfoil parameters (maximum camber, camber position, thickness ratio, chord),
\item three coordinates defining the leading‑edge position ($x$, $y$, $z$),
\item three angular rotations (twist, roll, yaw).
\end{itemize}
Due to symmetry constraints at the root section and relative leading edge displacement with the other sections, the total number of free design parameters $M$ for the half‑body amounts to 32, with only 2 for the root (section 1 thickness and chord), and all 10 for the other three.
The lofted surface generated from these parametric sections defines the external hydrodynamic shell used for CFD simulations. 
Further details on the construction and definition of the manta parametric model are described in Appendix \ref{appendix:cad}. 

This high‑dimensional parametrization captures the essential deformation mechanisms of the manta-inspired planform but leads to a large search space for direct optimization, even for the construction of a reliable surrogate model.
To address this challenge, the geometric design variables are later reduced into a low‑dimensional latent representation through PD‑PME, ensuring exploration of physically meaningful shape variations.

\subsection{Internal arrangement}
\begin{figure}[!h]
    \centering
    \includegraphics[width=0.5\columnwidth]{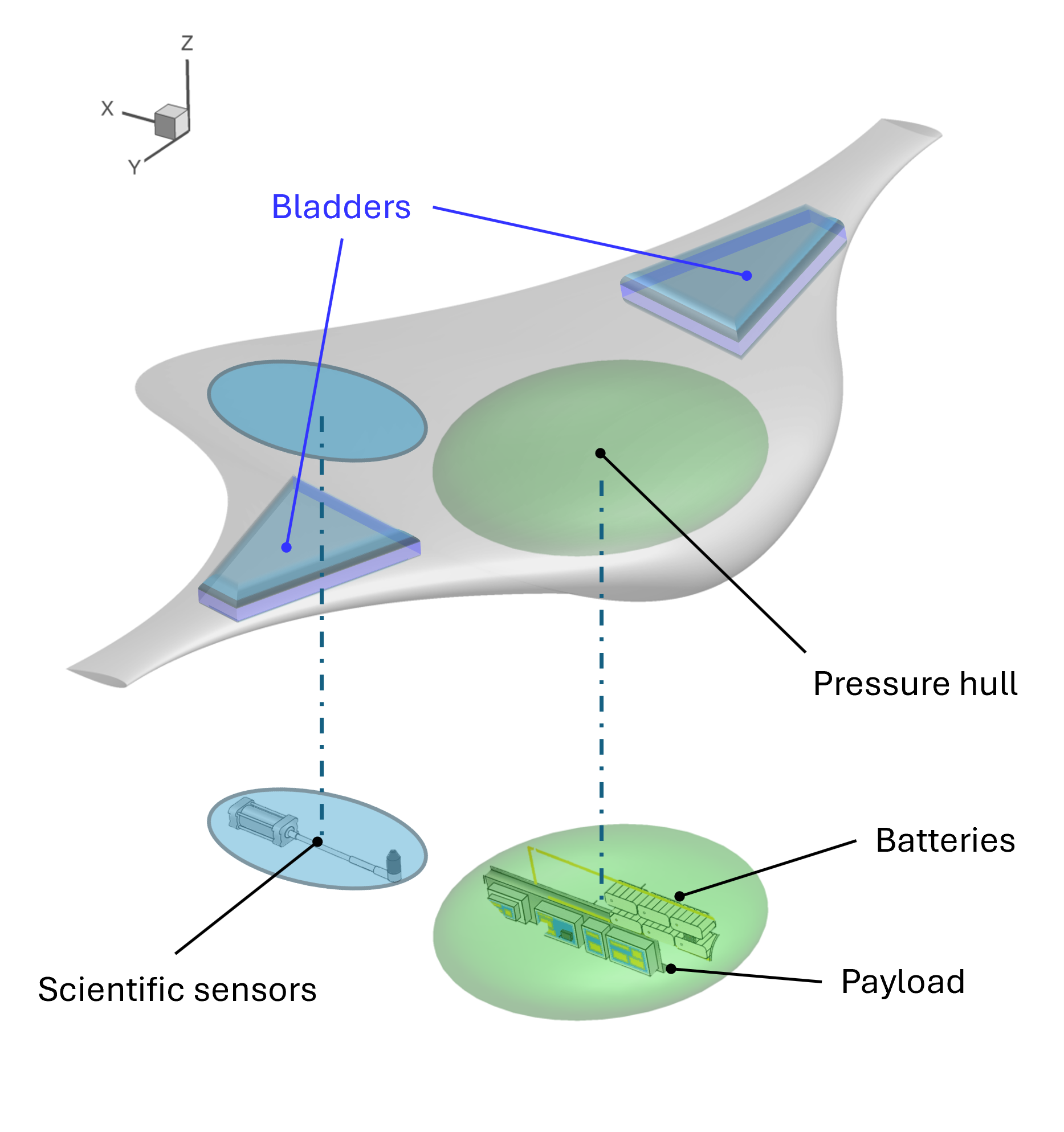}
    \caption{AUG sketch of internal arrangement}
    \label{fig:internal_arrangement}
\end{figure}

The AUG is assumed to be constructed of a solid polyethylene external shell, whose internal architecture is organized around a single ellipsoidal pressure hull located within the manta-shaped body (see Fig.~\ref{fig:internal_arrangement}). 
The external shell is treated as a high-density polyethylene (HDPE) structure (density $\rho_{\text{fill}}=950$~kg/m$^3$), while the pressure hull is modeled as a hollow aluminum vessel (made of 6061-T6 aluminum alloy, with density $\rho_{\text{ph}}=2700$~kg/m$^3$, Young's modulus $E_{\text{ph}}=69$~GPa, and Poisson's ratio $\nu_{\text{ph}}=0.33$) providing a dry environment for pressure-sensitive components. %The filling material is chosen as (HDPE), with density $\rho_{\text{fill}}=950$~kg/m$^3$, while the pressure hull is . The scientific payload is assumed to occupy a volume $V_{\text{sci}}=0.0068$~m$^3$.

The pressure hull contains the batteries, the onboard electronics, and the dry scientific payload, while pressure-tolerant sensing probes and the variable-buoyancy system are located outside the pressure hull in flooded compartments. 

For the purpose of optimization, the dimensions and weight of the pressure vessel are treated as design variables. The weight of the variable-buoyancy system is assumed fixed, while its volume is adjusted to meet hydrostatic requirements.  The weight and volume of the batteries, onboard electronics, scientific payload, and sensing probes are considered fixed.

\subsection{Pressure hull geometry and constraints}
The pressure hull is described by:
\begin{itemize}
\item center coordinates ($\xi_0$, $\zeta_0$),
\item semi‑axes ($a$, $b$, $c$),
\item wall thickness ($t$).
\end{itemize}

The lower‑level optimization problem determines these internal parameters to minimize empty weight while satisfying the following disciplinary constraints:
\begin{itemize}
\item \textbf{Structural constraint}, the pressure hull must resist external hydrostatic pressure without buckling. 
\item \textbf{Hydrostatic equilibrium}, the net difference between weight and buoyant force at mission depth must remain within a prescribed small tolerance for trimming.
\item \textbf{Positive surface buoyancy}, with empty bladders, the glider must be positively buoyant to enable resurfacing for communication.
\item \textbf{Geometric containment}, the internal ellipsoid must remain entirely enclosed within the external manta‑shaped shell, assessed via a pointwise constraint over the outer‑surface mesh.
\item \textbf{Payload accommodation}, the internal volume must exceed the total volume of the dry scientific payload and batteries.
\end{itemize}
These constraints strongly couple internal sizing variables with the external shape, justifying the bilevel formulation adopted in the optimization framework.

\subsection{Low- and high-fidelity simulations} \label{sec_designproblem_sim}
All hydrodynamic and hydrostatic simulations presented in this work are performed under nominal, constant environmental conditions representative of deep-water Mediterranean operations, while neglecting any sea currents. 
Specifically, temperature $T \approx 3^\circ$C, density $\rho \approx 1030$ kg/m$^3$, gravity $g = 9.804$ m/s$^2$, and dynamic viscosity $\mu \approx 0.0012$ kg/(ms) are assumed as fixed reference properties for all fidelity levels, unless otherwise stated, and are not intended to represent depth-dependent environmental variability.

Within this framework, hydrodynamic performance is evaluated through a hierarchical combination of low- and high-fidelity simulations, aimed at balancing computational efficiency and predictive accuracy during optimization.

Low-fidelity computations are carried out using the PUFFIn solver \cite{perali2024performance}, developed at ENSTA Bretagne. The solver is based on a boundary-element potential-flow formulation and provides pressure coefficients, integrated lift, and inviscid drag. Viscous effects are accounted for through corrective models based on local Reynolds numbers and flat-plate approximations. Thanks to its very low computational cost, this model is extensively employed within the surrogate-based optimization loop, enabling rapid exploration of the design space.

Importantly, the PUFFIn solver is exploited at multiple internal fidelity levels by varying the spatial resolution of the boundary-element discretization. A coarser grid configuration is employed as a low-cost data generator for the physics-driven dimensionality-reduction stage, where large ensembles of geometries must be evaluated to extract physically meaningful modes of variation. Conversely, a finer surface grid is adopted during the optimization phase to improve the accuracy of performance evaluations used for surrogate training and infill decisions. This multi-level usage further enhances the computational efficiency of the overall framework while preserving consistency across modeling stages.

High-fidelity simulations are introduced at different stages of the optimization procedure to improve prediction accuracy and to validate low-fidelity trends. Steady incompressible RANS simulations are performed using OpenFOAM \cite{jasak2007openfoam} (version~2212), adopting a $k$--$\omega$ SST turbulence model coupled with $\gamma$--$Re_{\theta}$ transition modeling. This choice is motivated by the relatively low Reynolds number, approximately $2.5\times10^{5}$ at the center body. The solver captures viscous drag contributions, skin-friction distributions, flow separation patterns, wake topology, and accurate lift and drag forces over a range of angles of attack.
Computational meshes include refined boundary-layer discretizations with $y^{+}\approx 1$, ensuring sufficient resolution to accurately represent the manta-shaped platform while avoiding the use of wall functions. High-fidelity RANS data are employed selectively to enhance surrogate-model accuracy in regions of interest and to verify the predictions obtained from low-fidelity simulations.

In addition, unsteady high-fidelity RANS computations are performed using the ISIS-CFD solver, developed at \'Ecole Centrale de Nantes, and are used to further verify both baseline and optimized geometries.

\begin{figure*}[!t]
    \centering
    \includegraphics[width=1\linewidth]{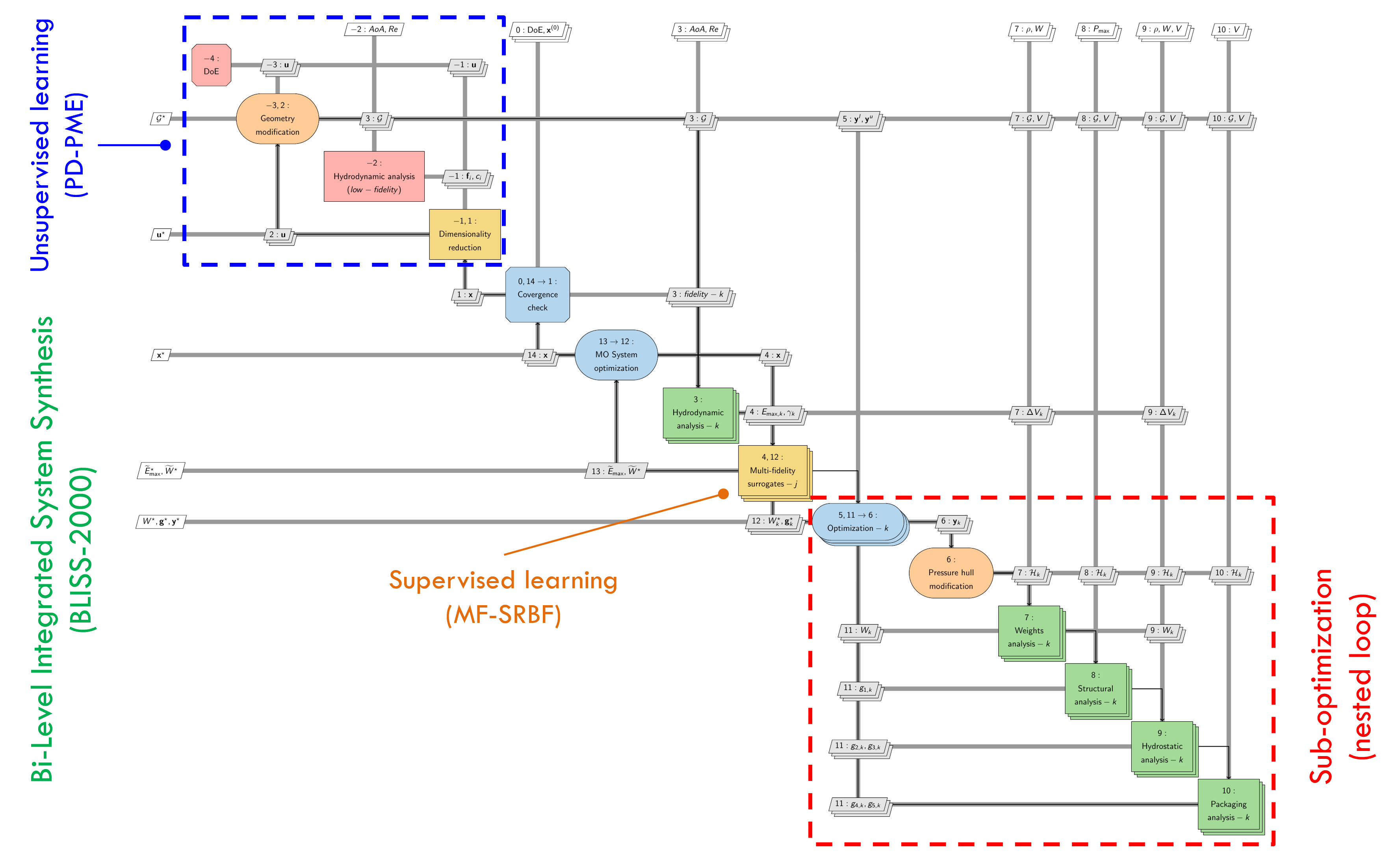}
    \caption{XDSM diagram for the multi-objective multi-fidelity BLISS-2000 workflow}
    \label{fig:xdsm}
\end{figure*}
Further details on the physical solvers, numerical schemes, boundary conditions, and computational grids are provided in Appendix~\ref{appendix:solvers}.

%--------------------------------------------------------------
\section{MDO problem formulation}
The MDO problem addressed in this work is formulated as a bi-level, bi-objective architecture that couples external-shape optimization with internal sizing of the pressure hull and buoyancy system. 
\change{The adoption of a bi-level architecture is motivated by the physical structure of the design problem. The external geometry primarily controls hydrodynamic efficiency, displaced volume, and buoyancy requirements, whereas the internal arrangement determines structural feasibility, hydrostatic balance, geometric containment, and payload accommodation. Although these two sets of variables are strongly coupled, they play fundamentally different roles in the design process. Treating the internal arrangement as a lower-level sizing problem enables the minimum admissible empty weight to be determined for each candidate external geometry while maintaining consistency among disciplines. This organization follows the BLISS philosophy of separating system-level performance exploration from disciplinary sizing and feasibility assessment, with the multidisciplinary coupling preserved through exchanged state variables.}
The upper level explores hydrodynamically efficient geometries in a reduced-order design space, whereas the lower level (solved for each upper-level candidate geometry) determines the minimum feasible empty weight for that geometry by solving a constrained internal sizing problem. Multi-fidelity surrogate models are employed throughout the framework to reduce computational cost while retaining sufficient accuracy. The overall BLISS-2000 architecture and workflow are summarized in Fig.~\ref{fig:xdsm} using an extended design structure matrix (XDSM) representation \cite{lambe2012extensions}.

%--------------------------------------------------------------

\subsection{Upper-level multi-objective problem} \label{sec:mod_upper}

Let $\mathbf{u} \in \mathbb{R}^{M}$ denote the full vector of geometric design variables and $\mathbf{x} \in \mathbb{R}^{N}$ the reduced-order coordinates obtained through PD-PME (see Sec.~\ref{sec:dimred}). For a given $\mathbf{x}$, the corresponding external geometry is reconstructed in the full design space and used to evaluate the displaced volume and the hydrodynamic response at the prescribed operating condition.

The upper-level design problem is formulated as a bi-objective optimization in the reduced-order design space $\mathbf{x} \in \mathcal{X}$, aiming to identify external geometries that maximize hydrodynamic efficiency while minimizing the corresponding minimum empty weight returned by the internal sizing problem (see Sec.~\ref{sec:mod_lower}):
\begin{align}
    \min_{\mathbf{x} \in \mathcal{X}}
    \quad & \left(-E_{\max}(\mathbf{x}), \, W^*_{\text{empty}}(\mathbf{x})\right), \\
    \text{s.t.} \quad & \mathbf{x}_{\min} \le \mathbf{x} \le \mathbf{x}_{\max},
\end{align}
where
\begin{equation}
E(\mathbf{x}) = \frac{L(\mathbf{x})}{D(\mathbf{x})}
\end{equation}
denotes the hydrodynamic efficiency, expressed as the lift-to-drag ratio at the operating condition, and $E_{\max}$ represents its maximum value over the considered range of angles of attack. 
It is evaluated using a multi-fidelity surrogate modeling framework that combines low- and high-fidelity simulations and provides both performance predictions and local uncertainty estimates. These uncertainty measures are exploited to guide adaptive refinement strategies, enabling efficient exploration of the reduced-order design space (see Sec.~\ref{sec:dimred}). A similar surrogate is constructed for $W^*_{\text{empty}}$. 

\change{A Pareto-based formulation is adopted because hydrodynamic efficiency and empty weight represent competing design objectives for which no unique a priori weighting is available at the preliminary design stage. Although single-objective scalarizations, such as weighted-sum or $\epsilon$-constraint formulations, could be used, the multi-objective formulation explicitly exposes the trade-off between performance and weight and provides designers with a family of feasible non-dominated solutions.}

The maximum hydrodynamic efficiency $E_{\max}$ is computed by evaluating the autonomous underwater glider (AUG) polar curve over angles of attack ranging from $-2^\circ$ to $12^\circ$, sampled at $1^\circ$ increments. 
For each design, \change{the 15 computed points are used to construct a smooth polar curve in the $(L,D)$ plane, which is then} post-processed by identifying the tangent passing through the origin of the coordinate system; the slope of this tangent corresponds to the maximum achievable lift-to-drag ratio. 
Although this procedure formally corresponds to a one-dimensional maximization with respect to the angle of attack, it is treated here as \change{a deterministic} post-processing step \change{for each geometry} rather than as an additional optimization level, due to its simplicity and lack of coupling with other design variables\change{: the angle of attack is not included among the MDO design variables, and the surrogate model is trained directly on the scalar response $E_{\max}(\mathbf{x})$.}

In addition to hydrodynamic performance, the upper-level formulation enforces steady-glide force equilibrium at constant speed through an equivalent buoyancy closure. The vehicle is assumed to be globally neutrally buoyant at the operating condition, \textit{i.e.}, the total buoyancy without use of the bladder $B$ and the weight $W$ are in hydrostatic equilibrium ($B - W = 0$), while a controlled buoyancy variation through the bladder is used to sustain steady gliding motion.

Under the assumption of steady motion along a straight glide path, the glide angle $\gamma(\mathbf{x})$ is defined by the balance of hydrodynamic forces, such that
\begin{equation}
\tan\gamma(\mathbf{x}) = -\frac{D(\mathbf{x})}{L(\mathbf{x})}.
\end{equation}

Given $\gamma(\mathbf{x})$, the baseline equivalent buoyancy variation $\Delta V_b(\mathbf{x})$ is defined as the minimum bladder volume variation required to balance the resultant hydrodynamic force during steady gliding at the operating condition,
\begin{equation}\label{eq:dvb}
\Delta V_b(\mathbf{x}) =
\frac{-D(\mathbf{x})\sin\gamma(\mathbf{x}) + L(\mathbf{x})\cos\gamma(\mathbf{x})}
{\rho_{\text{water}}\,g}.
\end{equation}

The quantity $\Delta V_b$ does not represent the total buoyancy capacity of the vehicle, but rather the buoyancy variation required with respect to neutral buoyancy to sustain uniform planing motion for a given external geometry. 
Under the assumed phase-switching strategy, the same magnitude of buoyancy variation is required for both descending and ascending legs, with opposite sign. 
Accordingly, the total buoyancy capacity of the variable-buoyancy system is $2\,\Delta V_b$.

The baseline value $\Delta V_b$ is not treated as an independent design variable, but as a state variable uniquely determined by the hydrodynamic response. 
It is subsequently used as an input to the lower-level internal sizing problem, where additional buoyancy margins are introduced to satisfy hydrostatic equilibrium and surfacing requirements.

%--------------------------------------------------------------

\subsection{Lower-level internal sizing problem} \label{sec:mod_lower}
For each external shape defined by $\mathbf{x}$, the internal design variables
\begin{equation}
    \mathbf{y} = (\xi_0, \zeta_0, a, b, c, t, v_{\text{buo}})
\end{equation}
denote the longitudinal and vertical coordinates of the pressure-hull center, its internal semi-axes, shell thickness, and the additional buoyancy volume required in the bladders to guarantee positive surface buoyancy. These variables $\mathbf{y}$ are optimized to minimize the empty weight $W_{\text{empty}}(\mathbf{x},\mathbf{y})$ while satisfying structural, hydrostatic, geometric, and packaging constraints.

The lower-level problem is formulated as:
\begin{align}
    \min_{\mathbf{y} \in \mathcal{Y}}
    \quad & W_{\text{empty}}(\mathbf{x},\mathbf{y}), \\
    \text{s.t.} \quad &
        g_{\text{struct}}(\mathbf{y}) \le 0, \\
    & g_{\text{hydro}}(\mathbf{x},\mathbf{y}) \le 0, \\
    & g_{\text{surf}}(\mathbf{x},\mathbf{y}) \le 0, \\
    & g_{\text{cont}}(\mathbf{x},\mathbf{y}) \le 0, \\
    & g_{\text{pay}}(\mathbf{y}) \le 0\,\\
    & \mathbf{y}_{\min} \le \mathbf{y} \le \mathbf{y}_{\max}.
\end{align}
The empty weight is estimated as the sum of the pressure hull weight and the weight of the filling material occupying the free volume between the manta shell, the pressure hull, and the buoyancy bladders:
\begin{equation}
W_{\text{empty}}(\mathbf{x},\mathbf{y})=W_{\rm ph}(\mathbf{y})+W_{\rm fill}(\mathbf{x},\mathbf{y}).
\end{equation}

The pressure hull weight is computed from the difference between its external and internal volumes,
\begin{align}
V_\text{ph}^\text{int}(\mathbf{y})&=\frac{4\pi}{3}\,a\,b\,c, \\
V_\text{ph}^\text{ext}(\mathbf{y})&=\frac{4\pi}{3}\,(a+t)(b+t)(c+t),\\
W_{\rm ph}(\mathbf{y})&=\big(V_\text{ph}^\text{ext}(\mathbf{y})-V_\text{ph}^\text{int}(\mathbf{y})\big)\rho_{\text{ph}}\,g,
\end{align}
while the filling material weight is given by
\begin{equation}
W_{\rm fill}(\mathbf{x},\mathbf{y})=\Big(V(\mathbf{x})-V^\text{int}(\mathbf{x},\mathbf{y})\Big)\rho_{\text{fill}}\,g.
\end{equation}
where $V$ denotes the external volume of the vehicle and $V^\text{int}$ is the internal volume occupied by the pressure hull, the bladders, and the scientific payload
\begin{equation}
    V^\text{int}(\mathbf{x},\mathbf{y}) = V_\text{ph}^\text{ext}(\mathbf{y}) + V_{\mathrm{bladd}}(\mathbf{x},\mathbf{y}) + V_\text{sci}
\end{equation}
with $V_{\mathrm{bladd}}$ the effective bladder volume engaged during steady gliding as
\begin{equation}
    V_{\mathrm{bladd}}(\mathbf{x},\mathbf{y}) = \Delta V_b(\mathbf{x}) + v_{\text{buo}}.
\end{equation}
Here, $\Delta V_b(\mathbf{x})$ denotes the buoyancy variation required with respect to neutral buoyancy to sustain steady gliding along a single leg, as determined at the upper level (as explained for Eq.~\ref{eq:dvb}). To enable bidirectional gliding, the buoyancy system must provide an overall stroke covering both descending and ascending phases; this requirement is accounted for through the additional variable $v_{\text{buo}}$, which complements $\Delta V_b$ and includes both the remaining stroke and operational margins associated with trim authority and surfacing capability.
%The HDPE density $\rho_{\text{fill}}=950$~kg/m$^3$, while the pressure hull is made of 6061-T6 aluminium alloy, with density $\rho_{\text{ph}}=2700$~kg/m$^3$, Young's modulus $E_{\text{ph}}=69$~GPa, and Poisson's ratio $\nu_{\text{ph}}=0.33$. 
The scientific payload is assumed to occupy a volume $V_{\text{sci}}=0.0068$~m$^3$.

The constraints $g_i$ enforce structural integrity, hydrostatic equilibrium, positive surface buoyancy, geometric containment, and payload accommodation, as detailed in the following subsections.

This lower-level optimization problem is solved using a single-objective deterministic particle swarm optimizer \cite{serani2016parameter}, enhanced with local search steps \cite{serani2014globally} to ensure efficient convergence in the presence of nonlinear constraints.

\subsubsection{Structural integrity}
The pressure hull must withstand the external hydrostatic pressure corresponding to the maximum operating depth of 1000~m. Structural integrity is assessed by imposing a buckling constraint under uniform external pressure, which represents the governing failure mode for thin-walled submerged shells.

Let denote the semi-axes $(a,b,c)$ of the oblate ellipsoidal pressure hull sorted in decreasing order. This ordering reflects the fact that the smallest radius of curvature $R_{\min}=\left\{\frac{b^2}{a},\frac{c^2}{a},\frac{a^2}{b},\frac{c^2}{b},\frac{a^2}{c},\frac{b^2}{c}\right\}$ governs the onset of elastic instability under external pressure. A closed-form approximation of the critical buckling pressure for an ellipsoidal shell is then given by
\begin{equation}
  P_{\rm cr}(\mathbf{y})=\frac{2E_{\text{ph}}}{\sqrt{3\,(1-\nu_{\text{ph}}^2)}}\left(\frac{t}{R_{\min}}\right)^2.
\end{equation}
This expression provides a conservative estimate of the elastic buckling pressure and is commonly adopted in preliminary design studies of pressure hulls, where computational efficiency and robustness are prioritized over high-fidelity structural analysis \cite{danielson1969buckling,ma2008buckling,smith2008buckling}.

The structural constraint enforces the requirement that the critical pressure exceeds the maximum hydrostatic pressure $P_{\max}$ at the target depth, leading to
\begin{equation}
g_{\rm struct}(\mathbf{y})=\frac{P_{\max}}{P_{\rm cr}(\mathbf{y})}-1\le 0,
\end{equation}
with $P_{\max}\approx 10$~MPa.

\subsubsection{Hydrostatic equilibrium}
The buoyancy force $B(\mathbf{x})$ at the operating depth is modeled in terms of the effective displaced volume contributing to hydrostatic support. 
In the present formulation, the baseline buoyancy variation $\Delta V_b(\mathbf{x})$ represents the minimum buoyancy variation required to sustain steady gliding, as determined at the upper level, while $v_{\text{buo}}$ accounts for additional buoyancy margins associated with trim authority and surfacing capability. 
Both quantities are treated as functional buoyancy requirements of the ballast system rather than as structural displacement volumes. Consequently, they are excluded from the effective displaced volume contributing to hydrostatic buoyancy, which is expressed as
\begin{equation}
 B(\mathbf{x},\mathbf{y})=\rho_{\text{water}}g\,\big(V(\mathbf{x})- V_{\mathrm{bladd}}(\mathbf{x},\mathbf{y})-V_{\text{sci}}\big).
\end{equation}

Hydrostatic equilibrium at the operating depth can be achieved when the difference between the total weight $W(\mathbf{x},\mathbf{y})$ and the buoyancy force vanishes within a prescribed tolerance $\tau$, leading to the constraint
\begin{equation}
g_{\text{hydro}}(\mathbf{x},\mathbf{y}) = \frac{W(\mathbf{x},\mathbf{y}) - B(\mathbf{x},\mathbf{y})}{\tau} - 1 \le 0.
\end{equation}
The total weight is given by
\begin{align}
W(\mathbf{x},\mathbf{y})=&\,W_{\text{empty}}(\mathbf{x},\mathbf{y}) + \\
&(m_\text{sci} + m_\text{pay} + m_\text{bat} + m_\text{buo})\,g, \nonumber
\end{align}
where the fixed masses correspond to scientific sensors ($m_{\text{sci}}=1.0$ kg), payload ($m_{\text{pay}}=2.5$ kg), batteries ($m_{\text{bat}}=8.0$ kg) housed within the pressure hull, and the buoyancy system ($m_{\text{buo}}=7.5$ kg), including bladders, actuators, and pumps.

\subsubsection{Positive surface buoyancy}
With empty bladders (i.e., $V_b=0$ and $v_{\text{buo}}=0$), positive buoyancy at the surface must be guaranteed to ensure resurfacing capability. This requirement is enforced through the constraint
\begin{equation}
g_{\text{surf}}(\mathbf{x},\mathbf{y})
    = \frac{W(\mathbf{x},\mathbf{y})}{\rho_{\text{water}}\,g\,(V(\mathbf{x}) - V_{\text{sci}})} - 1 \le 0.
\end{equation}

\subsubsection{Geometric containment}
The pressure hull must be fully enclosed within the external manta-shaped shell. The constraint is evaluated by computing, for each mesh point $p_i(\mathbf{x})$ of the outer shell and hull center $(\xi_0,\,0,\,\zeta_0)$, the normalized squared distance
\begin{align}
d_i^2(\mathbf{x},\mathbf{y})=&\frac{(p_{i,x}(\mathbf{x})-\xi_0)^2}{(a+t)^2}+\\
&\frac{p_{i,y}^2(\mathbf{x})}{(b+t)^2}+ \nonumber\\
&\frac{(p_{i,z}(\mathbf{x})-\zeta_0)^2}{(c+t)^2}. \nonumber
\end{align}
The pointwise violation is defined as
\begin{equation}
    v_i(\mathbf{x},\mathbf{y})=\max\!\big(0,\,(1+\varepsilon)/d_i^2(\mathbf{x},\mathbf{y})-1\big),
\end{equation}
and the overall containment constraint is obtained by summing violations over all mesh points:
\begin{equation}
g_{\rm cont}(\mathbf{x},\mathbf{y})=\sum_i v_i(\mathbf{x},\mathbf{y})\le 0.
\end{equation}

\subsubsection{Payload accommodation}
Payload and batteries must fit within the internal volume of the pressure hull $V_\text{ph}^\text{int}(\mathbf{y})$. This requirement is enforced through
\begin{equation}
    g_{\text{pay}}(\mathbf{y})
    = \frac{V_{\text{pay}} + V_{\text{bat}}}{V_\text{ph}^\text{int}(\mathbf{y})} - 1 \le 0,
\end{equation}
where the payload and battery volumes, taken from standard instrumentation specifications, are $V_{\text{pay}}=0.0026$ m$^3$ and $V_{\text{bat}}=0.005$ m$^3$, respectively.

\subsection{Coupling between levels}
The bi-level formulation is characterized by a two-way coupling between the external-shape optimization problem and the internal sizing problem, which ensures consistency between hydrodynamic performance, buoyancy requirements, and structural feasibility.

\textbf{Geometry $\rightarrow$ Internal sizing:} for a given reduced-order design vector $\mathbf{x}$, the reconstruction of the external geometry determines the displaced volume $V(\mathbf{x})$, wetted surface, and hydrodynamic forces $(L(\mathbf{x}),D(\mathbf{x}))$ at the prescribed operating condition. Through the steady-glide equilibrium closure enforced at the upper level, these quantities also uniquely define the baseline equivalent buoyancy volume $V_b(\mathbf{x})$, which represents the minimum buoyancy variation that must be provided by the ballast system to sustain steady gliding at constant speed, under the assumption of global hydrostatic equilibrium.

The set of quantities $\{V(\mathbf{x}),\,\,V_b(\mathbf{x})\}$, as well as the geometry, constitutes the input to the lower-level internal sizing problem, where they determine the available internal volume, the buoyancy demand to be accommodated by the ballast system, and the resulting hydrostatic and structural constraints.

\textbf{Internal sizing $\rightarrow$ Upper-level objectives:} for each external geometry, the solution of the lower-level problem provides the minimum feasible empty weight $W^*_{\text{empty}}(\mathbf{x})$, together with the additional buoyancy volume $v_{\text{buo}}$ required to guarantee positive surface buoyancy and trim authority. These quantities directly affect the second objective of the upper-level problem and contribute to shaping the Pareto-optimal trade-off between hydrodynamic efficiency and structural weight.

The coordination between levels is achieved through multi-fidelity surrogate models that approximate the objectives and constraints involved in both subproblems. By selectively combining low- and high-fidelity evaluations, the surrogate-based framework significantly reduces the number of expensive high-fidelity simulations required to explore the coupled design space while preserving consistency between the two optimization levels.

\section{Machine learning methods}

This section describes the machine learning methodologies adopted to enable efficient exploration and optimization of the high-dimensional geometric design space. The proposed framework combines unsupervised learning techniques for design-space reduction with supervised learning models for performance prediction and uncertainty quantification. An adaptive sampling strategy, driven by surrogate-model uncertainty and optimization needs, is employed to iteratively refine the learned models where accuracy is most critical.

%--------------------------------------------------------------
\subsection{Unsupervised learning for design-space reduction}\label{sec:dimred}
PD-PME \cite{serani2025extending} is employed to construct a reduced-order representation of the geometric design space that is driven by physically relevant variability rather than by purely geometric variance. From a machine learning perspective, PD-PME can be interpreted as an unsupervised representation-learning technique in which a low-dimensional latent space is inferred from unlabeled geometric samples augmented with associated physical information.

Given $\mathbf{u}\in\mathbb{R}^{M}$ as the vector of geometric design variables, for a set of $S$ sampled designs $\{\mathbf{u}^{(j)}\}_{j=1}^{S}$, low-fidelity simulations are used to compute distributed physical fields $\mathbf{f}^{(j)}\in\mathbb{R}^{n_f}$ (with $n_f$ related to the fields discretizations) and lumped physical quantities $\mathbf{c}^{(j)}\in\mathbb{R}^{n_c}$, such as forces. These data are assembled into a physics-augmented data matrix
\begin{equation} \label{eq:dimred:augmented_mx}
\mathbf{P} =
\begin{bmatrix}
\mathbf{U} \\
\mathbf{F} \\
\mathbf{C}
\end{bmatrix}
\in \mathbb{R}^{(M+n_f+n_c)\times S},
\end{equation}
where $\mathbf{U}$, $\mathbf{F}$, and $\mathbf{C}$ collect the design variables, distributed fields, and lumped quantities, respectively. Each block is centered by subtracting its empirical mean.

To account for the heterogeneous nature and different physical scales of the data, PD-PME introduces a block-diagonal weighting matrix
\begin{equation}
\mathbf{W} =
\mathrm{diag}
\big(
\mathbf{0},\,
\mathbf{W}_F,\,
\mathbf{W}_C
\big),
\end{equation}
where the geometric block $\mathbf{U}$ is assigned zero weight, ensuring that the embedding is driven exclusively by physical information. The diagonal matrices $\mathbf{W}_F$ and $\mathbf{W}_C$ are constructed from the inverse empirical variances of the corresponding physical observables, thereby enforcing a variance-normalized contribution of each physical component to the embedding.

The reduced embedding is obtained by solving the weighted covariance eigenvalue problem
\begin{equation}
\mathbf{A}\,\mathbf{W}\,\mathbf{Z}
=
\mathbf{Z}\,\boldsymbol{\Lambda},
\qquad
\mathbf{A}=\frac{1}{S}\mathbf{P}\mathbf{P}^{\mathsf{T}},
\end{equation}
where $\boldsymbol{\Lambda}$ contains the eigenvalues $\lambda_k$, and the columns of $\mathbf{Z}$ are the eigenvectors associated with decreasing physically weighted variance $\mathbf{z}_k = [\mathbf{v}_k^\mathsf{T}, \boldsymbol{\phi}_k^\mathsf{T}, \boldsymbol{\pi}_k^\mathsf{T}]^{\mathsf{T}}$ with $\mathbf{v}_k$, $\boldsymbol{\phi}_k$, and $\boldsymbol{\pi}_k$ the eigenvector components associated to original design variable and distributed and lumped physical quantities.

The reduced dimensionality $N$ is selected such that a prescribed fraction $\eta$ of the total weighted variance is retained,
\begin{equation}
\frac{\sum_{k=1}^{N}\lambda_k}{\sum_{k}\lambda_k}\ge\eta,
\end{equation}
providing a principled and data-driven criterion for truncation.
The eigenvector block associated with the geometric variables defines the reduced-order basis $\mathbf{V}\in\mathbb{R}^{M\times N}$, which enables the analytical back-mapping from the reduced to the original design space,
\begin{equation}
\mathbf{u}\approx \langle\mathbf{u}\rangle + \mathbf{V}\mathbf{x},
\qquad
\mathbf{x}\in\mathbb{R}^{N}.
\end{equation}
and guarantees geometric consistency and interpretability of the reduced representation.

In the present work, PD-PME is used to define the reduced design space explored by the upper-level optimization, ensuring that surrogate modeling and adaptive sampling are concentrated along physically relevant directions. 
Although the required number of samples remains moderate and does not scale prohibitively with the original design-space dimension, a sufficiently large ensemble of geometries is nevertheless needed to robustly identify the dominant embedded directions. For this reason, PUFFIn simulations with coarse surface discretizations are employed to efficiently generate the data required by PD-PME  (Sec.~\ref{sec_designproblem_sim}). 
Further details on the mathematical formulation of PD-PME can be found in \cite{serani2023parametric,serani2025extending,gaggero2026physics}.

\change{The PD-PME workflow adopted in this work can be reproduced using the PME-toolkit, a Python implementation of PME workflows publicly available on Zenodo~\cite{serani_2026_19068340} and distributed through PyPI.}

%--------------------------------------------------------------
%-----------------------------------------------------------
\subsection{Supervised multi-fidelity surrogate modeling}

The performance quantities $E_{\max}$ and $W^{\star}_\text{empty}$ (Sec.~\ref{sec:mod_upper}) are approximated using supervised multi-fidelity surrogate models that combine low-cost potential-flow simulations with selective high-fidelity RANS evaluations. The objective is to learn accurate and uncertainty-aware mappings between the reduced design variables $\mathbf{x}$ and the performance metrics of interest, while minimizing the overall computational cost.

Let $T_\ell = \{(\mathbf{x}_j, f_\ell(\mathbf{x}_j))\}_{j=1}^{N_\ell}$ denote the training dataset available at fidelity level $\ell$, with $\ell=1$ corresponding to the lowest-fidelity model and $\ell=2$ corresponding to the high-fidelity. The multi-fidelity surrogate is constructed as
\begin{equation}
\hat{f}(\mathbf{x}) =
\widetilde{f}_1(\mathbf{x}) +
\widetilde{\varepsilon}(\mathbf{x}),
\end{equation}
where $\widetilde{f}_1$ and $\widetilde{\varepsilon}$ are independent SRBF \cite{volpi2015development} regressors for the low-fidelity and discrepancy between fidelity levels, respectively.
The discrepancy is defined as
\begin{equation}
\varepsilon(\mathbf{x}_j) =
f_2(\mathbf{x}_j) - \widetilde{f}_1(\mathbf{x}_j).
\end{equation}
This additive formulation enables systematic correction of low-fidelity trends while preserving their global structure.

Each surrogate employs an RBF kernel of the form
\begin{equation}
\phi(\|\mathbf{x}-\mathbf{x}_j\|;\,\epsilon) =
\|\mathbf{x}-\mathbf{x}_j\|^{\epsilon},
\end{equation}
where the shape parameter $\epsilon$ is treated as a stochastic variable. Sampling $\epsilon$ from a prescribed distribution $\sim \mathcal{U}(1,3)$ allows the surrogate to capture epistemic uncertainty associated with kernel selection and data sparsity.

To mitigate overfitting and ensure numerical robustness, Tikhonov regularization \cite{golub1999tikhonov} is introduced by solving the penalized least-squares problem
\begin{equation}
\min_{\mathbf{w},\mathbf{c}}
\;
\|\mathbf{A}\mathbf{w}+\mathbf{P}\mathbf{c}-\mathbf{y}\|^2
+\mu\|\mathbf{w}\|^2,
\end{equation}
which leads to the augmented linear system, to be solved in the least-squares sense:
\begin{equation}
\begin{bmatrix}
\mathbf{A} & \mathbf{P} \\
\sqrt{\mu}\mathbf{I} & \mathbf{0}
\end{bmatrix}
\begin{bmatrix}
\mathbf{w} \\ \mathbf{c}
\end{bmatrix}
=
\begin{bmatrix}
\mathbf{y} \\ \mathbf{0}
\end{bmatrix}.
\end{equation}

Uncertainty estimates are computed independently for each fidelity level and combined assuming uncorrelated contributions,
\begin{equation}
U_{\hat{f}}(\mathbf{x}) =
\sqrt{
U_{\widetilde{f}_1}^2(\mathbf{x}) +
U_{\widetilde{\varepsilon}}^2(\mathbf{x})
}.
\end{equation}
These uncertainty fields play a central role in driving adaptive sampling, fidelity allocation, and multi-objective optimization.

Further details on the SRBF formulation and multi-fidelity architecture can be found in \cite{volpi2015development,pellegrini2023multi}.

%-----------------------------------------------------------
\subsection{Batch multi-objective Bayesian optimization}
The exploration of the reduced design space is driven by a batch multi-objective Bayesian optimization strategy based on the expected hypervolume improvement (EHVI) \cite{abdolshah2018expected,yang2019multi}. Given the current approximation of the Pareto set $P_t$ computed from the surrogate, and a reference point $\mathbf{r}$, the EHVI at a candidate design $\mathbf{x}$ is defined as the potential for improvement, given the estimated probability of the objective functions in that point:
\begin{equation}
\mathrm{EHVI}(\mathbf{x}) =
\mathbb{E}_{\mathbf{y}\sim\mathcal{N}(\hat{\mathbf{f}}(\mathbf{x}),\boldsymbol{\Sigma}(\mathbf{x}))}
\Big[
H(P_t \cup \{\mathbf{y}\}) - H(P_t)
\Big],
\end{equation}
where $\hat{\mathbf{f}}(\mathbf{x})$ denotes the surrogate mean prediction, $\boldsymbol{\Sigma}(\mathbf{x})$ is a diagonal covariance matrix assembled from surrogate uncertainty estimates, and $H(\cdot)$ is the hypervolume indicator \cite{zitzler1999multiobjective}.

At each optimization iteration $t$, candidate points are generated by solving multiple optimization problems on the surrogate models. First, the surrogate predictions of the objective functions are used to compute an approximation of the current Pareto front in the reduced space,
\begin{equation}
{P}_t = \Big\{ \mathbf{x} \in \mathcal{X} \, \mid \, \nexists\, \mathbf{x}' : \hat{\mathbf{f}}(\mathbf{x}') \prec \hat{\mathbf{f}}(\mathbf{x})\Big\},
\end{equation}
where $\hat{\mathbf{f}}(\mathbf{x})$ denotes the vector of surrogate-predicted objectives and $\prec$ indicates Pareto dominance.

Then, to promote diversity among the selected infill points, the predicted Pareto front $\mathcal{P}_t$ is subdivided. For this, it is embedded into an augmented feature space that combines design variables and objective values, $\big[{\mathbf{x}},\;\hat{\mathbf{f}}(\mathbf{x})\big]$,
where both design variables and objective functions are normalized to ensure comparable weighting between design-space and objective-space information. Clustering is then performed directly on $\mathcal{P}_t$ using a $k$-means algorithm \cite{burkardt2009k}. The number of clusters $k_t$ is selected by maximizing the average Silhouette score \cite{rousseeuw1987silhouettes} and defines the batch size at iteration $t$.

Finally, for each cluster $C_j$, a representative infill point is selected by solving
\begin{equation}
\mathbf{x}_j^\star =
\arg\max_{\mathbf{x} \in C_j}
\big\{
\mathrm{EHVI}(\mathbf{x})
\big\},
\end{equation}
This strategy ensures a balanced trade-off between exploitation of high-performing regions of the design space and exploration across the full extent of the predicted Pareto front.

%-----------------------------------------------------------
\subsection{Adaptive fidelity allocation}
Each candidate point selected by the batch Bayesian optimization procedure is assigned a simulation fidelity level based on an uncertainty-to-cost criterion. Specifically, the optimal fidelity level for a given design $\mathbf{x}_j^*$ is chosen as
\begin{equation}
\ell^\star(\mathbf{x}_j^*) =
\arg\max_{\ell}
\frac{
U^{(\ell)}_{\mathrm{agg}}(\mathbf{x}_j^*)
}{
c_\ell
},
\end{equation}
where $c_\ell$ denotes the computational cost associated with fidelity level $\ell$.

The aggregated uncertainty at fidelity $\ell$ is defined as
\begin{equation}
U^{(\ell)}_{\mathrm{agg}}(\mathbf{x}) =
\sqrt{
\sum_{m}
\big(U^{(\ell)}_m(\mathbf{x})\big)^2
},
\end{equation}
where the index $m$ spans all objective functions. This strategy allocates high-fidelity simulations only where they are expected to provide the largest information gain relative to their computational expense, while relying on low-fidelity evaluations elsewhere.

Overall, the combination of supervised multi-fidelity learning, uncertainty-aware Bayesian optimization, and adaptive fidelity allocation enables an efficient and scalable exploration of the reduced design space.

% -----------------------------------------------------------
\subsection{Workflow Summary}

The complete optimization workflow is:

\begin{enumerate}
    \item Generate a large initial dataset via Sobol sampling in the 
            reduced PD-PME space.
    \item Evaluate low-fidelity simulations for all samples; evaluate
            high-fidelity simulations selectively.
    \item Train multi-fidelity SRBF surrogates of objectives and constraints.
    \item Perform batch Bayesian optimization using EHVI and clustering.
    \item Assign fidelity to each candidate using uncertainty-to-cost ratios.
    \item Update the surrogate model and repeat until Pareto convergence.
\end{enumerate}

\change{Within this adaptive process, surrogate reliability in the Pareto-relevant region is improved iteratively by enriching the training set where new evaluations are expected to be most informative for the multi-objective search. At each iteration, the current surrogate models provide both mean predictions and uncertainty estimates for the objectives and constraints. Candidate points are selected on the predicted Pareto front using the EHVI-based batch strategy described above, while the fidelity level is assigned according to the uncertainty-to-cost criterion. The newly evaluated samples are then added to the training set and the multi-fidelity SRBF models are retrained. The enrichment process is terminated when the uncertainty associated with the predicted Pareto front becomes sufficiently narrow, particularly in the region of the selected optimum.}

This integrated framework provides a computationally efficient means to explore
highly non-linear, multi-physics design spaces while maintaining reliable
uncertainty quantification.

\section{Numerical results}

This section presents the numerical evaluation of the proposed multi-fidelity,
multi-objective optimization framework. The analysis covers four main aspects:
(i) assessment of the reduced-order design space generated by PD--PME,
(ii) surrogate accuracy across fidelity levels,
(iii) convergence of the multi-objective search, and
(iv) characterization of the obtained optimal configuration relative to the
baseline manta-inspired geometry.

\begin{figure*}[!b]
    \centering
    \subfigure[Lift force]{\includegraphics[width=0.32\linewidth]{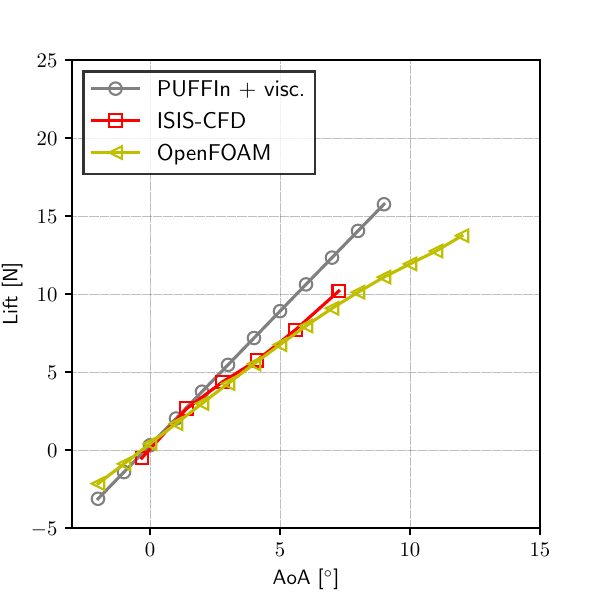}}
    \subfigure[Drag force]{\includegraphics[width=0.32\linewidth]{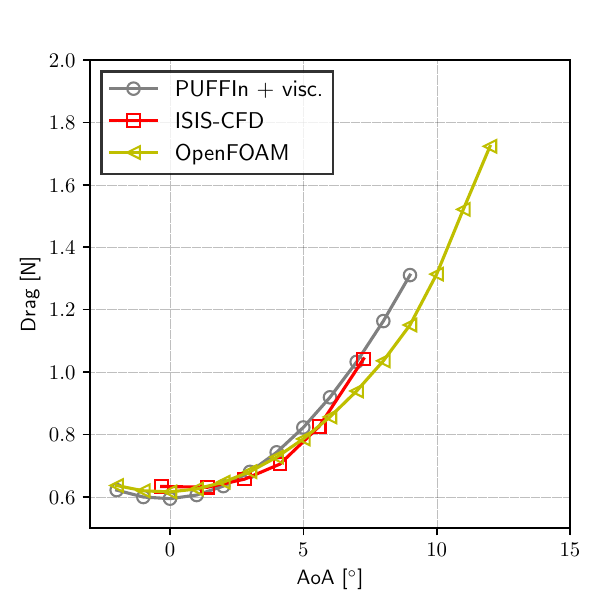}}
    \subfigure[Polar plot]{\includegraphics[width=0.32\linewidth]{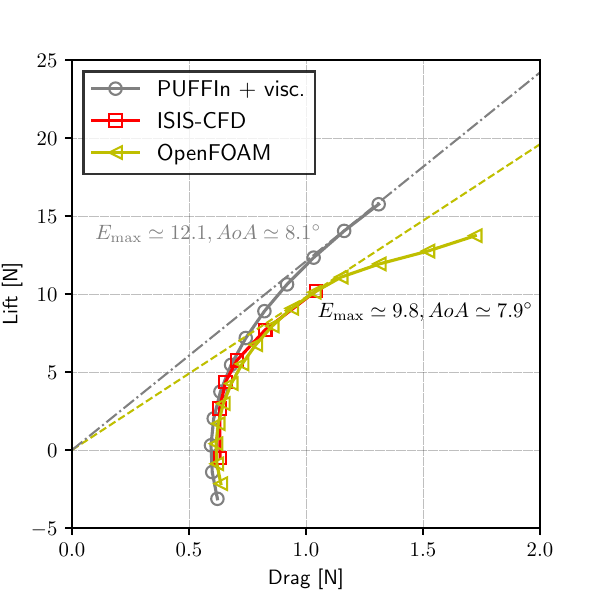}}
    \caption{Comparison of hydrodynamic performance for manta-ray AUG configuration}
    \label{fig:manta_perform}
\end{figure*}

\begin{figure*}[!t]
    \centering
    \includegraphics[width=0.19\linewidth]{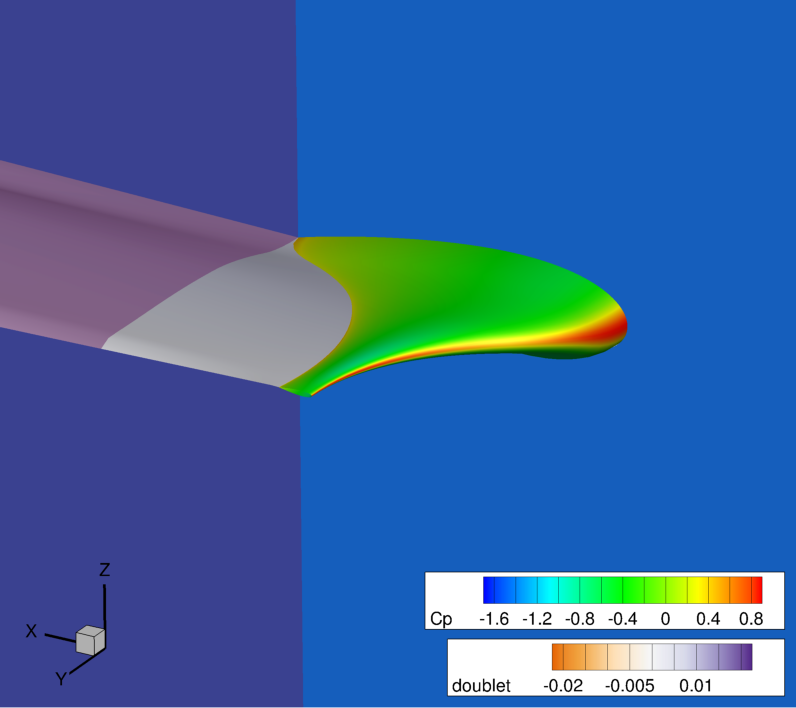}
    \includegraphics[width=0.19\linewidth]{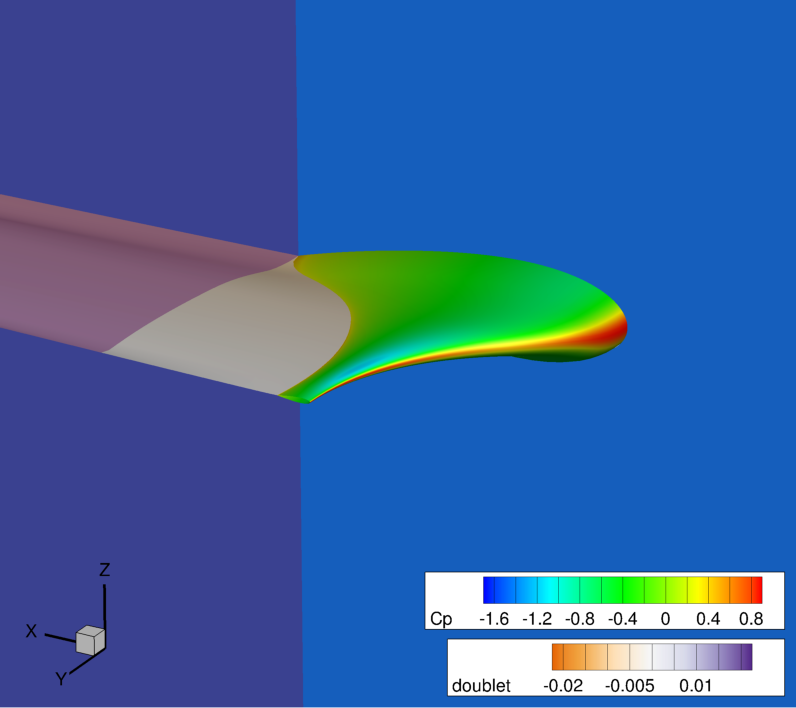}
    \includegraphics[width=0.19\linewidth]{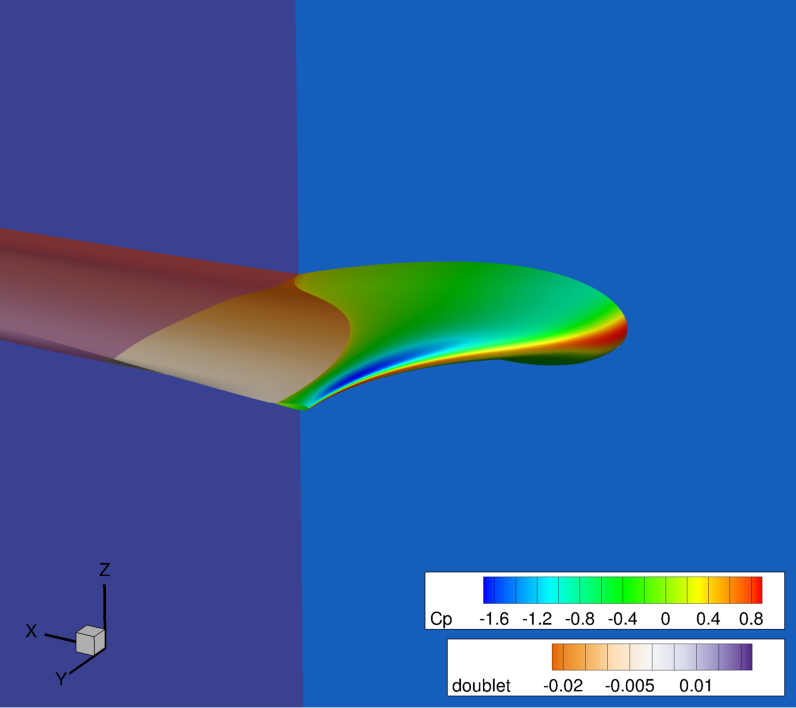}
    \includegraphics[width=0.19\linewidth]{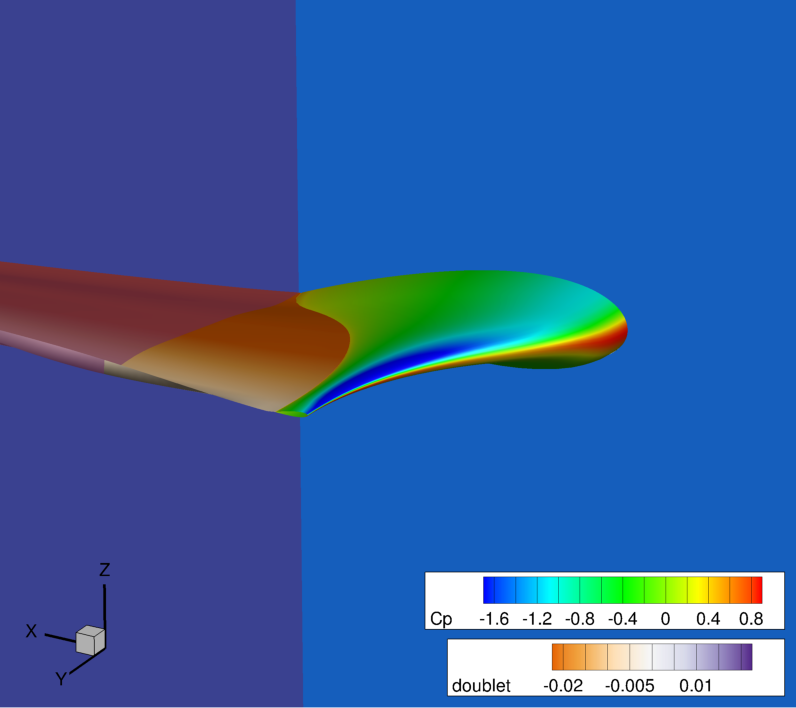}
    \includegraphics[width=0.19\linewidth]{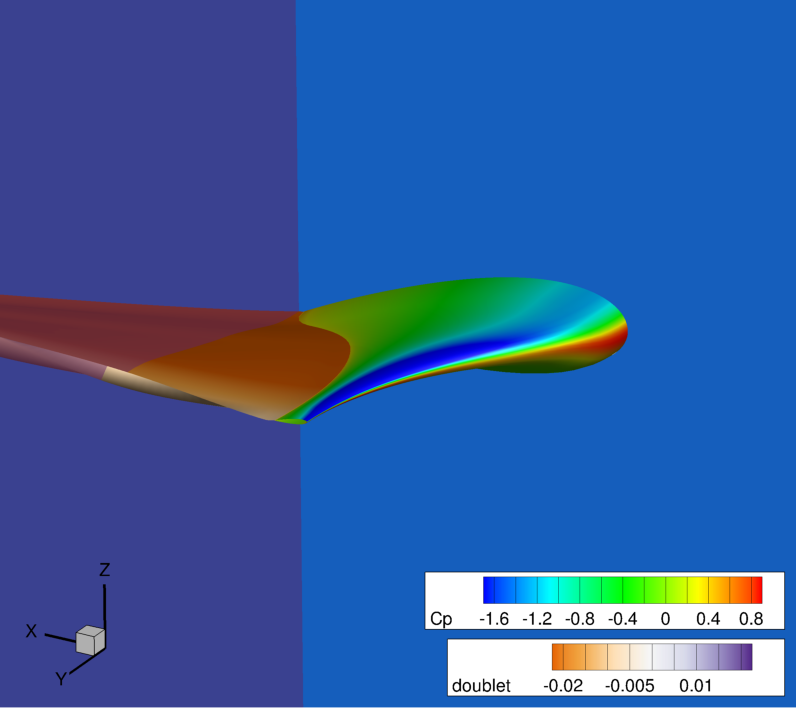}
    \subfigure[$AoA=-2^\circ$]{\includegraphics[width=0.19\linewidth]{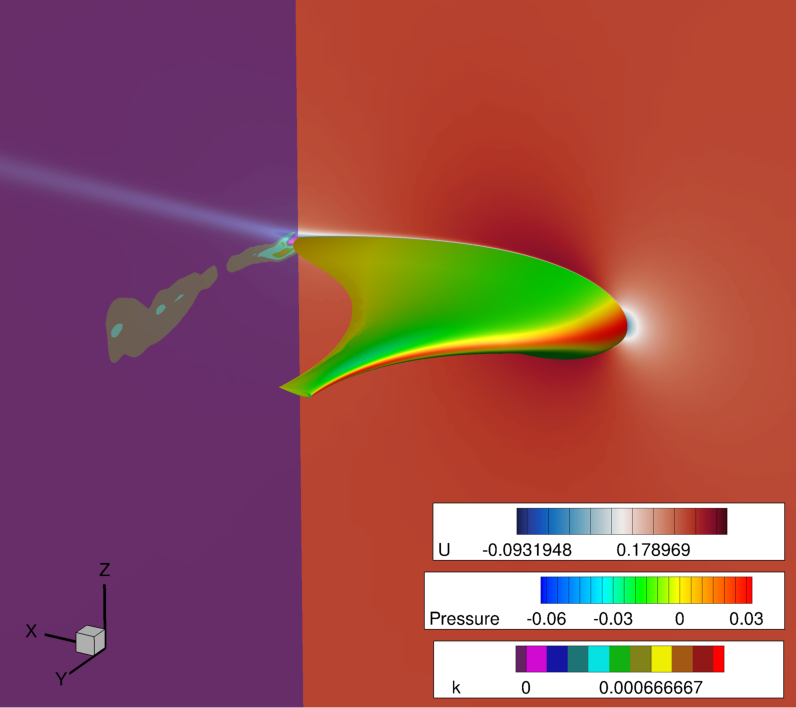}}
    \subfigure[$AoA=1^\circ$]{\includegraphics[width=0.19\linewidth]{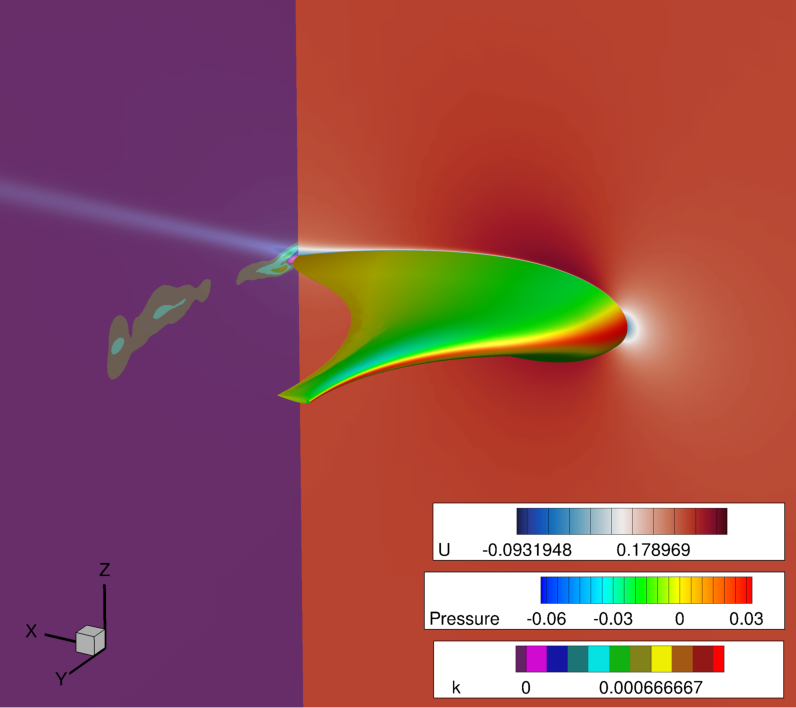}}
    \subfigure[$AoA=4^\circ$]{\includegraphics[width=0.19\linewidth]{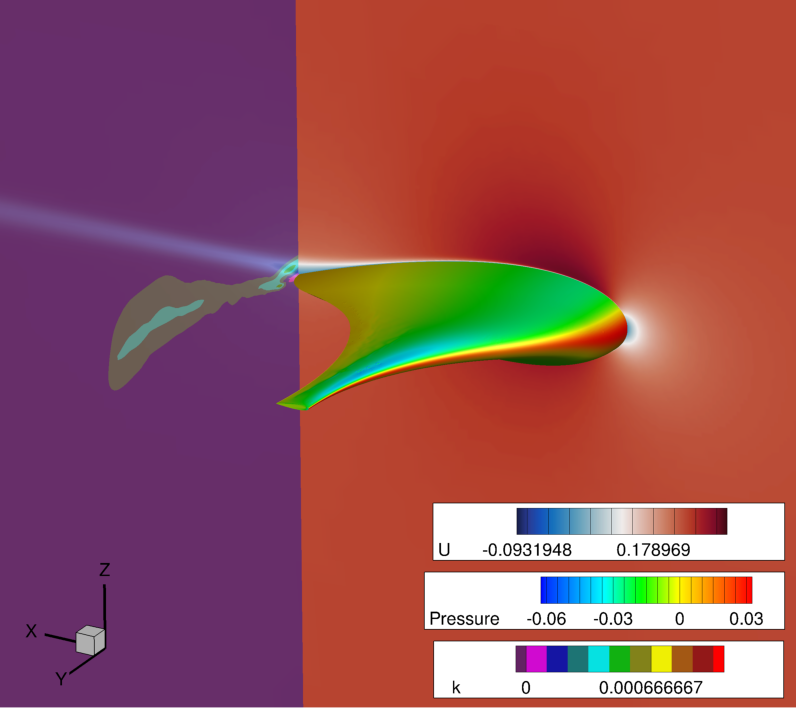}}
    \subfigure[$AoA=7^\circ$]{\includegraphics[width=0.19\linewidth]{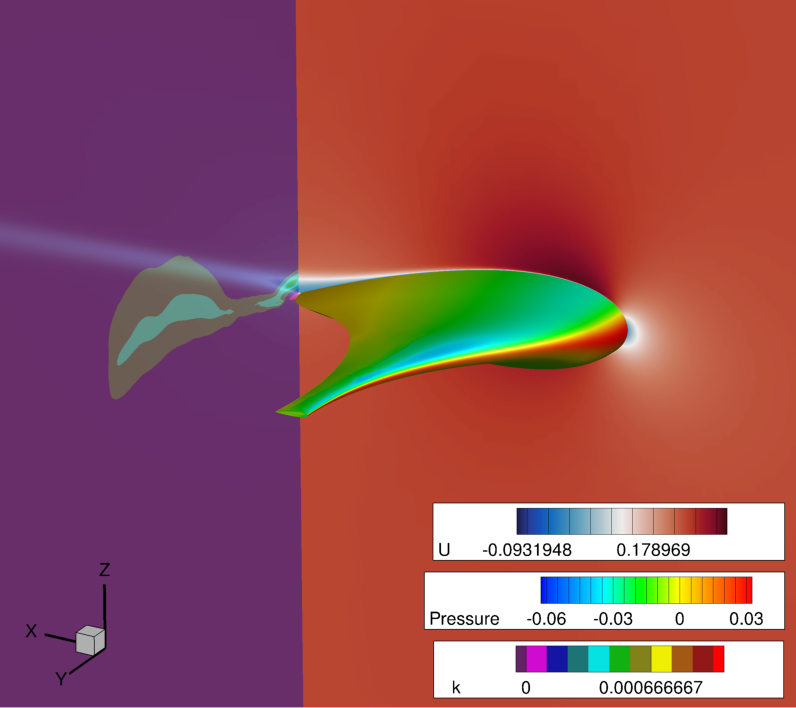}}
    \subfigure[$AoA=10^\circ$]{\includegraphics[width=0.19\linewidth]{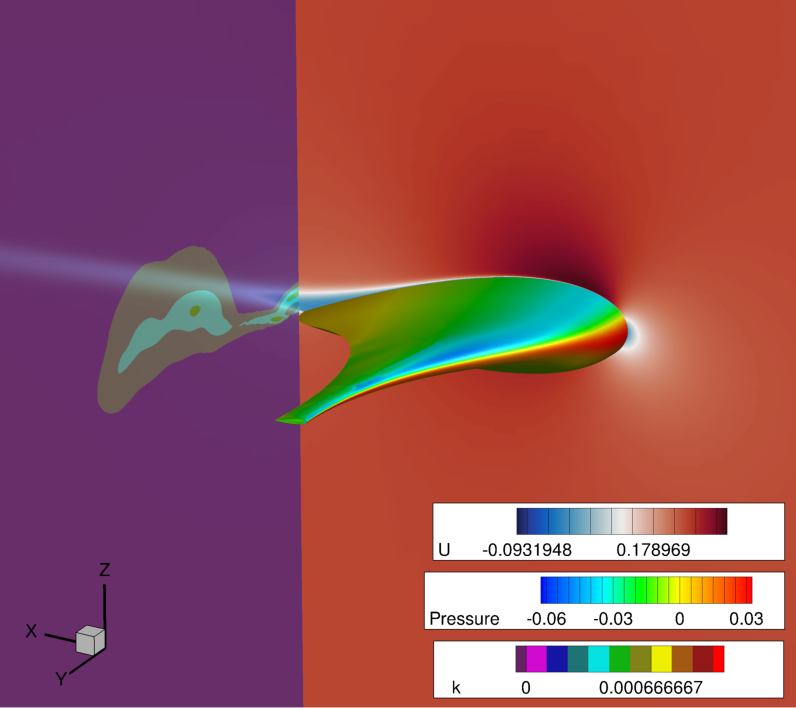}}\\
    \caption{Comparison of pressure distribution and wake for manta-ray AUG configuration: (top) PUFFIn and (bottom) OpenFOAM results}
    \label{fig:manta_fields}
\end{figure*}

\subsection{Baseline hydrodynamic assessment}
Before performing dimensionality reduction, a preliminary hydrodynamic analysis was conducted on the baseline manta-ray-inspired AUG configuration to identify representative operating conditions and assess the consistency among the numerical solvers employed in the multi-fidelity framework. The baseline geometry corresponds to the bio-inspired blended-wing-body configuration introduced in previous work \cite{serani2025preliminary}, featuring a wingspan of approximately $2$~m and a center-body length of $1$~m. All simulations were performed at a nominal cruise speed of $U_\infty=0.25$~m/s, corresponding to a Reynolds number $Re\approx2.5\times10^{5}$ based on the center-body chord.

Three solvers with increasing modeling fidelity and computational cost were used to assess the hydrodynamic performance of the baseline configuration. The low-fidelity solver PUFFIn computes the potential flow solution with a nonlinear Kutta condition and includes a panel-wise viscous correction based on flat-plate friction correlations. Medium- and high-fidelity simulations were performed using incompressible RANS solvers: OpenFOAM, employing a $k$--$\omega$ SST turbulence model coupled with $\gamma$--$Re_{\theta}$ transition modeling, and ISIS--CFD, performing unsteady RANS (URANS) simulations with the same turbulence closure but without transition modeling, on adaptively refined meshes.

Figure~\ref{fig:manta_perform} reports the lift, drag, and hydrodynamic efficiency curves obtained over an angle-of-attack range from $-2^\circ$ to $12^\circ$. PUFFIn and OpenFOAM provide closely matching predictions at low angles of attack ($-2^\circ \leq \mathrm{AoA} \leq 3^\circ$), where the flow remains largely attached and viscous effects are moderate. At higher incidence, the potential solver progressively overpredicts lift, as expected, due to the absence of explicit separation modeling. The OpenFOAM results are corroborated by the URANS simulations performed with ISIS--CFD up to approximately $8^\circ$, beyond which unsteady separation effects increasingly affect convergence and solution stability.

The lift-to-drag ratio exhibits a clear maximum at an angle of attack of approximately $\mathrm{AoA}^*\approx8^\circ$ for all solvers. The corresponding maximum efficiency is estimated to be around $L/D\approx12$ for the potential solver and $L/D\approx10$ for the RANS simulations. This operating condition is therefore selected as the reference point for subsequent analyses, as it corresponds to a physically meaningful steady-gliding regime and is representative of high-efficiency operation during both descending and ascending phases of the mission profile.

The surface pressure distributions and wake velocity fields shown in Fig.~\ref{fig:manta_fields} further support this choice. At low angles of attack, PUFFIn and OpenFOAM predict very similar pressure patterns and wake structures. As the incidence increases, the RANS simulations reveal thicker wakes and the onset of flow separation in the wing–body junction and outer-wing regions, which are not captured by the potential-flow model. 

All physical quantities employed for physics-driven dimensionality reduction are therefore evaluated at the reference angle of attack $\mathrm{AoA}^*\approx8^\circ$. Although this choice restricts the physical information used by PD-PME to a single operating condition, it ensures consistency across the dataset and focuses the embedding on variations that are most directly correlated with peak hydrodynamic efficiency. The implications of this assumption are further discussed in the context of the optimization results.

Finally, the computational cost associated with the full polar evaluation of the baseline configuration provides a quantitative measure of the fidelity--cost trade-off exploited in the optimization framework. The complete polar computation requires approximately $680$~s on $128$ CPUs for PUFFIn and about $7300$~s for OpenFOAM, corresponding to a cost ratio of roughly $1{:}10$. This ratio motivates the adoption of a multi-fidelity strategy in which low-fidelity simulations are used extensively for exploration and surrogate training, while high-fidelity evaluations are reserved for selective refinement and validation.

% -----------------------------------------------------------
\subsection{Design space dimensionality reduction results}
The PD–PME dimensionality reduction was trained using an initial dataset of $S = 16,385$ low-fidelity samples generated via Sobol sequences in the original 32-dimensional geometric design space.
For each sampled geometry, hydrodynamic quantities were evaluated using the PUFFIn potential-flow solver with viscous correction, providing distributed pressure coefficients on the glider surface, together with the integrated lift and drag forces as lumped quantities. All physical quantities used to construct the PD-PME data matrix were evaluated at the reference angle of attack of $8^\circ$, corresponding to the maximum hydrodynamic efficiency of the baseline configuration identified in the preliminary analysis. Extending the PD-PME formulation to multiple operating conditions is straightforward, but was deemed unnecessary for the present performance-driven optimization. \change{The dataset used for the dimensionality-reduction analysis is publicly available through Zenodo~\cite{serani_2026_18936594}.}

Prior to dimensionality reduction, the dataset was filtered to remove unfeasible or non-physical configurations. Geometries leading to degenerate shapes, solver divergence, or undefined outputs were discarded. This preliminary screening already provides a first validation of the chosen parameterization, highlighting geometric variations that are incompatible with the underlying physics or numerically unstable.
A second filtering stage was applied to exclude extreme outliers in the physical responses, retaining only samples whose outputs fell within the range $[Q_1-3\mathrm{IQR},Q_3+3\mathrm{IQR}]$, with $Q_1$ and $Q_3$ denoting the first and third quartiles, respectively. This distribution-agnostic criterion was intentionally permissive, ensuring that only clearly pathological samples were removed while preserving a broad variability across the design space.

The retained geometric and physical data were assembled into the augmented data matrix required by the PD–PME formulation (Eq.~\eqref{eq:dimred:augmented_mx}), where distributed pressure fields and lumped force coefficients drive the embedding, while the original design variables are embedded with zero weight to preserve analytical back-mapping. This construction ensures that the resulting reduced coordinates represent directions of maximum physical variance, rather than purely geometric variability.

\begin{figure}[!b]
    \centering
\includegraphics[width=0.7\linewidth]{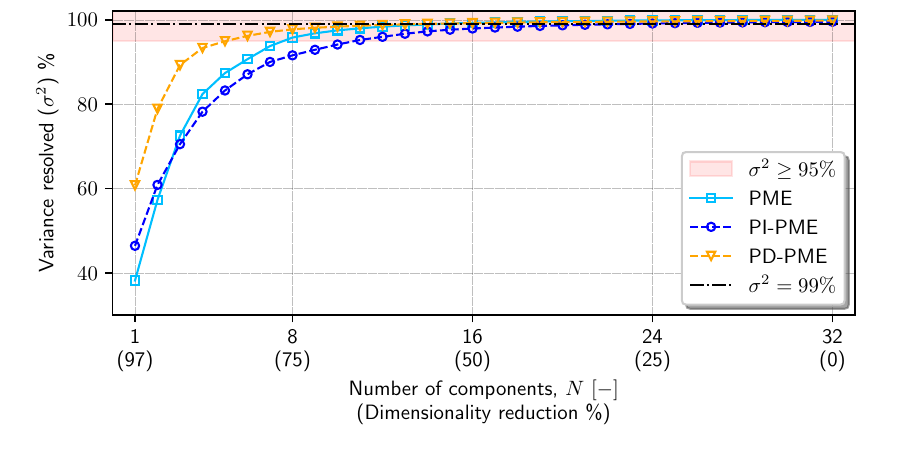}\\
    \caption{Variance resolved as a function of the reduced design variables}
    \label{fig:eigsums}
\end{figure}
\begin{figure*}[!t]
    \centering
    \includegraphics[width=0.49\linewidth]{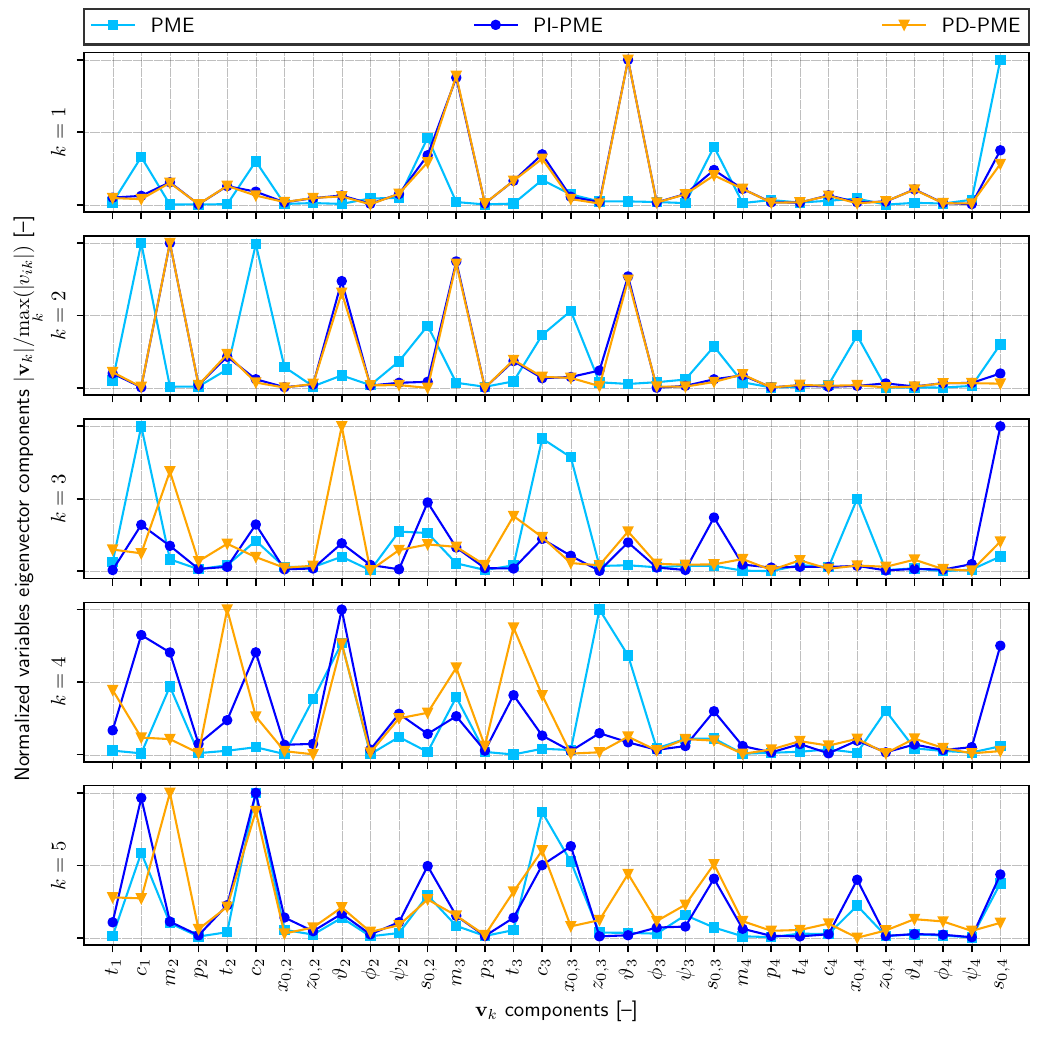}
    \includegraphics[width=0.49\linewidth]{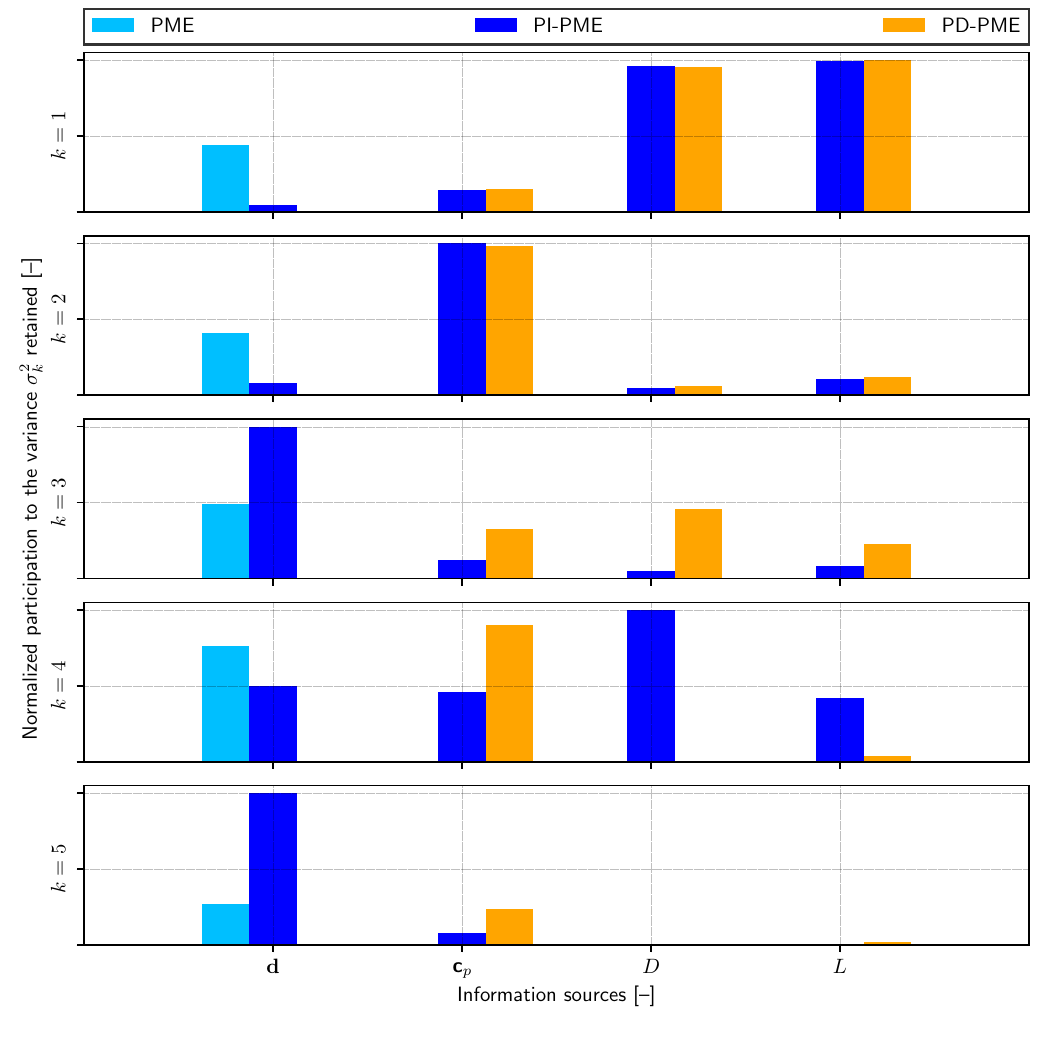}
    \caption{Design-space dimensionality reduction (left) eigenvectors $\mathbf{v}_k$ that embed the original design variables and (right) participation to the variance retained by each eigenvector of geometrical and physical information.}
    \label{fig:glider_eigs}
\end{figure*}
Figure~\ref{fig:eigsums} compares the cumulative variance captured by three PME formulations \cite{serani2025extending}, namely standard PME, physics–informed PME (PI-PME), and PD–PME, as a function of the number of retained modes. Standard PME, which is based only on geometric information,  reproduces 95\% of the geometric variance with approximately $N \approx 8$ modes, reflecting the richness of the manta-shaped geometry and its section-wise parameterization. When geometric and physical quantities are combined in PI–PME, the number of modes required to reach the same variance threshold increases to $N \approx 11$, indicating a partial decoupling between geometric variability and hydrodynamic response. In contrast, the purely physics-based PD–PME achieves 95\% of the physical variance with only $N = 5$ modes.

This strong compression highlights a key characteristic of the underwater glider design space: large geometric variations do not necessarily induce proportional changes in lift, drag, or pressure distribution. While PME primarily captures span-wise and sectional shape changes that dominate geometric variance, PD–PME filters out these directions when they are weakly correlated with performance. As a result, the reduced PD–PME space isolates a small number of directions that are most influential for hydrodynamic efficiency.

Inspection of the PD–PME eigenvectors (see Fig.~\ref{fig:glider_eigs}) reveals that the leading modes are strongly associated with global force coefficients and pressure-distribution patterns, rather than with localized geometric fluctuations. In particular, variations in camber and pitch in the transition region between the center body and the outer wing, which modify the spanwise lift distribution,  dominate the first modes, consistent with classical hydrofoil theory and with the physical interpretation of lift generation in manta-inspired platforms. Purely geometric modes associated with tip-span variations or local thickness perturbations appear only in higher-order components and are therefore naturally filtered out when a performance-driven reduction is sought.

Based on these results, the subsequent optimization was carried out exclusively in the five-dimensional PD–PME latent space. Analytical back-mapping was used to reconstruct full-order geometries from reduced coordinates, enabling consistent evaluation of hydrodynamic performance and structural constraints. This choice represents a deliberate trade-off: by sacrificing a small fraction of geometric variance, the optimization focuses on directions that are most relevant for hydrodynamic efficiency, substantially reducing the effective search space without compromising physical fidelity.

\begin{figure*}[!t]
    \centering
    \subfigure[Iter = 0]{\includegraphics[width=0.329\linewidth]{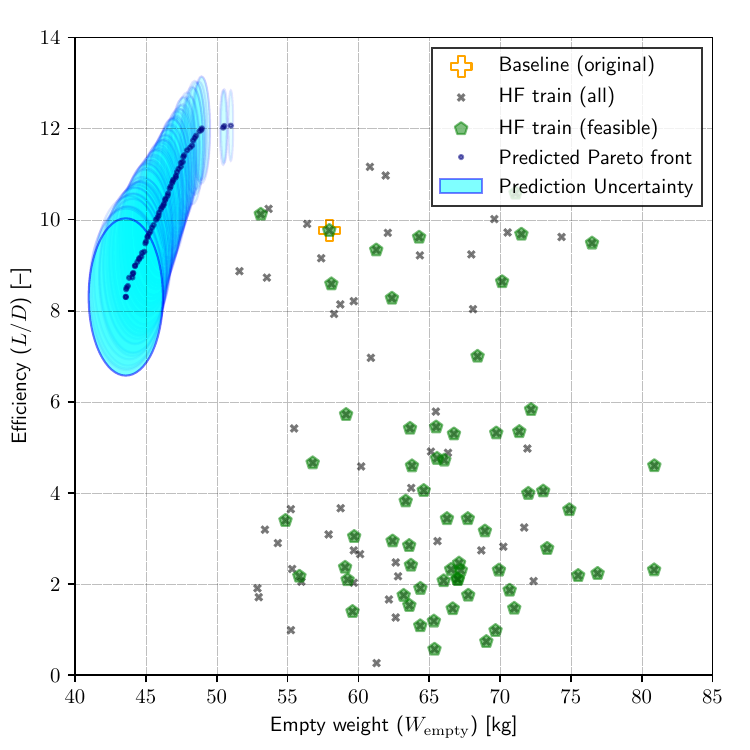}}
    \subfigure[Iter = 2]{\includegraphics[width=0.329\linewidth]{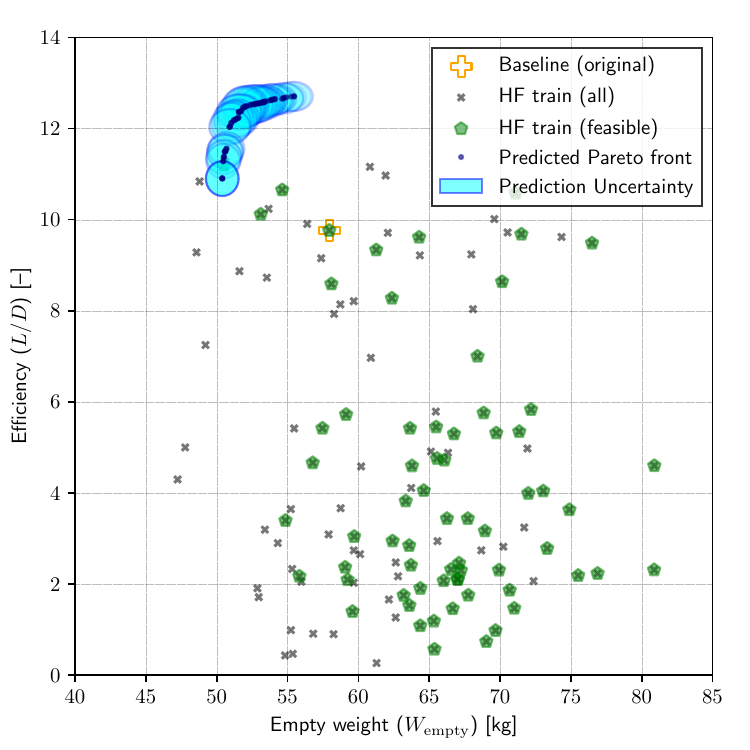}}
    \subfigure[Iter = 3]{\includegraphics[width=0.329\linewidth]{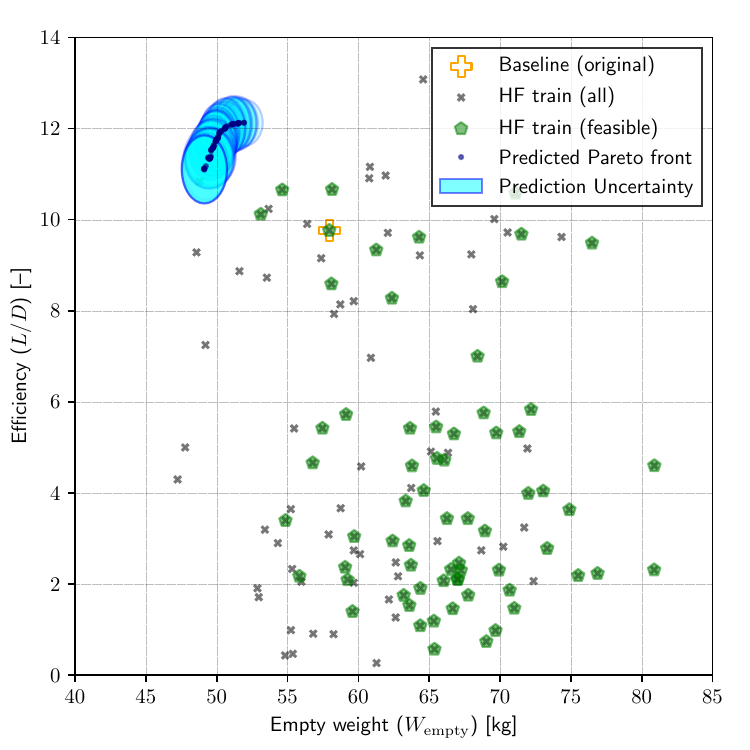}}\\
    \subfigure[Iter = 5]{\includegraphics[width=0.329\linewidth]{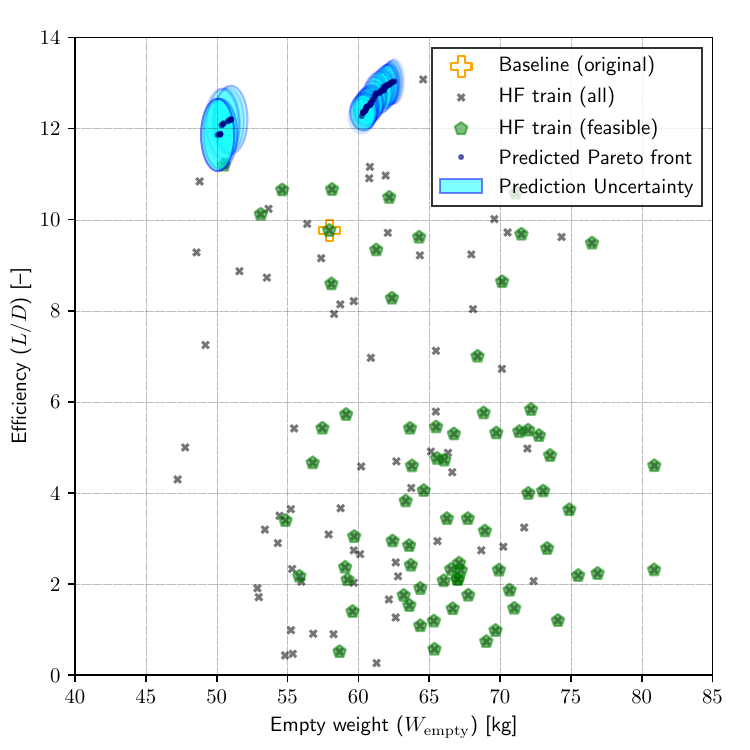}}
    \subfigure[Iter = 6]{\includegraphics[width=0.329\linewidth]{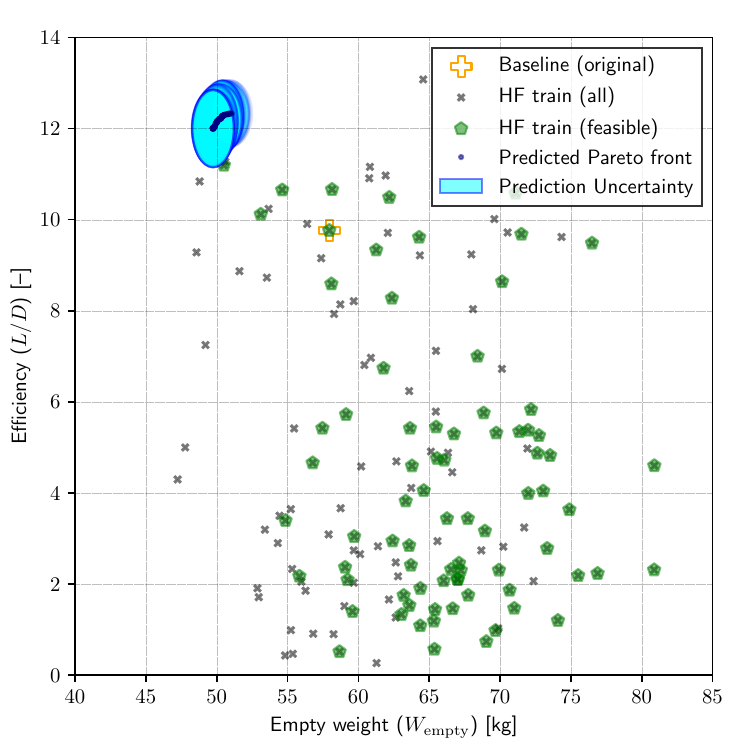}}
    \subfigure[Iter = 9]{\includegraphics[width=0.329\linewidth]{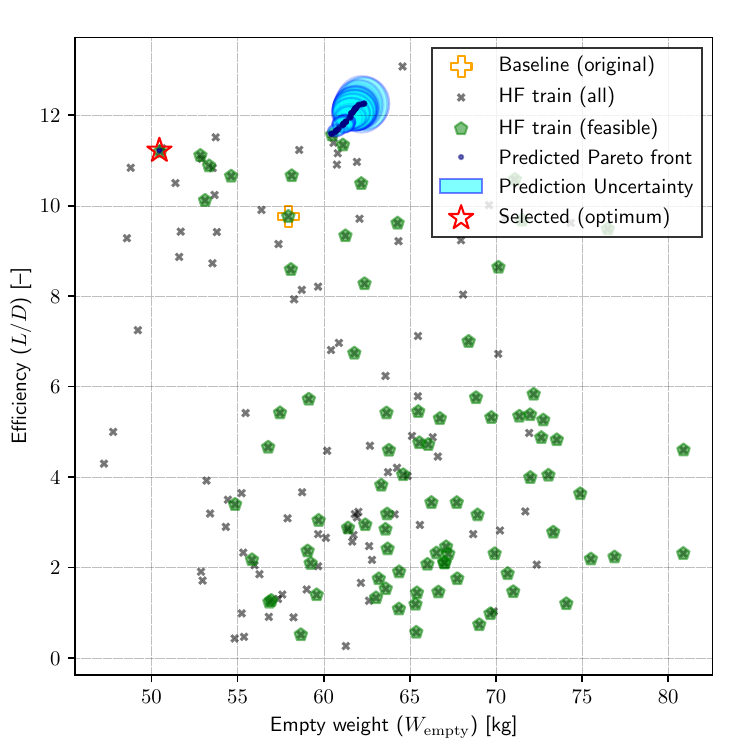}}\\
\caption{Evolution of the surrogate-based Pareto set over selected active-learning iterations in the objective space.}
\label{fig:pareto}
\end{figure*}

% -----------------------------------------------------------

\begin{figure*}[!t]
    \centering
    \includegraphics[width=0.495\linewidth]{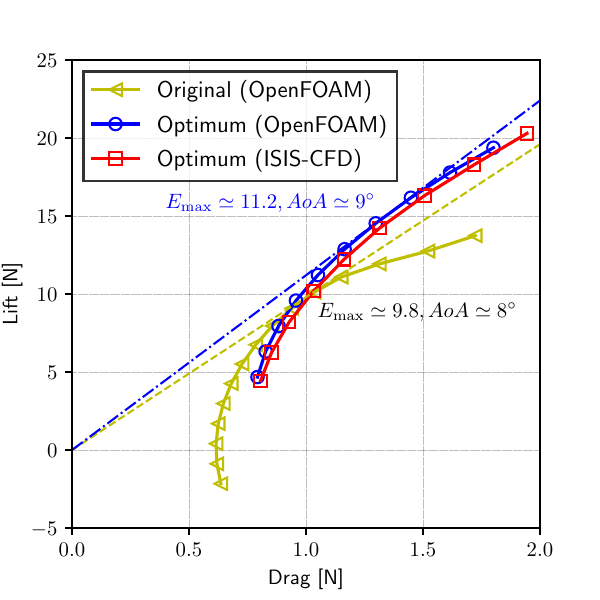}
    \includegraphics[width=0.495\linewidth]{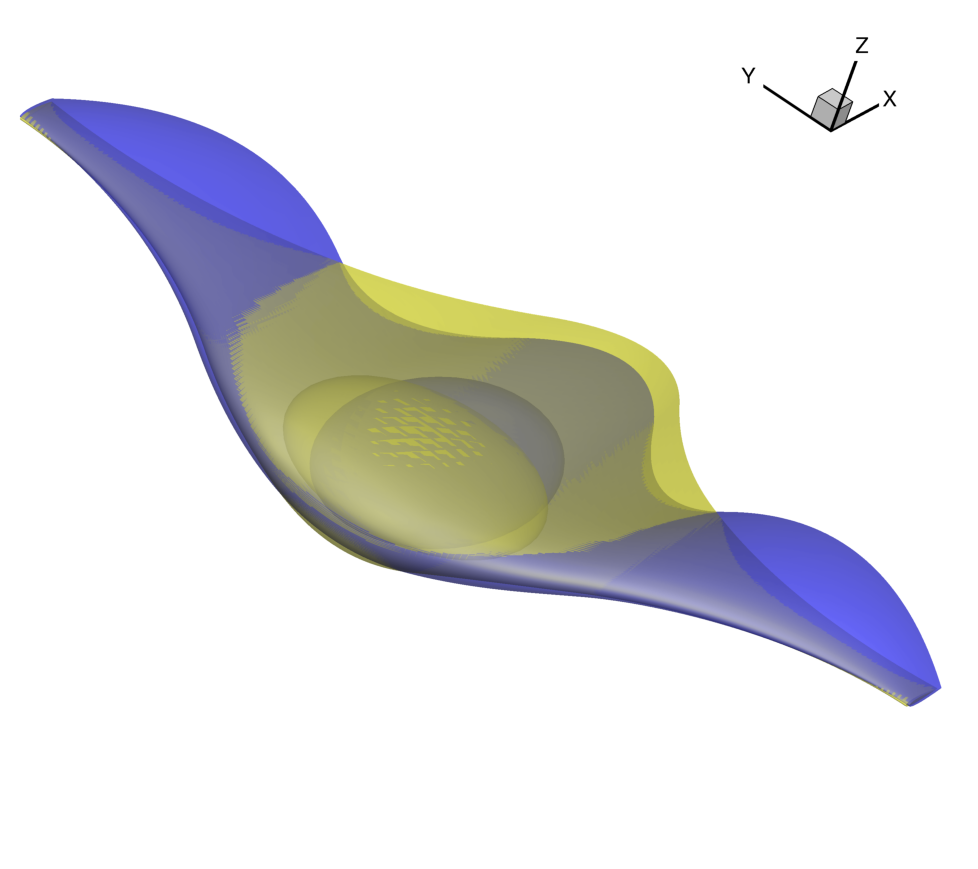}
    \caption{Multi-objective multi-fidelity MDO outcome (on left) and comparison (on right) between original (yellow) and selected optimum (blue) AUG external shape and internal pressure hull} 
    \label{fig:optimum}
\end{figure*}
\subsection{Surrogate-based optimization results}
Figure~\ref{fig:pareto} reports the evolution of the surrogate-based multi-objective optimization process through selected active-learning iterations, shown in the objective space $(W^*_{\text{empty}},\,E_{\max})$. After PD-PME dimensionality reduction, the search is performed in the reduced design space, while each candidate solution is back-mapped to the full geometric representation to evaluate hydrodynamic performance and to solve the internal sizing problem.

The initial multi-fidelity training set is constructed from two nested Sobol designs comprising 512 low-fidelity (LF) and 128 high-fidelity (HF) samples. For each geometry, hydrodynamic polar curves are computed over $AoA\in[-2^\circ,12^\circ]$ with $1^\circ$ increments, and the maximum hydrodynamic efficiency $E_{\max}$ is extracted using the tangent-to-origin criterion. $W^*_{\text{empty}}$ for the samples is determined with the lower-level sizing problem. These initial samples provide the first surrogate approximation of the objectives and constraints.

Subsequent iterations progressively enrich the training set by selecting new candidate designs on the predicted Pareto front using the batch Bayesian strategy based on EHVI with clustering.
Simulation fidelity is assigned adaptively based on the uncertainty-to-cost ratio, resulting in a systematic prevalence of low-fidelity evaluations while reserving high-fidelity simulations for the most informative regions of the design space.

A convergence of the Pareto set is achieved after nine iterations, for a total budget of 578 low-fidelity and 193 high-fidelity evaluations.
The progressive stabilization of the non-dominated set can be observed in Fig.~\ref{fig:pareto}, together with the reduction of the associated uncertainty bands. In the figure, the predicted Pareto solutions are shown together with a local uncertainty representation in the objective plane: at each Pareto point, an ellipse is drawn with axes equal to the predicted uncertainty of $W^*_{\text{empty}}$ (horizontal axis) and $E_{\max}$ (vertical axis). At the final iteration, the maximum uncertainty associated with the predicted Pareto front, for both objectives, falls below $3\%$ when normalized by the range of the corresponding objective values computed over the predicted Pareto set at the same iteration. This indicates that the relative ranking of non-dominated solutions is supported by sufficiently narrow uncertainty levels.

The baseline configuration (yellow cross in Fig.~\ref{fig:pareto}) lies well inside the dominated region of the final Pareto set.
The selected optimal solution (star at the final iteration) achieves a
$14.7\%$ increase in maximum hydrodynamic efficiency and a $12.8\%$ reduction in empty weight with respect to the baseline configuration, while satisfying all structural, hydrostatic, geometric, payload, and surfacing-buoyancy constraints. Finally, Fig.~\ref{fig:optimum} (left) compares the polars of the original and optimized AUG, while Fig.~\ref{fig:optimum} (right) highlights the geometric differences between the baseline and optimized configurations, including the corresponding pressure-hull arrangements. \change{The main performance metrics of the baseline and selected optimal configurations are summarized in Tab.~\ref{tab:comparison}.}

A detailed hydrodynamic interpretation of the optimized configuration and its further verification are discussed in the following subsection.

\begin{table*}[!t] 
\centering 
\textcolor{black}{\caption{Comparison between the baseline and selected optimal configuration.} \label{tab:comparison} 
\begin{tabular}{lccccc} 
\toprule 
Metric & Symbol & Units & Baseline & Optimum & Variation \\ 
\midrule 
Maximum efficiency & $E_{\max}$      & --      & 9.76 & 11.2 & $+14.7\%$ \\ 
Empty weight       & $W^\star_{\rm empty}$ & kg      & 57.9 & 50.5 & $-12.8\%$ \\ 
Angle of attack    & $AoA$           & $^\circ$ & 8    & 9    & $+1$ \\ 
\bottomrule 
\end{tabular} 
}
\end{table*}

\subsection{Discussion}
The optimization results reveal clear and physically interpretable modifications of the hydrodynamic behavior of the manta-inspired AUG. While the previous section quantified the improvements in terms of objective values, the following discussion focuses on the flow mechanisms underlying the observed increase in efficiency and on their relation to the optimized geometry.
\begin{figure}[!t]
    \centering
    \includegraphics[width=0.5\columnwidth]{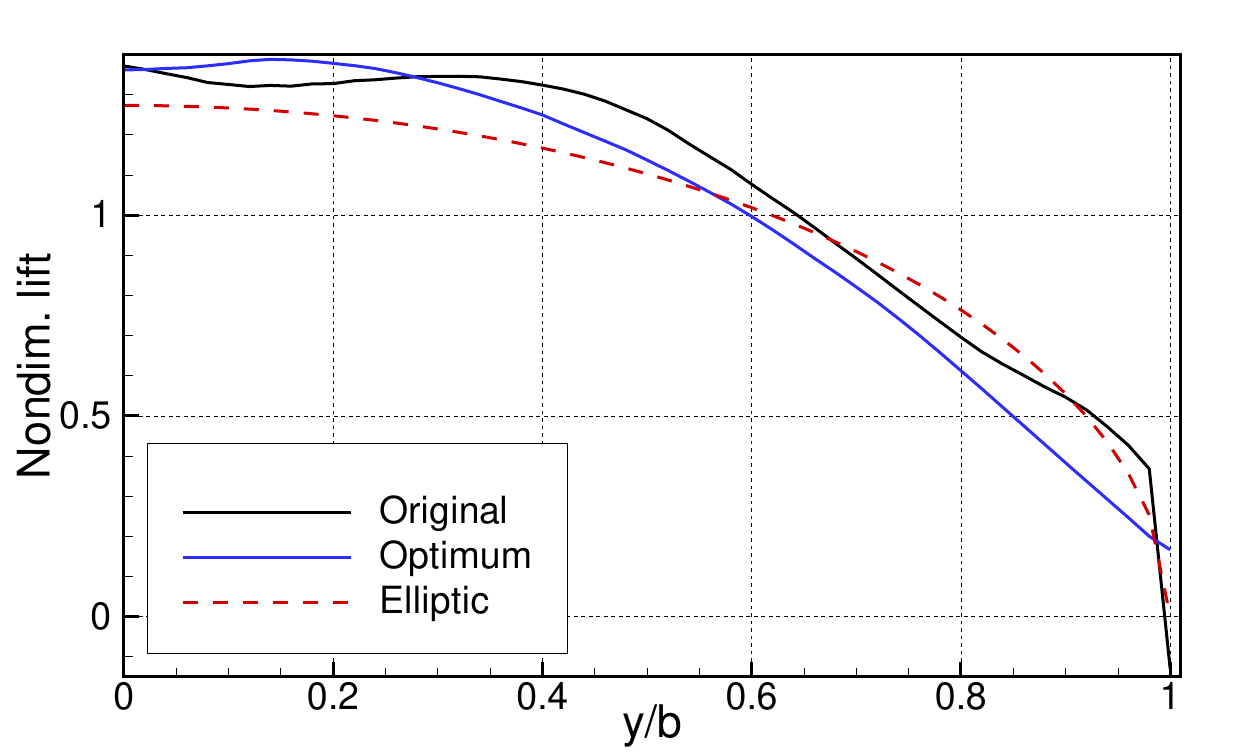}
    \caption{Lift distribution in spanwise direction for the original and selected optimum} 
    \label{fig:isis:liftdistr}
\end{figure}
\begin{figure*}[!t]
    \centering
    \includegraphics[width=0.485\linewidth]{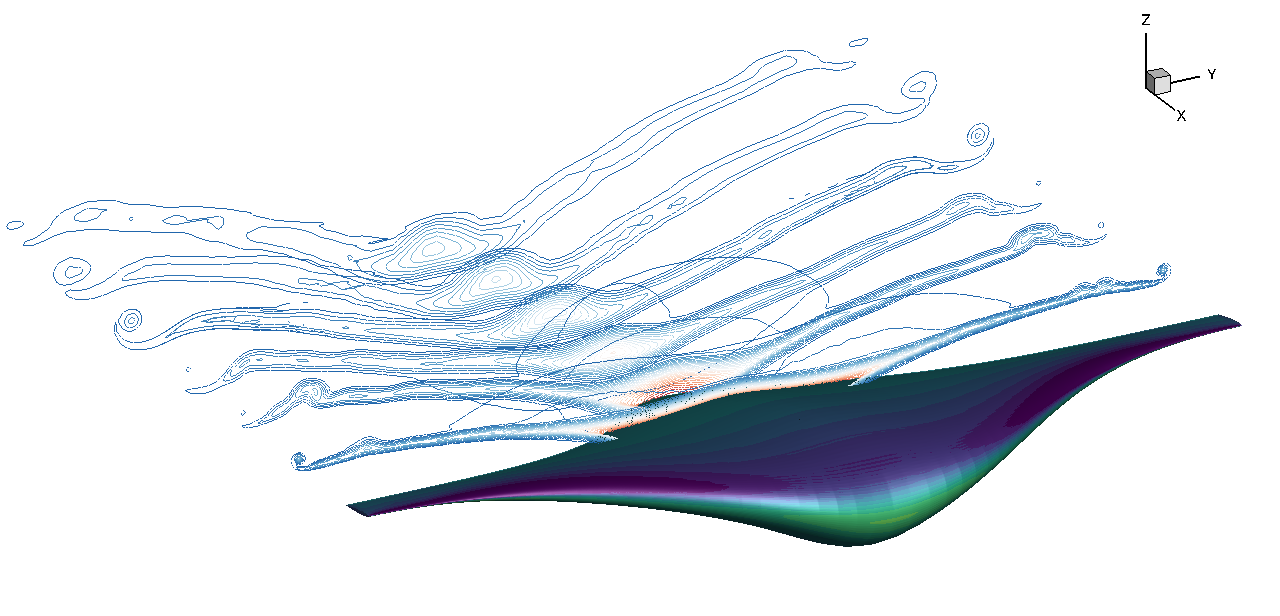} \hfill
    \includegraphics[width=0.485\linewidth]{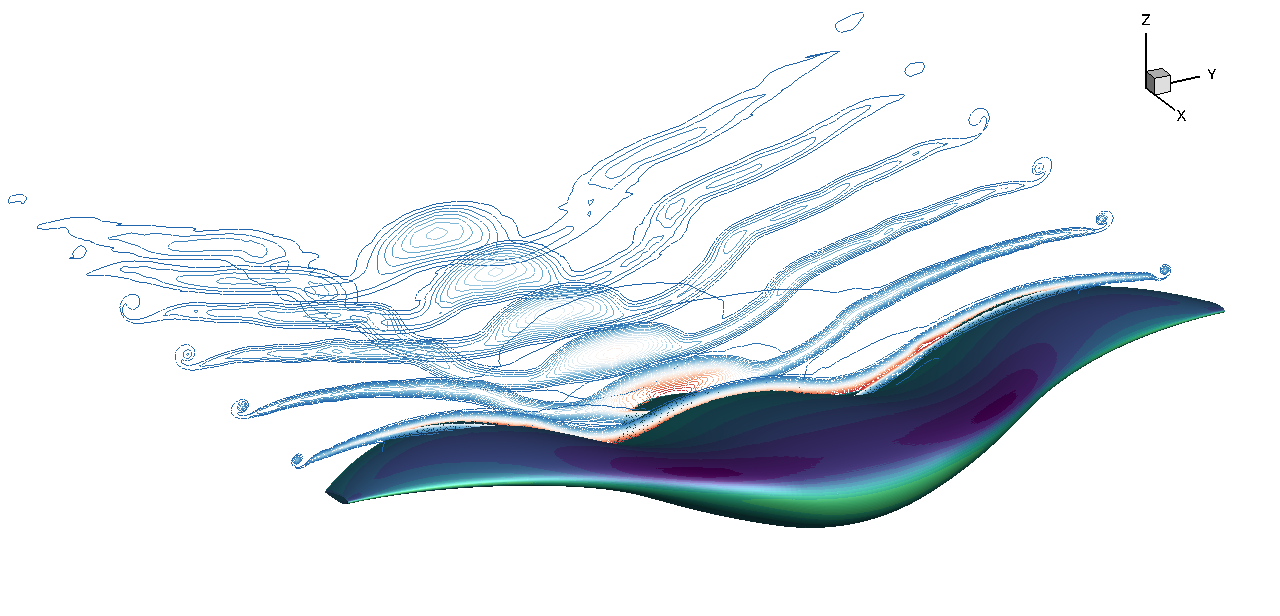} 
     \caption{Surface pressure and axial velocity in the wake, at the optimal L/D, for the base geometry (left) and the optimum (right)}
    \label{fig:isis:wake}
\end{figure*}
\begin{figure*}[!t]
    \centering
    \includegraphics[width=0.32\linewidth]{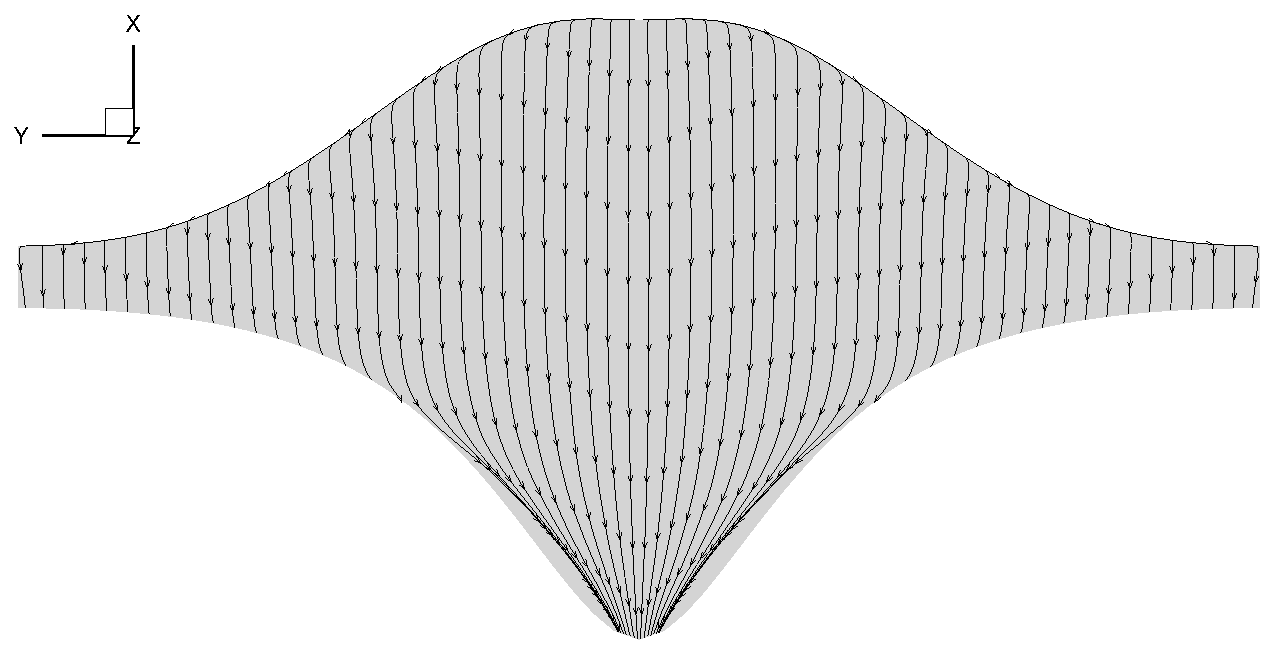} \hfill
    \includegraphics[width=0.32\linewidth]{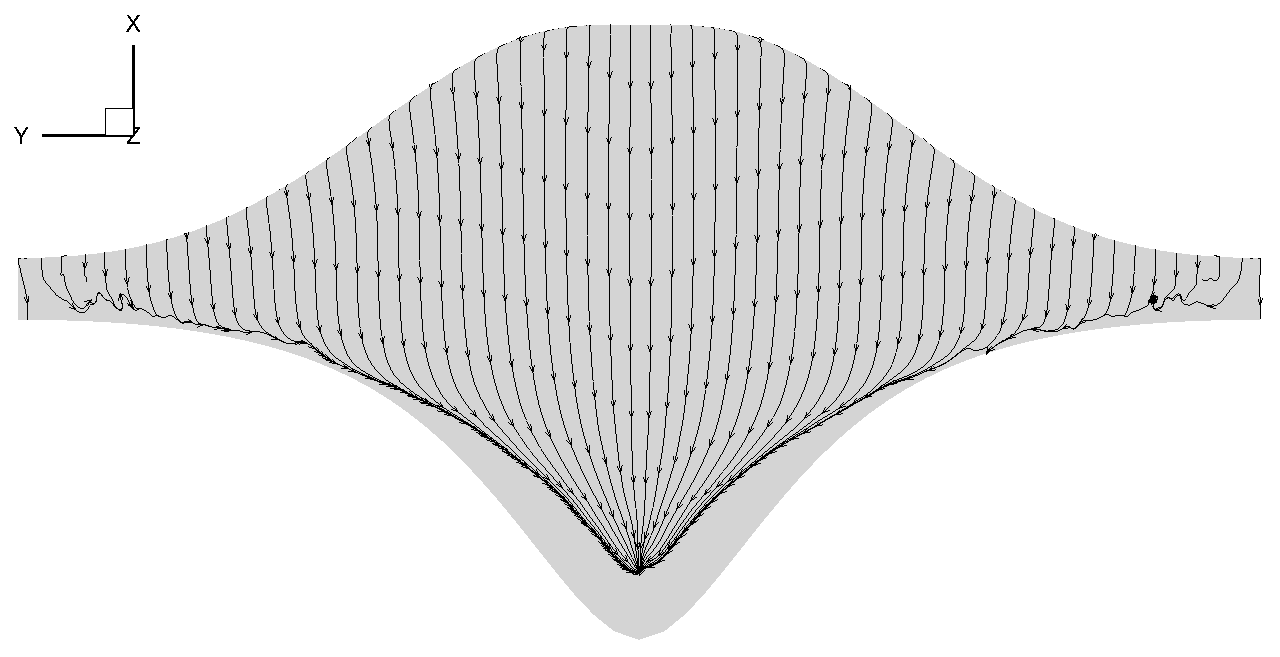} \hfill
     \includegraphics[width=0.32\linewidth]{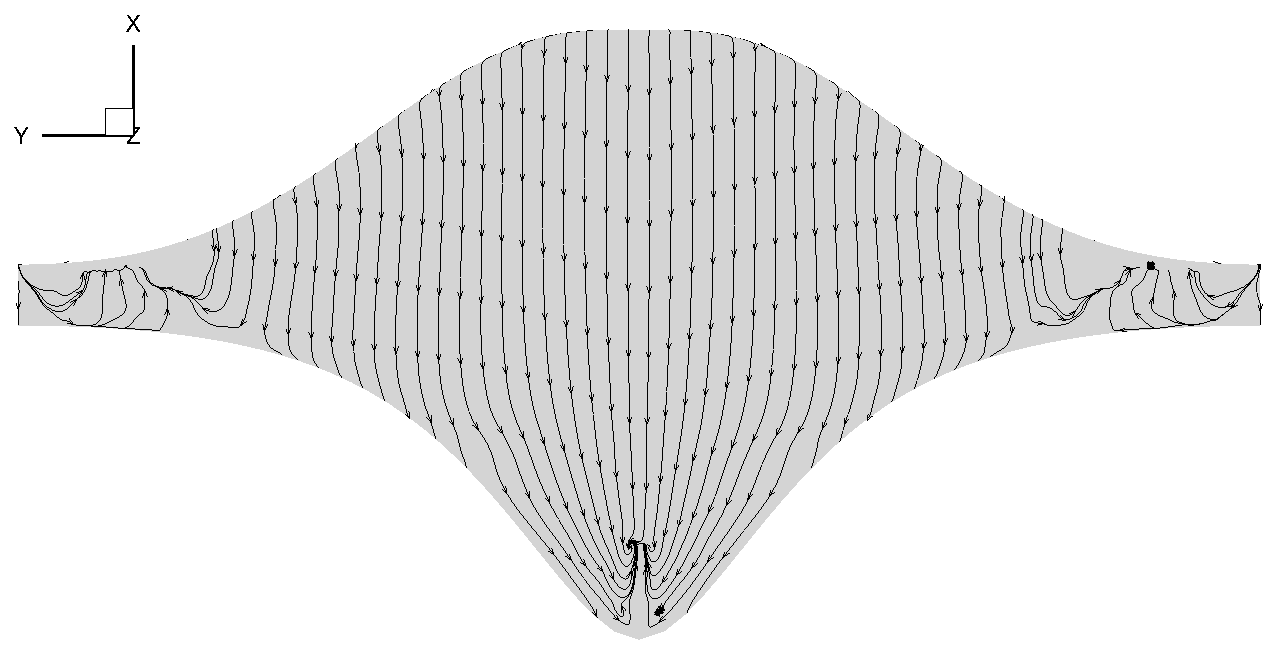} \\
    \includegraphics[width=0.32\linewidth]{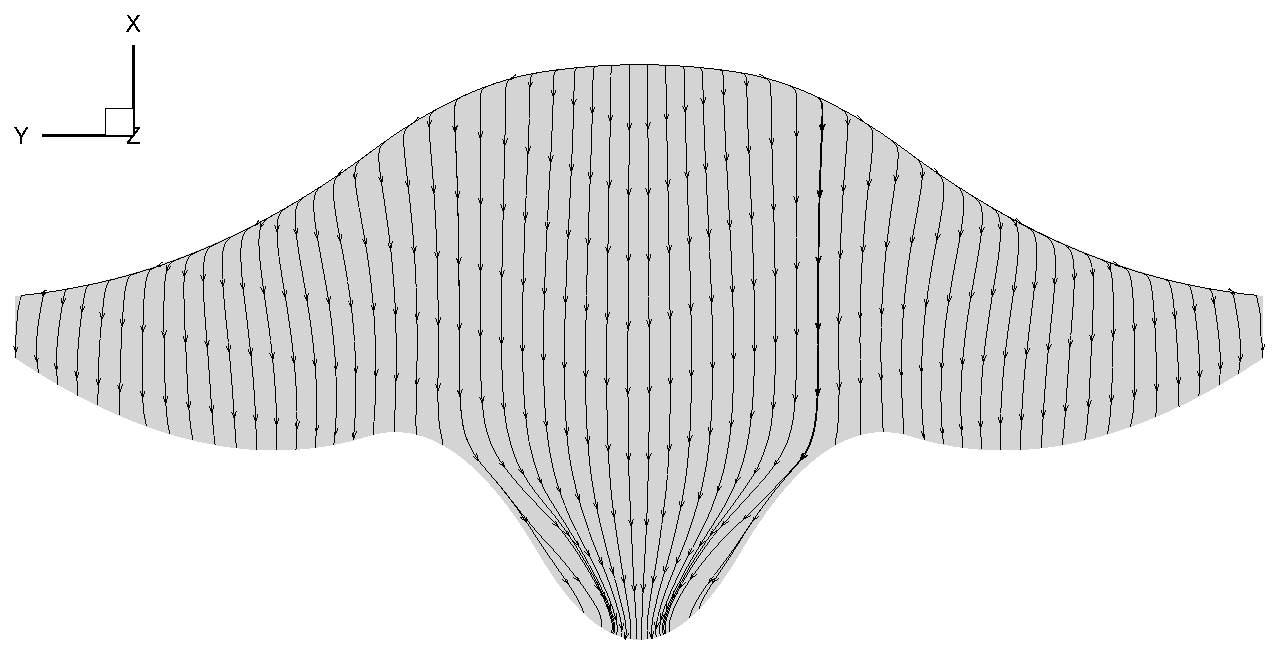} \hfill
    \includegraphics[width=0.32\linewidth]{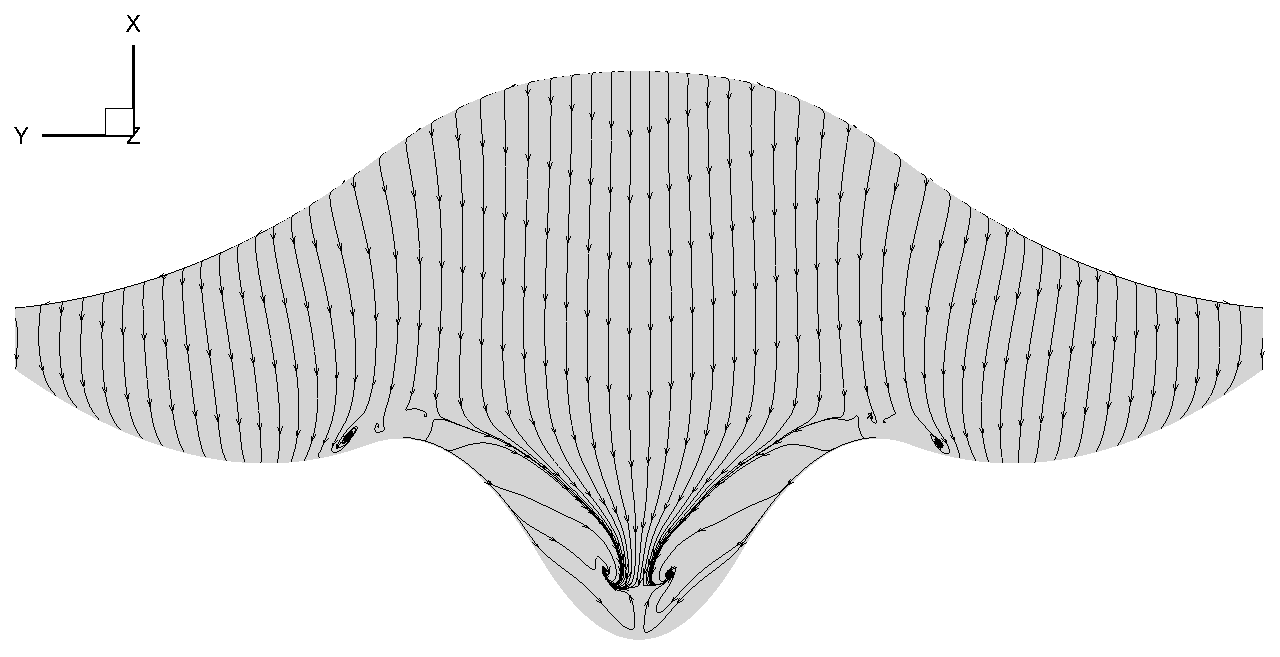} \hfill
     \includegraphics[width=0.32\linewidth]{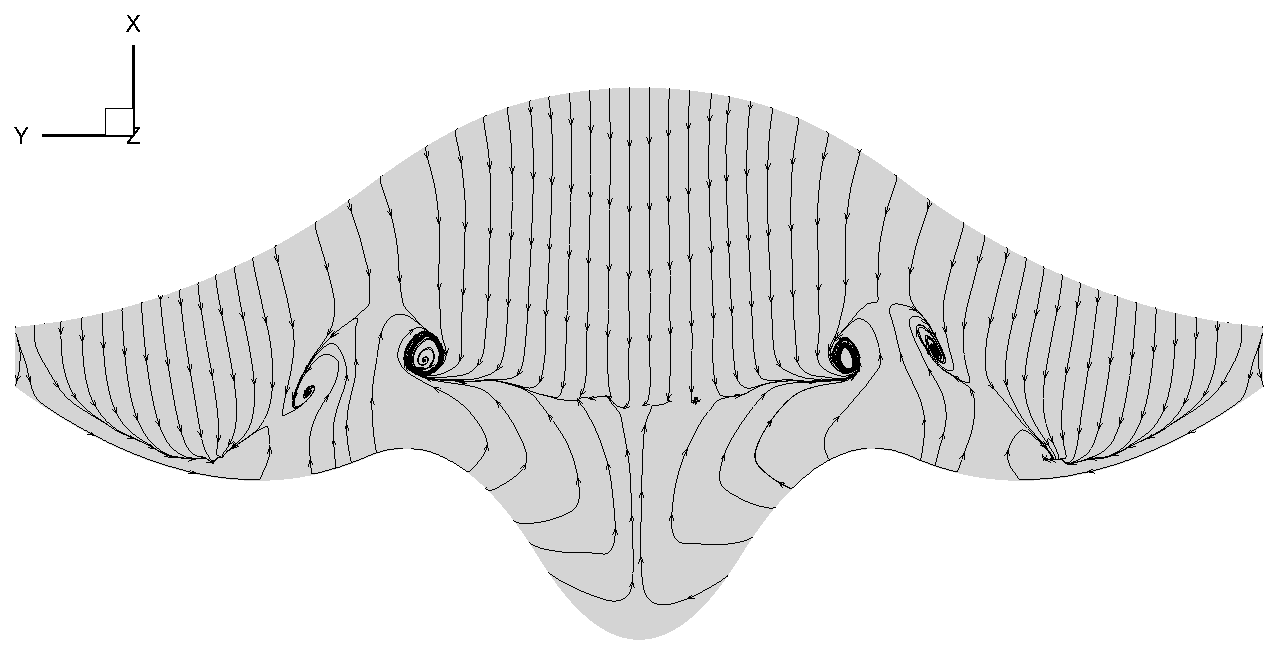} 
     \caption{Top surface streamlines for the base geometry (top) and the optimum (bottom). Left to right: zero lift, optimum incidence, and highest obtained lift.}
    \label{fig:isis:streamlines}
\end{figure*}

A primary outcome of the optimization is the modification of the spanwise lift distribution. Figure~\ref{fig:isis:liftdistr} reports the normalized spanwise lift distributions at the angle of maximum efficiency, obtained from ISIS-CFD simulations. 
Due to the presence of the pressure hull, neither the baseline nor the optimized configuration achieves an ideal elliptic lift distribution; both exhibit a local lift peak near the outer edge of the pressure vessel. However, the optimized wing shows a significantly smoother lift variation, particularly toward the outer wing sections, where load oscillations are reduced. This smoother spanwise behavior, rather than proximity to an elliptic distribution, is identified as the dominant
factor driving the efficiency improvement.

The wake topology further corroborates this interpretation.
Figure~\ref{fig:isis:wake} compares surface pressure and axial velocity fields in the wake at the optimal operating condition for the baseline and optimized geometries. The optimized configuration exhibits fewer localized wake perturbations behind the outer wing sections, indicating a reduction in induced drag and a more coherent downstream flow structure.

The evolution of flow separation with angle of attack is illustrated in
Fig.~\ref{fig:isis:streamlines}, which reports top-surface streamlines for both geometries at representative conditions. At zero lift, the optimized geometry exhibits slightly larger separated regions than the baseline, including separation on the lower wing surface (not shown). This behavior is consistent with the higher drag observed for the optimized shape at zero lift and reflects the fact that the optimization is not targeted at off-design conditions. At the angle of maximum efficiency, flow separation on the optimized geometry is confined to the inner aft region of the wing, whereas the baseline configuration displays extended separation along the entire trailing edge, particularly near the wing tips. At higher angles of attack, the optimized geometry preserves attached flow over a larger fraction of the outer wing, a feature that is beneficial for roll stability.%, even though this aspect was not explicitly included as an optimization objective.

The effect of these flow modifications is also evident when analyzing the
hydrodynamic polars reported in Fig.~\ref{fig:optimum}  (left). The efficiency gain achieved by the optimized configuration is primarily associated with an upward shift of the polar curve in the $(L,D)$ plane, rather than a uniform reduction of drag at fixed lift. This observation is consistent with the problem formulation, which targets the maximization of lift-to-drag ratio under steady-glide conditions, without explicitly constraining drag reduction at fixed lift. As a result, the optimization favors geometries that improve lift generation and load distribution under steady-glide conditions, even at the expense of slightly higher drag in low-lift regimes.

Finally, analyzing the geometric differences between the baseline and optimized configurations (see Fig.~\ref{fig:optimum}, right), it can be noted that the optimized external shape exhibits a planform that is visually closer to a bird than to the original manta-inspired geometry. This observation motivates a final remark on biomimicry: unlike AUGs, real fish maintain neutral buoyancy through their swim bladder and rely on muscular propulsion, and manta rays are therefore optimized for efficient swimming rather than lift generation \cite{Fish2016Hydrodynamic}. Conceptually, an underwater glider operating
through buoyancy-driven motion is closer to a soaring bird than to a swimming fish. The emergence of a bird-like planform in the optimized design can thus be interpreted as an indication of the effectiveness of the automatic optimization process in identifying geometries that are consistent with the underlying buoyancy-driven propulsion mechanism of underwater gliders.

\section{Conclusions}
This work presented a comprehensive multi-fidelity, machine-learning–assisted MDO framework for the preliminary design of a bio-inspired autonomous underwater glider. The proposed methodology combines physics-driven dimensionality reduction, multi-fidelity surrogate modeling, and batch Bayesian optimization within a bi-level formulation that simultaneously addresses external shape design and internal pressure-hull sizing.

The physics-driven parametric model embedding proved effective in reducing the original high-dimensional geometric parameterization to a compact five-dimensional latent space while preserving the physical variability most relevant to hydrodynamic performance. This reduction enabled efficient surrogate construction and focused the optimization process on physically meaningful design directions, mitigating the curse of dimensionality typically encountered in shape-based MDO problems.

A multi-fidelity stochastic radial basis function surrogate was employed to combine low-cost potential-flow evaluations with high-fidelity RANS simulations. The associated uncertainty quantification supported an adaptive, batch Bayesian optimization strategy that systematically prioritized informative regions of the design space while controlling computational cost. Convergence of the Pareto set was achieved after nine enrichment iterations, requiring a total of 578 low-fidelity and 193 high-fidelity evaluations, with high-fidelity analyses representing a small fraction of the overall computational budget.

The final Pareto front is characterized by narrow uncertainty levels: at the last iteration, the maximum uncertainty associated with both objectives along the predicted Pareto set falls below 3\% when normalized by the corresponding objective ranges. This indicates that the relative ranking of non-dominated solutions is supported by sufficiently robust surrogate predictions. The baseline configuration lies well inside the dominated region of the final Pareto set, confirming the benefits of the proposed optimization framework over the original manta-inspired design.

The selected optimal configuration achieves a 14.7\% improvement in maximum hydrodynamic efficiency and a 12.8\% reduction in empty weight relative to the baseline, while satisfying all structural, hydrostatic, geometric, payload, and surfacing-buoyancy constraints. Independent RANS and URANS analyses confirm that the efficiency gain is primarily associated with a smoother spanwise lift distribution and reduced induced drag, rather than with a uniform drag reduction at fixed lift. These flow features emerge naturally from the optimization process, despite not being imposed explicitly as design objectives.

Beyond the specific application to underwater gliders, the results highlight the importance of treating external geometry, internal arrangement, and feasibility constraints in a fully coupled manner during early-stage design. The observed disconnected structure of the Pareto set reflects the strong interaction between hydrodynamic performance, buoyancy requirements, and geometric containment, and underscores the limitations of sequential or single-fidelity conceptual-design approaches for such systems.

Overall, this study demonstrates that integrating physics-driven dimensionality reduction with multi-fidelity surrogate modeling and Bayesian multi-objective optimization provides an effective and computationally tractable strategy for complex multidisciplinary design problems. The proposed framework is readily generalizable to other marine and aerospace applications characterized by high-dimensional design spaces, multiple fidelity levels, and tightly coupled disciplinary constraints.

Future work will focus on a more systematic assessment of the proposed methodological components beyond the specific application considered here. In particular, further investigation is needed to analyze the behavior of batch multi-objective Bayesian optimization strategies under strong feasibility constraints \cite{lin2023parallel}, including the impact of clustering criteria, batch size selection, and fidelity-allocation policies on convergence and robustness of the Pareto set.

In parallel, extensions of the dimensionality-reduction strategy will be considered. While the present work relies on physics-driven linear embeddings, the integration of nonlinear dimensionality-reduction techniques, such as autoencoder-based or manifold-learning approaches \cite{serani2026nonlinear}, represents a promising direction for capturing more complex geometry–physics interactions in highly nonlinear design spaces \cite{serani2025survey}.

Finally, the combination of adaptive dimensionality reduction, multi-fidelity surrogate modeling, and uncertainty-aware optimization will be further explored in a broader range of marine and aerospace design problems, with the objective of establishing general guidelines for the efficient and explainable use of machine-learning tools in early-stage multidisciplinary design optimization.

%%%%%%%%%%%%%%%%%%%%%%%%%%%%%%%%%%%%%%%%%%%%%%%%%%%%%%%%%%%%%%%%%%%%%%%%%%%

\section*{Acknowledgments}
This work has been conducted within the NATO-AVT-404 Research Task Group on “Enhanced Design Processes of Military Vehicles through Machine Learning Methods”. CNR-INM is grateful to the Italian Ministry of University and Research through PRIN 2022 program, project BIODRONES, 20227JNM52 - CUP B53D23005560006, and acknowledges the CINECA award under the ISCRA-B grant HP10BCZZBF and ISCRA-C grant HP10CJM3DH initiatives, for the availability of high-performance computing resources and support.

\section*{Declarations}
\paragraph{Replication of results} \change{The PD-PME dimensionality-reduction workflow can be reproduced using the PME-toolkit, which is publicly available on Zenodo~\cite{serani_2026_19068340} and distributed as a Python package through PyPI (\texttt{pip install pme-toolkit}). The other approaches are described in detail to facilitate replication of the results.}

\paragraph{Data Availability} \change{The dataset supporting the dimensionality-reduction analysis presented in this work is openly available on Zenodo (\url{https://zenodo.org/records/18936594}). Additional datasets generated during the multi-fidelity optimization campaign are available from the corresponding author upon reasonable request.}

\appendix
\section{CAD Parametric Modeling}
\label{appendix:cad}
This appendix provides additional details on the geometric parameterization and CAD generation procedure used to construct the manta-inspired external shape of the autonomous underwater glider. 

The outer geometry is defined through four transverse sections (root, main-body end, transition, wing tip). Each section is parameterized using ten variables: four NACA 4-digit airfoil parameters (maximum camber, camber position, thickness ratio, chord), three coordinates defining the leading-edge location $(x_0,y_0,z_0)$, and three rotation angles (twist $\theta$, roll $\phi$, yaw $\psi$). At the symmetry plane, eight degrees of freedom are constrained to enforce geometric and kinematic continuity.
The manta shape is then built using the OpenCASCADE CAD kernel \cite{OpenCascade}, and the Gmsh v4.12.1 software \cite{geuzaine2009gmsh}. 

Due to the inability of the CAD kernel to handle a single closed B-spline with a trailing-edge cusp, each airfoil is built from:
\begin{itemize}
\item an upper trailing-edge straight segment,
\item a $C^2$ B-spline through all airfoil points except the trailing edge,
\item a lower trailing-edge straight segment.
\end{itemize}
This yields a closed $C^0$ loop, with curvature discontinuities made negligible by sufficiently dense sampling.

The external skin is generated using the \texttt{ThruSections} operation of OpenCASCADE. A smooth loft surface is produced by enforcing:
\begin{itemize}
\item tangency and direction continuity at the symmetry plane,
\item repeated wing-tip sections to regularize curvature toward the tip,
\item final trimming via Boolean intersection with a bounding volume is used to obtain half-span geometry.
\end{itemize}
%
%The loft is constructed only for the half-span and mirrored across the plane $y=0$. 
Additional spanwise tangency is enforced at the mid-plane to avoid kinks in the flow-aligned direction.

\section{Numerical Solvers}
\label{appendix:solvers}

This appendix summarizes the numerical settings, mesh generation procedures, and solver configurations used for the three flow solvers employed in the multi-fidelity framework: PUFFIn (low-fidelity), OpenFOAM (high-fidelity steady RANS), and ISIS--CFD (high-fidelity unsteady RANS for verification).

\subsection{PUFFIn}
PUFFIn is an incompressible potential-flow solver designed for hydrodynamic analysis involving lifting surfaces such as hydrofoils \cite{perali2024performance}. PUFFIn solves the Laplace equation for the velocity perturbation potential, employing a boundary integral equation discretized through the boundary element method. The discretization uses quadrilateral elements with constant strength distributions of sources and doublets.

\begin{figure}[!t]
    \centering
    \includegraphics[width=0.5\columnwidth]{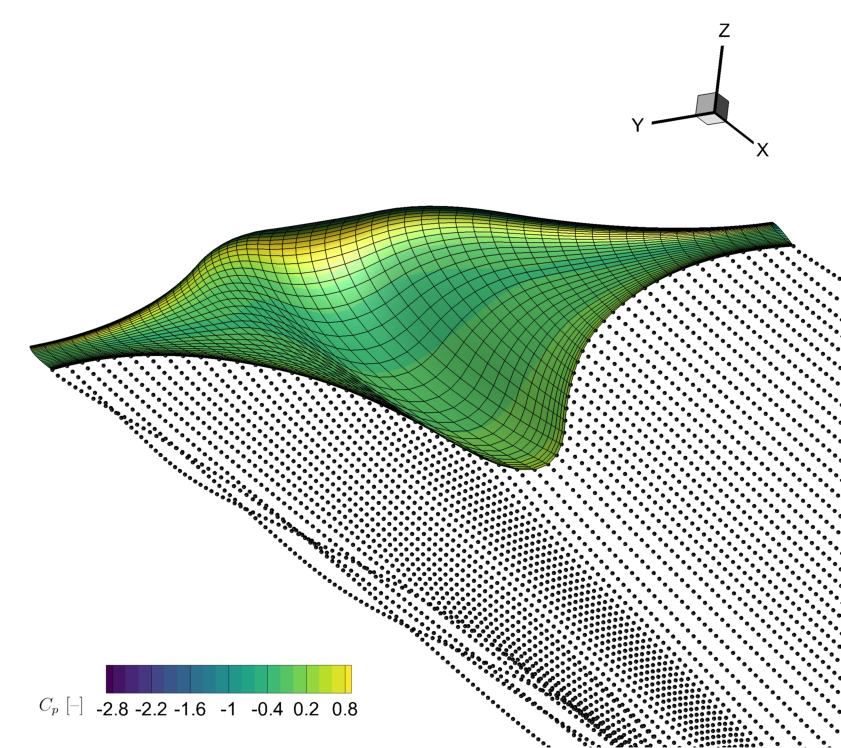}
    \caption{Example of PUFFIn output with pressure distribution on the body of the Manta-ray baseline}
    \label{fig:manta_puffin}
\end{figure}
The solver inherently assumes inviscid, irrotational, and incompressible flow. The wake (see, e.g., Fig.~\ref{fig:manta_puffin}) is represented as a surface starting from the foil's trailing edge, enforcing the Kutta condition to ensure finite velocities at the trailing edge. Wake convection is modeled using a Lagrangian approach, where wake panel positions are updated by convecting trailing-edge nodes with the local flow velocity obtained at each time step. For numerical stability, a vortex viscous core model is used to handle wake-body interactions effectively.

The solver settings utilized for this study involved steady-state simulations whose solution was accelerated by enabling the steady option in the solver configuration. The incoming flow was specified with a uniform velocity of 0.25 m/s along the $x$-axis. The fluid density was set to 1030 kg/m$^3$. Wake convection incorporated self-induced velocities with a first-order Euler integration scheme, and the Kutta strip model employed was aligned with the trailing edge bisector to enhance numerical stability.
A wake limitation was activated, restricting the number of wake panels for computational efficiency. 

\begin{figure*}[!b]
    \centering
    \includegraphics[width=0.325\linewidth]{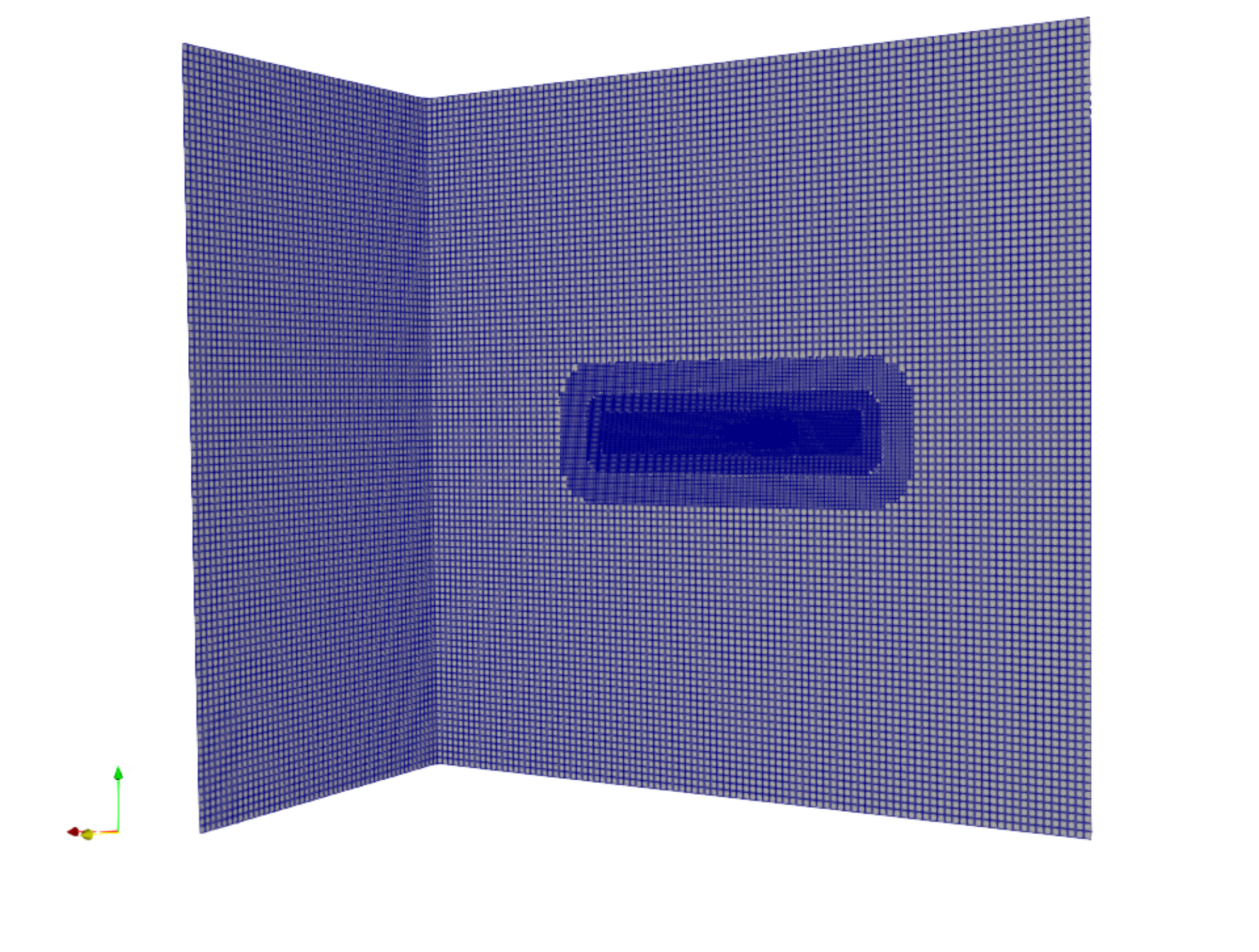}
    \includegraphics[width=0.325\linewidth]{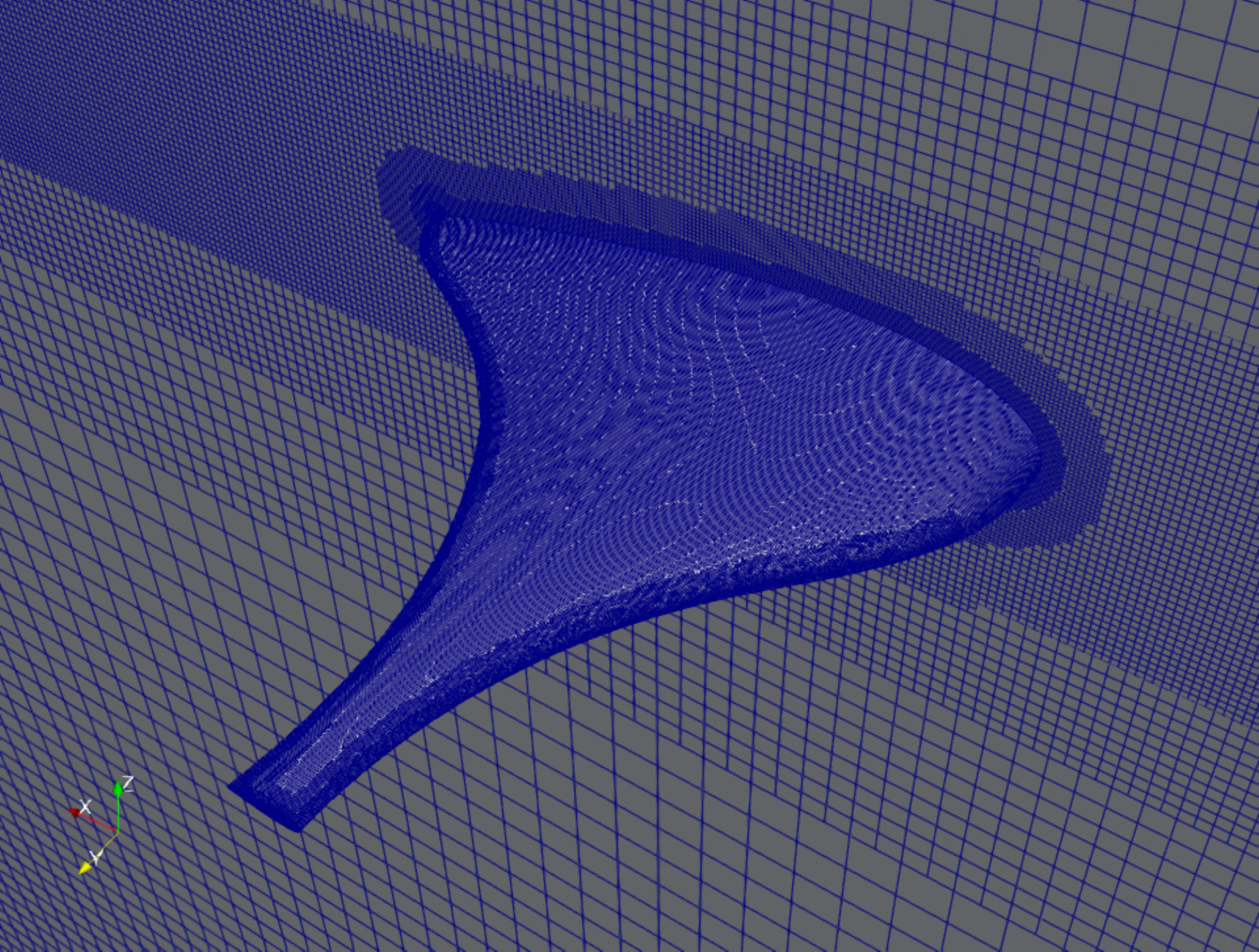}
    \includegraphics[width=0.325\linewidth]{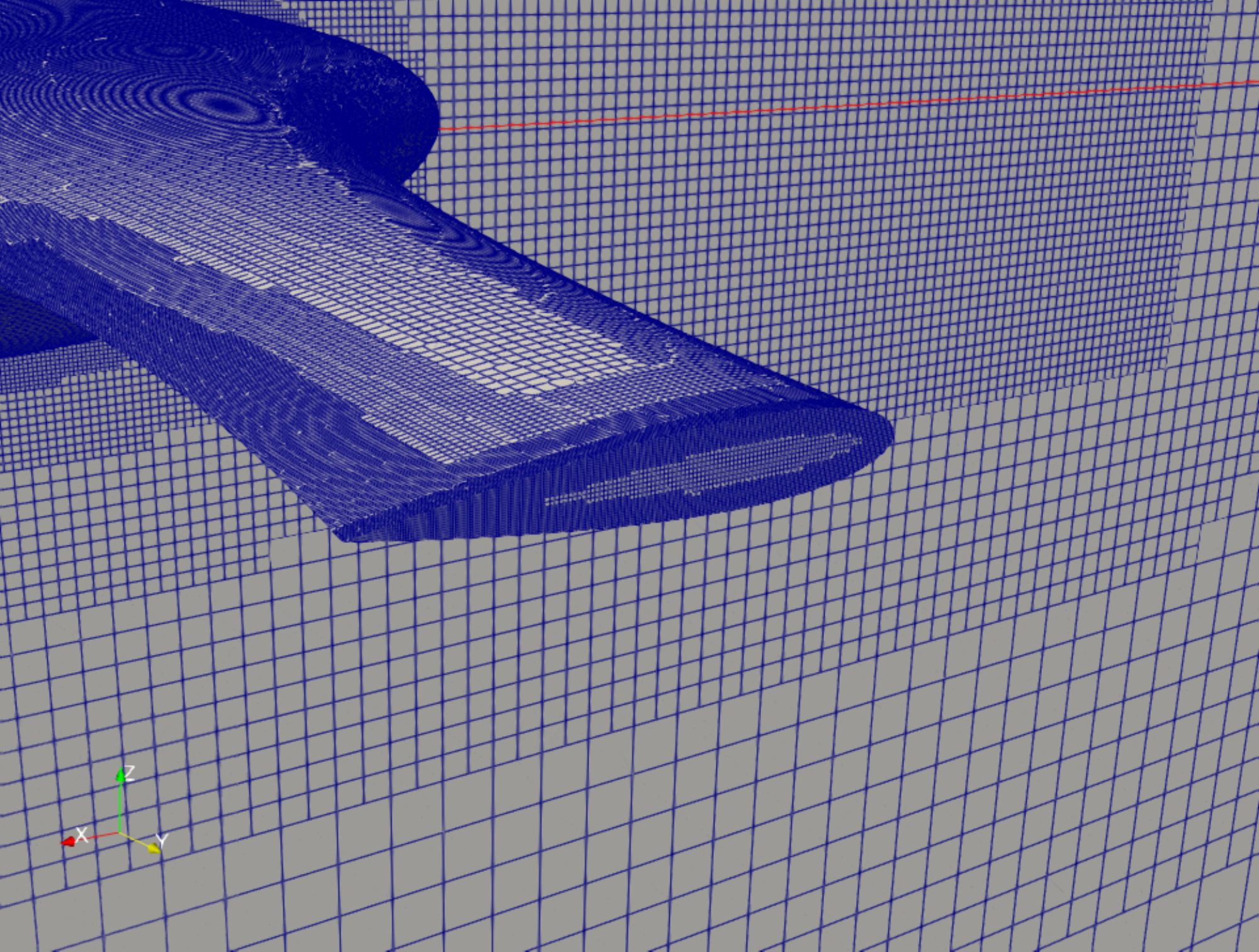}
    \caption{OpenFOAM (left) background and (center) surface grids, with detail (right) on tip of the AUG}
    \label{fig:manta_grid-of}
\end{figure*}
Finally, to augment PUFFIn’s inviscid flow predictions, a viscous drag estimation module was integrated into the workflow. This module estimates the viscous resistance employing a flat-plate approximation based on the Prandtl boundary-layer theory with laminar-to-turbulent transition. Specifically, local Reynolds numbers were computed based on local flow velocities and characteristic panel lengths derived from the surface discretization. The corresponding friction coefficients were calculated using Prandtl correlations suitable for both laminar and turbulent boundary layers \cite{schlichting2016bl}. The viscous drag components were then obtained by integrating local friction forces over the foil surface, providing a comprehensive resistance analysis complementary to the inviscid solution.

This hybrid computational approach, combining the efficiency of potential flow methods with a boundary-layer-based viscous correction, enabled accurate and computationally efficient predictions of AUG performance within the multi-fidelity optimization framework.

An example of solver output with pressure distribution on the body of the Manta-ray baseline configuration along with the wake is shown in Fig.~\ref{fig:manta_puffin}.

Computations were run on parallel CPUs using OpenMP parallelization. The computational grid was generated externally and input into PUFFIn as structured Tecplot-formatted data files.
The grid is composed of 56$\times$56 panels for a total of 3192 grid nodes and shown in Fig.~\ref{fig:manta_puffin}.

\subsection{OpenFOAM}
OpenFOAM is an open-source CFD software package based on the finite volume method for incompressible turbulent flows. In the present work, the \texttt{simpleFoam} solver was employed, tailored for steady-state incompressible turbulent flows. Turbulence was modeled employing the $k-\omega$ SST model, chosen for its effectiveness in capturing flow separation and accurately resolving near-wall turbulence structures. Additionally, a $\gamma-Re_\theta$ transition model was used.
The boundary conditions applied are: the velocity at the inlet fixed at a uniform magnitude of 0.25 m/s, with a no-slip condition imposed on all solid surfaces representing the AUG geometry; at the outlet, a zero-gradient condition allowed the flow to adapt freely; pressure boundary conditions were set as zero-gradient at the inlet and fixed-value at zero at the outlet, while solid walls and symmetry boundaries were treated with zero-gradient conditions.
Turbulence quantities such as turbulent kinetic energy ($k$) and specific dissipation rate ($\omega$) were initialized based on calculated inlet turbulence intensities and length scales. The turbulent viscosity ($\nu_t$) was computed according to the turbulence model, enforcing no-slip conditions at the solid surfaces. Transition modeling parameters, including intermittency ($\gamma$) and transition Reynolds number ($Re_{\theta t}$), were carefully initialized to ensure realistic laminar-to-turbulent transition predictions.

Numerically, a SIMPLE algorithm was utilized for pressure-velocity coupling, featuring optimized under-relaxation factors (0.3 for pressure, 0.7 for velocity, and 0.5 for turbulence-related quantities) to achieve stable convergence. Gradient schemes adopted were Gauss linear, divergence schemes employed bounded Gauss linear-upwind for improved numerical stability, and the Laplacian schemes utilized Gauss linear-corrected discretization. All interpolations between cell centers used linear interpolation to maintain accuracy.

Parallel computations employed OpenFOAM's \texttt{decomposePar} utility, dividing the computational domain into 112 subdomains using the Scotch decomposition method, significantly enhancing computational efficiency.

The computational mesh generation process employed OpenFOAM’s built-in tools \texttt{blockMesh} and \texttt{snappyHexMesh}. Initially, \texttt{blockMesh} generated a structured background domain, spanning from -4 m upstream to 6 m downstream, 5 m in height, and 10 m laterally, discretized uniformly into 100$\times$50$\times$100 cells.
Subsequently, \texttt{snappyHexMesh} refined the mesh near the geometric surfaces defined through STL files of the AUG geometry, ensuring accurate geometry representation with refinement levels ranging between 5 and 7. Explicit edge refinement was performed on feature edges extracted with an included angle threshold of 10$^\circ$.
Additional refinement was performed within a specified wake region extending from -1 m to 3 m in the streamwise direction and $\pm$0.3 m vertically, achieving uniform mesh refinement at level 3. A total of 10 boundary cell layers with a final thickness of $4 \times 10^{-4}$ m and expansion ratio of 1.2 were added around the manta geometry, ensuring detailed resolution ($y^+\leq 1$) of the near-wall turbulence phenomena.
The total number of grid points is about 15M (see Fig.~\ref{fig:manta_grid-of}).
Rigorous mesh quality control ensured maximum orthogonality, while skewness and aspect ratios were kept within stringent thresholds to facilitate stable numerical calculations.

\begin{figure}[!b]
    \centering
    \includegraphics[width=0.495\columnwidth]{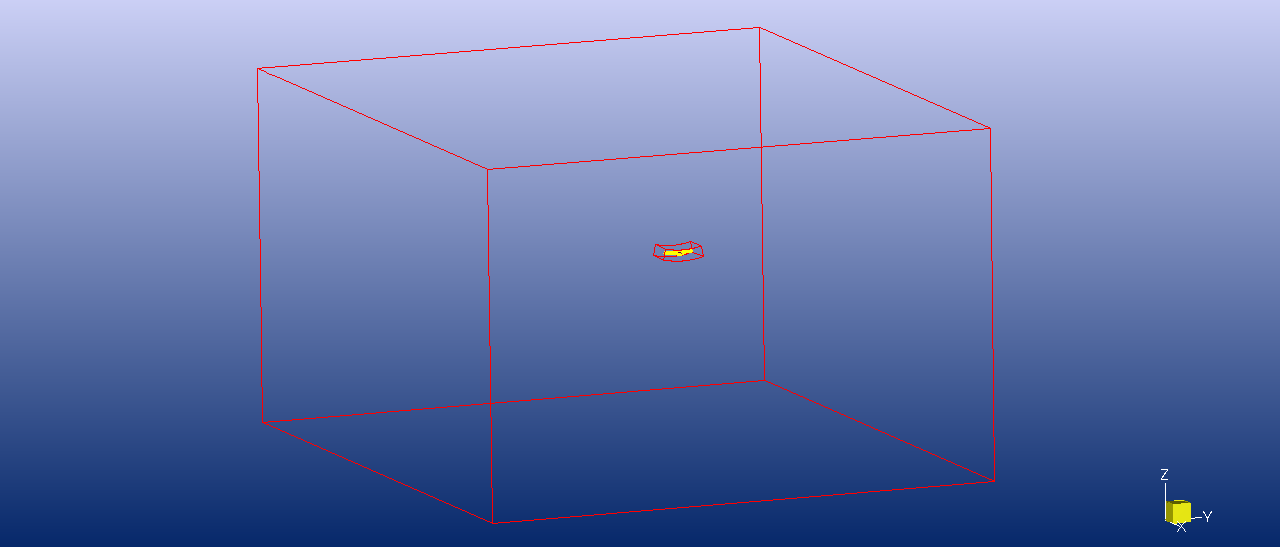} 
    \includegraphics[width=0.495\columnwidth]{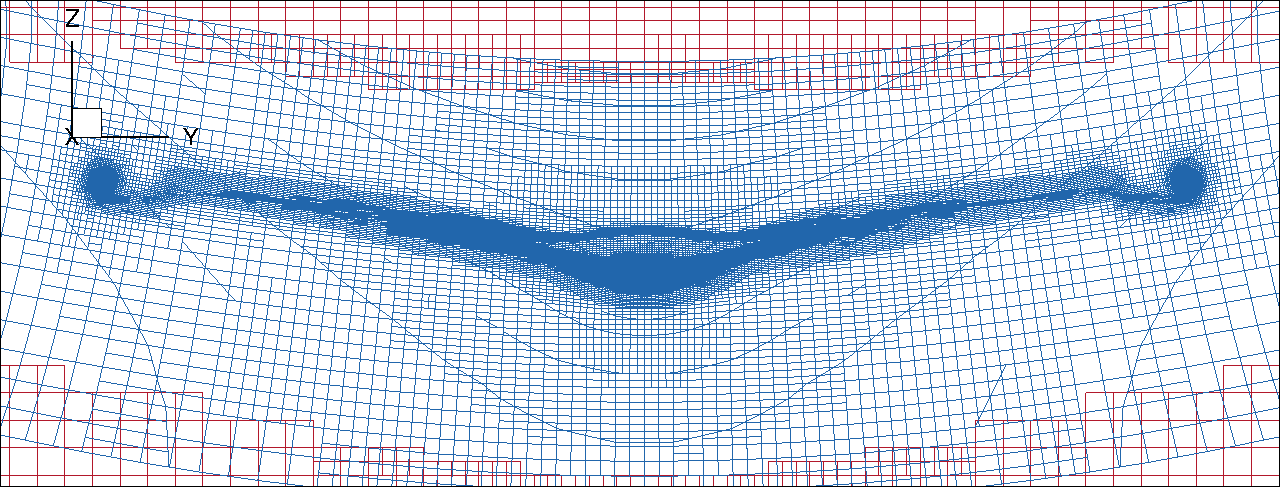} \\
    \caption{Example of ISIS-CFD sock mesh topology, and $x$-constant cut through the adaptively refined sock mesh (blue: overset domain, red: background domain)}
    \label{fig:manta_isis}
\end{figure}
\subsection{ISIS-CFD} \label{sect:solvers_isis}
Additional RANS simulations for verification have been performed with ISIS-CFD, developed at ECN--CNRS and distributed in the FINE/Marine computing suite from Cadence Design Systems. ISIS-CFD is an incompressible unstructured finite-volume solver for multifluid flow \cite{queutey2007interface,wackers2011free} based on a SIMPLE-like formulation.  
The unstructured discretization is face-based. While all unknown state variables are cell-centered, the systems of equations are constructed face by face. Therefore, cells with an arbitrary number of arbitrarily-shaped constitutive faces are accepted, which enables, for example, adaptive mesh refinement. 

For these simulations, the convective flux reconstruction uses the AVLSMART scheme \cite{przulj:01} while diffusive fluxes use a centered Gauss scheme with misalignment corrections. Turbulence is modeled using $k-\omega$ SST without a transition model; boundary conditions are fixed velocity everywhere except at the outflow, where the pressure is imposed. Full details of the simulation procedure for hydrofoils can be found in \cite{wackers2022adaptivegrid}.
 
ISIS-CFD simulates hydrofoils at an imposed lift force. For this, the pitch angle is adjusted in a quasi-static positioning method, to find the equilibrium position that generates the target force. 
In this procedure, corrections to the angles are regularly computed. The foil is rotated to the new position over several time steps, then remains in position for a few time steps to enable the forces to stabilize. This procedure is stable even for large time steps, since no unsteady motion has to be resolved.

An adaptive grid refinement (AGR) method is used, based on anisotropic refinement of unstructured hexahedral meshes  \cite{wackers2022adaptivegrid}. The refinement procedure is fully parallel and is called at regular intervals during the simulation. Once the grid is adapted to the flow, the refinement procedure stops modifying the mesh. 
The refinement criterion, which determines where the mesh should be refined, is derived from the second derivatives of the flow. 

To represent the wing, a `sock mesh' is used (see Fig.~\ref{fig:manta_isis}): a curved domain that follows the shape of the foil with a body-aligned mesh, embedded in a larger domain by the overset mesh (or chimera) technique. Since the overset mesh is aligned with the foil, it produces surface meshes that locally resemble structured body-aligned grids.  Especially for hydrofoils, this approach provides better quality cells on the leading and trailing edges. Also, it reduces the total required number of cells.

%Bibliography
\bibliographystyle{unsrt}  
\bibliography{references}

@article{serani2026nonlinear,
  title={A nonlinear extension of parametric model embedding for dimensionality reduction in parametric shape design},
  author={Serani, Andrea and Palma, Giorgio and Diez, Matteo},
  journal={arXiv preprint arXiv:2605.11759},
  year={2026}
}

@software{serani_2026_19068340,
  author       = {Serani, Andrea},
  title        = {PME-toolkit},
  month        = mar,
  year         = 2026,
  publisher    = {Zenodo},
  version      = {v1.2.0},
  doi          = {10.5281/zenodo.19068340},
  swhid        = {swh:1:dir:c40dc151b83e72fbc9701e805a3bc22cef64cfe4
                   ;origin=https://doi.org/10.5281/zenodo.18962859;vi
                   sit=swh:1:snp:5adc326542f91273346220d97470fe29960f
                   9cc2;anchor=swh:1:rel:8d2461f58fc026ad9f5fdcc991fe
                   47c2f38c674d;path=cnr-inm-mao-pme-toolkit-b8ed16c
                  },
}

@dataset{serani_2026_18936594,
  author       = {Serani, Andrea and
                  Palma, Giorgio},
  title        = {Design-Space Dimensionality Reduction Benchmark
                   Dataset – Bio-Inspired Underwater Glider
                  },
  month        = feb,
  year         = 2026,
  publisher    = {Zenodo},
  version      = {1.1},
  doi          = {10.5281/zenodo.18936594},
}

@article{beachy2025adaptive,
  title={Adaptive Selection of Decomposed Function Information Sources for Rapid Neural Networks},
  author={Beachy, Atticus and Bae, Harok and Camberos, Jose A and Grandhi, Ramana V},
  journal={AIAA Journal},
  volume={63},
  number={6},
  pages={2332--2344},
  year={2025},
  publisher={American Institute of Aeronautics and Astronautics}
}

@article{beachy2021emulator,
  title={Emulator embedded neural networks for multi-fidelity conceptual design exploration of hypersonic vehicles: A. Beachy et al.},
  author={Beachy, Atticus and Bae, Harok and Boyd, Ian and Grandhi, Ramana},
  journal={Structural and Multidisciplinary Optimization},
  volume={64},
  number={5},
  pages={2999--3016},
  year={2021},
  publisher={Springer}
}

@article{beachy2024epistemic,
  title={Epistemic modeling uncertainty of rapid neural network ensembles for adaptive learning},
  author={Beachy, Atticus and Bae, Harok and Camberos, Jose A and Grandhi, Ramana V},
  journal={Finite Elements in Analysis and Design},
  volume={228},
  pages={104064},
  year={2024},
  publisher={Elsevier}
}

@article{wang2024model,
  title={A model-based shape conceptual design framework of blend-wing-body underwater gliders with curved wings},
  author={Wang, Wenxin and Wang, Xinjing and Dong, Huachao and Wang, Peng and Shen, Jiangtao},
  journal={Ships and offshore structures},
  volume={19},
  number={4},
  pages={497--508},
  year={2024},
  publisher={Taylor \& Francis}
}

@article{sun2017shape,
  title={Shape optimization of blended-wing-body underwater glider by using gliding range as the optimization target},
  author={Sun, Chunya and Song, Baowei and Wang, Peng and Wang, Xinjing},
  journal={International Journal of Naval Architecture and Ocean Engineering},
  volume={9},
  number={6},
  pages={693--704},
  year={2017},
  publisher={Elsevier}
}

@article{sun2015parametric,
  title={Parametric geometric model and shape optimization of an underwater glider with blended-wing-body},
  author={Sun, Chunya and Song, Baowei and Wang, Peng},
  journal={International Journal of Naval Architecture and Ocean Engineering},
  volume={7},
  number={6},
  pages={995--1006},
  year={2015},
  publisher={Elsevier}
}

@article{burkardt2009k,
  title={K-means clustering},
  author={Burkardt, John},
  journal={Virginia Tech, Advanced Research Computing, Interdisciplinary Center for Applied Mathematics},
  volume={5},
  year={2009}
}

@article{golub1999tikhonov,
  title={Tikhonov regularization and total least squares},
  author={Golub, Gene H and Hansen, Per Christian and O'Leary, Dianne P},
  journal={SIAM journal on matrix analysis and applications},
  volume={21},
  number={1},
  pages={185--194},
  year={1999},
  publisher={SIAM}
}

@book{schlichting2016bl,
  author    = {Schlichting, Hermann and Gersten, Klaus},
  title     = {Boundary-Layer Theory},
  edition   = {9},
  year      = {2016},
  publisher = {Springer},
  address   = {Berlin, Heidelberg},
  isbn      = {978-3-662-52918-1},
  doi       = {10.1007/978-3-662-52919-8}
}

@article{rousseeuw1987silhouettes,
  title={Silhouettes: a graphical aid to the interpretation and validation of cluster analysis},
  author={Rousseeuw, Peter J},
  journal={Journal of Computational and Applied Mathematics},
  volume={20},
  pages={53--65},
  year={1987},
  publisher={Elsevier}
}

@incollection{Stephen09,
author = {Wood and Stephen},
title = {Autonomous Underwater Gliders},
booktitle = {Underwater Vehicles},
publisher = {IntechOpen},
address = {London},
year = {2009},
editor = {Alexander V. Inzartsev},
chapter = {26},
doi = {10.5772/6718}
}

@article{sun2021design,
  title={Design and optimization of a bio-inspired hull shape for {AUV} by surrogate model technology},
  author={Sun, Tongshuai and Chen, Guangyao and Yang, Shaoqiong and Wang, Yanhui and Wang, Yanzhe and Tan, Hua and Zhang, Lianhong},
  journal={Engineering Applications of Computational Fluid Mechanics},
  volume={15},
  number={1},
  pages={1057--1074},
  year={2021},
  publisher={Taylor \& Francis}
}

@article{lambe2012extensions,
  title={Extensions to the design structure matrix for the description of multidisciplinary design, analysis, and optimization processes},
  author={Lambe, Andrew B. and Martins, Joaquim R. R. A.},
  journal={Structural and Multidisciplinary Optimization},
  volume={46},
  number={2},
  pages={273--284},
  year={2012},
  publisher={Springer}
}

@article{honaryar2018design,
  title={Design of a bio-inspired hull shape for an {AUV} from hydrodynamic stability point of view through experiment and numerical analysis},
  author={Honaryar, Amir and Ghiasi, Mahmoud},
  journal={Journal of Bionic Engineering},
  volume={15},
  number={6},
  pages={950--959},
  year={2018},
  publisher={Springer}
}

@article{gafurov2015autonomous,
  title={Autonomous unmanned underwater vehicles development tendencies},
  author={Gafurov, Salimzhan A and Klochkov, Evgeniy V},
  journal={Procedia Engineering},
  volume={106},
  pages={141--148},
  year={2015},
  publisher={Elsevier}
}

@article{fattepur2024bio,
  title={Bio-inspired designs: Leveraging biological brilliance in mechanical engineering—An overview},
  author={Fattepur, Gururaj and Patil, Arun Y and Kumar, Piyush and Kumar, Anil and Hegde, Chandrashekhar and Siddhalingeshwar, IG and Kumar, Raman and Khan, TM Yunus},
  journal={3 Biotech},
  volume={14},
  number={12},
  pages={312},
  year={2024},
  publisher={Springer}
}

@article{danielson1969buckling,
  title={Buckling and initial postbuckling behavior of spheroidal shells under pressure.},
  author={Danielson, Donald Alfred},
  journal={AIAA journal},
  volume={7},
  number={5},
  pages={936--944},
  year={1969}
}

@article{smith2008buckling,
  title={Buckling of externally pressurized prolate ellipsoidal domes},
  author={Smith, P and B{\l}achut, J},
journal = {Journal of Pressure Vessel Technology},
  volume={1},
  number={130},
  pages={011210},
  year={2008},
  publisher={ASME}
}

@article{ma2008buckling,
  title={Buckling of super ellipsoidal shells under uniform pressure},
  author={Ma, YQ and Wang, CM and Ang, KK and Xiang, Yang},
  journal={The IES Journal Part A: Civil \& Structural Engineering},
  volume={1},
  number={3},
  pages={218--225},
  year={2008},
  publisher={Taylor \& Francis}
}

@article{lin2023parallel,
  title={Parallel multi-objective {B}ayesian optimization approaches based on multi-fidelity surrogate modeling},
  author={Lin, Quan and Hu, Jiexiang and Zhou, Qi},
  journal={Aerospace Science and Technology},
  volume={143},
  pages={108725},
  year={2023},
  publisher={Elsevier}
}

@article{serani2025survey,
  title={A survey on design-space dimensionality reduction methods for shape optimization},
  author={Serani, Andrea and Diez, Matteo},
  journal={Archives of Computational Methods in Engineering},
  volume={33},
  pages={1671–1698},
  year={2026},
  publisher={Springer},
  doi={10.1007/s11831-025-10349-x}
}

@article{serani2019adaptive,
  title={Adaptive multi-fidelity sampling for {CFD}-based optimisation via radial basis function metamodels},
  author={Serani, Andrea and Pellegrini, Riccardo and Wackers, Jeroen and Jeanson, C-E and Queutey, Patrick and Visonneau, Michel and Diez, Matteo},
  journal={International Journal of Computational Fluid Dynamics},
  volume={33},
  number={6-7},
  pages={237--255},
  year={2019},
  publisher={Taylor \& Francis}
}

@article{serani2024scoping,
  title={A scoping review on simulation-based design optimization in marine engineering: Trends, best practices, and gaps},
  author={Serani, Andrea and Scholcz, Thomas P and Vanzi, Valentina},
  journal={Archives of Computational Methods in Engineering},
  volume={31},
  pages={4709–4737},
  year={2024}
}

@article{pellegrini2023multi,
  title={A multi-fidelity active learning method for global design optimization problems with noisy evaluations},
  author={Pellegrini, Riccardo and Wackers, Jeroen and Broglia, Riccardo and Serani, Andrea and Visonneau, Michel and Diez, Matteo},
  journal={Engineering with Computers},
  volume={39},
  number={5},
  pages={3183--3206},
  year={2023},
  publisher={Springer}
}

@inproceedings{zitzler1999multiobjective,
  title={Multiobjective evolutionary algorithms: A comparative case study and the strength {Pareto} approach},
  author={Zitzler, Eckart and Thiele, Lothar},
  booktitle={Evolutionary Computation},
  year={1999}
}

@article{yang2019multi,
  title={Multi-objective Bayesian global optimization using expected hypervolume improvement gradient},
  author={Yang, Kaifeng and Emmerich, Michael and Deutz, Andr{\'e} and B{\"a}ck, Thomas},
  journal={Swarm and evolutionary computation},
  volume={44},
  pages={945--956},
  year={2019},
  publisher={Elsevier}
}

@article{gaggero2026physics,
  title={Physics-informed dimensionality reduction for propeller shape optimization},
  author={Gaggero, Stefano and Serani, Andrea},
  journal={Applied Ocean Research},
  volume={166},
  pages={104932},
  year={2026},
  publisher={Elsevier}
}

@inproceedings{jasak2007openfoam,
  title={Open{FOAM}: A {C}++ library for complex physics simulations},
  author={Jasak, Hrvoje and Jemcov, Aleksandar and Tukovic, Zeljko and others},
  booktitle={International workshop on coupled methods in numerical dynamics},
  volume={1000},
  pages={1--20},
  year={2007},
  address={Dubrovnik, Croatia}
}

@article{geuzaine2009gmsh,
  title={Gmsh: A 3-{D} finite element mesh generator with built-in pre-and post-processing facilities},
  author={Geuzaine, Christophe and Remacle, Jean Fran{\c{c}}ois},
  journal={International Journal for Numerical Methods in Engineering},
  volume={79},
  number={11},
  pages={1309--1331},
  year={2009},
  publisher={Wiley Online Library}
}

@misc{OpenCascade,
  author = {{Open Cascade}},
  title = {Open Cascade Technology: The Open Source Platform for {3D CAD, CAM, CAE}},
  howpublished = {\url{https://www.opencascade.com/}}
}

@inproceedings{abdolshah2018expected,
  title={Expected hypervolume improvement with constraints},
  author={Abdolshah, Majid and Shilton, Alistair and Rana, Santu and Gupta, Sunil and Venkatesh, Svetha},
  booktitle={2018 24th International Conference on Pattern Recognition (ICPR)},
  pages={3238--3243},
  year={2018},
  organization={IEEE}
}

@inproceedings{serani2025preliminary,
  title={Preliminary Design Optimization for Internal Arrangement and Hull Geometry of a Bio-Inspired Autonomous Underwater Glider through Machine Learning},
  author={Serani, Andrea and Palma, Giorgio and Wackers, Jeroen and Diez, Matteo},
  booktitle={MARINE 2025},
  year={2025}
}

@article{serani2022hull,
  title={Hull-form stochastic optimization via computational-cost reduction methods},
  author={Serani, Andrea and Stern, Frederick and Campana, Emilio F and Diez, Matteo},
  journal={Engineering with Computers},
  volume={38},
  number={Suppl 3},
  pages={2245--2269},
  year={2022},
  publisher={Springer}
}

@book{martins2021engineering,
  title     = {Engineering Design Optimization},
  author    = {Martins, Joaquim R. R. A. and Ning, Andrew},
  year      = {2021},
  publisher = {Cambridge University Press},
  address   = {Cambridge, UK}
}

@article{gray2013standard,
  title={Standard platform for benchmarking multidisciplinary design analysis and optimization architectures},
  author={Gray, Justin and Moore, Kenneth T and Hearn, Tristan A and Naylor, Bret A},
  journal={AIAA journal},
  volume={51},
  number={10},
  pages={2380--2394},
  year={2013},
  publisher={American Institute of Aeronautics and Astronautics}
}

@inproceedings{sobieszczanski2002bilevel,
  title={Bilevel integrated system synthesis {(BLISS)} for concurrent and distributed processing},
  author={Sobieszczanski, JS and Altus, TD and Philips, M and Sandusky, RJ},
  booktitle={9th AIAA/ISSMO symposium on multidisciplinary analysis and optimisation},
  year={2002}
}

@article{hernandez2023design,
  title={Design of a bioinspired underwater glider for oceanographic research},
  author={Hern{\'a}ndez-Jaramillo, Diana C and V{\'a}squez, Rafael E},
  journal={Biomimetics},
  volume={8},
  number={1},
  pages={80},
  year={2023},
  publisher={MDPI}
}

@article{liu2026large,
  title={A large manta ray-inspired bio-robotic platform for deep-sea exploration from laboratory to 2000-m depths},
  author={Liu, Qimeng and He, Qu and Li, Weikun and Guo, Pengming and You, Jie and Cui, Weicheng and Fan, Dixia},
  journal={Ocean Engineering},
  volume={347},
  pages={123772},
  year={2026},
  publisher={Elsevier}
}

@article{bellman1966dynamic,
  title={Dynamic programming},
  author={Bellman, Richard},
  journal={Science},
  volume={153},
  number={3731},
  pages={34--37},
  year={1966},
  publisher={American Association for the Advancement of Science}
}

@article{serani2025extending,
  title={Extending parametric model embedding with physical information for design-space dimensionality reduction in shape optimization},
  author={Serani, Andrea and Palma, Giorgio and Wackers, Jeroen and Quagliarella, Domenico and Gaggero, Stefano and Diez, Matteo},
  journal={Engineering with Computers},
  volume={41},
  pages={4643–4663},
  year={2025},
  publisher={Springer},
  doi={10.1007/s00366-025-02211-2},
}

@article{serani2023parametric,
  title={Parametric model embedding},
  author={Serani, Andrea and Diez, Matteo},
  journal={Computer Methods in Applied Mechanics and Engineering},
  volume={404},
  pages={115776},
  year={2023},
  publisher={Elsevier}
}

@inproceedings{d2017nonlinear,
  title={Nonlinear methods for design-space dimensionality reduction in shape optimization},
  author={D’Agostino, Danny and Serani, Andrea and Campana, Emilio F and Diez, Matteo},
  booktitle={International Workshop on Machine Learning, Optimization, and Big Data},
  pages={121--132},
  year={2017},
  organization={Springer}
}

@article{liu2025cnn,
  title={A {CNN-PINN-DRL} driven method for shape optimization of airfoils},
  author={Liu, Ying Yuan and Shen, Jian Xiong and Yang, Ping Ping and Yang, Xin Wen},
  journal={Engineering Applications of Computational Fluid Mechanics},
  volume={19},
  number={1},
  pages={2445144},
  year={2025},
  publisher={Taylor \& Francis}
}

@article{seo2024study,
  title={A study on ship hull form transformation using convolutional autoencoder},
  author={Seo, Jeongbeom and Kim, Dayeon and Lee, Inwon},
  journal={Journal of Computational Design and Engineering},
  volume={11},
  number={1},
  pages={34--48},
  year={2024},
  publisher={Oxford University Press}
}

@article{d2020design,
  title={Design-space assessment and dimensionality reduction: An off-line method for shape reparameterization in simulation-based optimization},
  author={D’Agostino, Danny and Serani, Andrea and Diez, Matteo},
  journal={Ocean Engineering},
  volume={197},
  pages={106852},
  year={2020},
  publisher={Elsevier}
}

@article{ccelik2021reduced,
  title={A reduced order data-driven method for resistance prediction and shape optimization of hull vane},
  author={{\c{C}}elik, Cihad and Dan{\i}{\c{s}}man, Devrim B{\"u}lent and Khan, Shahroz and Kaklis, Panagiotis},
  journal={Ocean Engineering},
  volume={235},
  pages={109406},
  year={2021},
  publisher={Elsevier}
}

@article{li2022low,
  title={Low-{Reynolds}-number airfoil design optimization using deep-learning-based tailored airfoil modes},
  author={Li, Jichao and Zhang, Mengqi and Tay, Chien Ming Jonathan and Liu, Ningyu and Cui, Yongdong and Chew, Siou Chye and Khoo, Boo Cheong},
  journal={Aerospace Science and Technology},
  volume={121},
  pages={107309},
  year={2022},
  publisher={Elsevier}
}

@article{difiore2024active,
  title={Active Learning and {B}ayesian Optimization: A Unified Perspective to Learn with a Goal},
  author={Di Fiore, F. and Nardelli, M. and Mainini, L},
  journal={Archives of Computational Methods in Engineering},
  volume={31},
  pages={2985–3013},
  year={2024},
  publisher={Springer}
}

@article{li2022machine,
  title={Machine learning in aerodynamic shape optimization},
  author={Li, Jichao and Du, Xiaosong and Martins, Joaquim R. R. A. },
  journal={Progress in Aerospace Sciences},
  volume={134},
  pages={100849},
  year={2022},
  publisher={Elsevier}
}

@article{li2021shape,
  title={Shape optimization for a conventional underwater glider to decrease average periodic resistance},
  author={Li, Jing and Wang, Xin and Wang, Peng and Dong, Hua and Chen, Cai},
  journal={China Ocean Engineering},
  volume={35},
  number={5},
  pages={724--735},
  year={2021},
  publisher={Springer}
}

@article{wang2025wing,
  title={Wing shape optimization for an air-launched underwater glider considering impact loads and gliding performance},
  author={Wang, Qiang and Wu, Xiangcheng and Zhang, Tianxiang and Xu, Yuxin},
  journal={International Journal of Naval Architecture and Ocean Engineering},
  pages={100683},
  year={2025},
  publisher={Elsevier}
}

@inproceedings{he2024hydrodynamic,
  title={Hydrodynamic Optimization Design of Underwater Glider Wings Based on Biomimetic Principles},
  author={He, Hong and Chen, Tianding and Wang, Hui and Chen, Ao and Yu, Xin and Fan, Yimei},
  booktitle={Proceedings of the 2024 4th International Conference on Computational Modeling, Simulation and Data Analysis},
  pages={1110--1117},
  year={2024}
}

@article{peherstorfer2018survey,
  title={Survey of multifidelity methods in uncertainty propagation, inference, and optimization},
  author={Peherstorfer, Benjamin and Willcox, Karen and Gunzburger, Max},
  journal={Siam Review},
  volume={60},
  number={3},
  pages={550--591},
  year={2018},
  publisher={SIAM}
}

@article{martins2013multidisciplinary,
  title={Multidisciplinary design optimization: a survey of architectures},
  author={Martins, Joaquim R. R. A. and Lambe, Andrew B.},
  journal={AIAA journal},
  volume={51},
  number={9},
  pages={2049--2075},
  year={2013},
  publisher={American Institute of Aeronautics and Astronautics}
}

@article{sherman2002autonomous,
  title={The autonomous underwater glider {Spray}},
  author={Sherman, Jeff and Davis, Russ E and Owens, WB and Valdes, J},
  journal={IEEE Journal of oceanic Engineering},
  volume={26},
  number={4},
  pages={437--446},
  year={2002},
  publisher={IEEE}
}

@article{eriksen2001seaglider,
  title={Seaglider: A long-range autonomous underwater vehicle for oceanographic research},
  author={Eriksen, Charles C and Osse, T James and Light, Russell D and Wen, Timothy and Lehman, Thomas W and Sabin, Peter L and Ballard, John W and Chiodi, Andrew M},
  journal={IEEE Journal of oceanic Engineering},
  volume={26},
  number={4},
  pages={424--436},
  year={2001},
  publisher={IEEE}
}

@article{petritoli2024autonomous,
  title={Autonomous underwater glider: A comprehensive review},
  author={Petritoli, Enrico and Leccese, Fabio},
  journal={Drones},
  volume={9},
  number={1},
  pages={21},
  year={2024},
  publisher={MDPI}
}

@book{crawley2015system,
  title     = {System Architecture: Strategy and Product Development for Complex Systems},
  author    = {Crawley, Edward and Cameron, Bruce and Selva, Daniel},
  year      = {2015},
  publisher = {Prentice Hall Press},
  address   = {Boston, MA},
  isbn = {9780133975345}
}

@misc{ioc-goos,
  author = {UNESCO},
  title = {Intergovernmental Oceanographic Commission website},
  note = {Last accessed 22 April 2022},
  url = {https://www.ioc.unesco.org/en/global-ocean-observing-system}
}

@article{volpi2015development,
  title={Development and validation of a dynamic metamodel based on stochastic radial basis functions and uncertainty quantification},
  author={Volpi, Silvia and Diez, Matteo and Gaul, Nicholas J and Song, Hyeongjin and Iemma, Umberto and Choi, KK and Campana, Emilio F and Stern, Frederick},
  journal={Structural and Multidisciplinary Optimization},
  volume={51},
  pages={347--368},
  year={2015},
  publisher={Springer}
}

@incollection{serani2014globally,
  title     = {Globally convergent hybridization of particle swarm optimization using line search-based derivative-free techniques},
  author    = {Serani, Andrea and Diez, Matteo and Campana, Emilio Fortunato and Fasano, Giovanni and Peri, Daniele and Iemma, Umberto},
  booktitle = {Recent Advances in Swarm Intelligence and Evolutionary Computation},
  pages     = {25--47},
  year      = {2014},
  publisher = {Springer},
  address   = {Berlin, Heidelberg}
}

@article{serani2016parameter,
  title={Parameter selection in synchronous and asynchronous deterministic particle swarm optimization for ship hydrodynamics problems},
  author={Serani, Andrea and Leotardi, Cecilia and Iemma, Umberto and Campana, Emilio F and Fasano, Giovanni and Diez, Matteo},
  journal={Applied Soft Computing},
  volume={49},
  pages={313--334},
  year={2016},
  publisher={Elsevier}
}

@article{perali2024performance,
  title={Performance prediction of a hydrofoil near the free surface using low {(BEM)} and high {(RANS)} fidelity methods},
  author={Perali, Paolo and Sacher, Matthieu and Leroux, Jean-Baptiste and Wackers, Jeroen and Augier, Beno{\^\i}t and Hauville, Fr{\'e}d{\'e}ric and Bot, Patrick},
  journal={Applied Ocean Research},
  volume={151},
  pages={104157},
  year={2024},
  publisher={Elsevier}
}

@Article{Fish2016Hydrodynamic,
AUTHOR = {Fish, Frank E. and Schreiber, Christian M. and Moored, Keith W. and Liu, Geng and Dong, Haibo and Bart-Smith, Hilary},
TITLE = {Hydrodynamic Performance of Aquatic Flapping: Efficiency of Underwater Flight in the Manta},
JOURNAL = {Aerospace},
VOLUME = {3},
YEAR = {2016},
NUMBER = {3},
pages={20}
}

@article{queutey2007interface,
  title={An interface capturing method for free-surface hydrodynamic flows},
  author={Queutey, Patrick and Visonneau, Michel},
  journal={Computers \& Fluids},
  volume={36},
  number={9},
  pages={1481--1510},
  year={2007},
  publisher={Elsevier}
}

@article{wackers2011free,
  title={Free-surface viscous flow solution methods for ship hydrodynamics},
  author={Wackers, Jeroen and Koren, Barry and Raven, Hoyte Christiaan and Van der Ploeg, A and Starke, A R and Deng, G B and Queutey, Patrick and Visonneau, Michel and Hino, Takanori and Ohashi, Kunihide},
  journal={Archives of Computational Methods in Engineering},
  volume={18},
  pages={1--41},
  year={2011},
  publisher={Springer}
}

@article{wackers2022adaptivegrid,
  title={Adaptive grid refinement for ship resistance computations},
  author={Wackers, Jeroen and Deng, Gan Bo and Raymond, Cl{\'e}mence and Guilmineau, Emmanuel and Leroyer, Alban and Queutey, Patrick and Visonneau, Michel},
  journal={Ocean Engineering},
  volume={250},
  pages={110969},
  year={2022},
  publisher={Elsevier}
}

@InProceedings{przulj:01,
  author = 	 {V. Pr\v{z}ulj and B. Basara},
  title = 	 {Bounded convection schemes for unstructured grids},
  booktitle = 	 {15th AIAA Computational Fluid Dynamics Conference},
  series = 	 {AIAA paper 2001-2593},
  year = 	 {2001},
  month = 	 {11-14 June},
  address = 	 {Anaheim, CA}
}

\end{document}